\documentclass[useAMS,usenatbib,usegraphicx]{mn2e}
\usepackage{fixltx2e}

\newcommand{\lya}{{\rm Ly}\alpha}

\newcommand{\hkpc}{h^{-1}{\rm kpc}}
\newcommand{\hmpc}{h^{-1}{\rm Mpc}}

\newcommand{\kms}{\;{\rm km}\,{\rm s}^{-1}}
\newcommand{\cms}{\;{\rm cm}^{-2}}

\newcommand{\Zsolar}{\;{\rm Z}_{\odot}}
\newcommand{\msolar}{\;{\rm M}_{\odot}}
\newcommand{\logrm}{{\rm log}}

\newcommand{\vw}{{v_{\rm wind}}}

\newcommand{\gad}{{\sc Gadget-2}}
\newcommand{\CI}{{\hbox{C\,{\sc i}}}}
\newcommand{\CII}{{\hbox{C\,{\sc ii}}}}
\newcommand{\CIII}{{\hbox{C\,{\sc iii}}}}
\newcommand{\CIV}{\hbox{C\,{\sc iv}}}
\newcommand{\CV}{\hbox{C\,{\sc v}}}
\newcommand{\SiI}{{\hbox{Si\,{\sc i}}}}
\newcommand{\SiII}{{\hbox{Si\,{\sc ii}}}}
\newcommand{\SiIII}{{\hbox{Si\,{\sc iii}}}}
\newcommand{\SiIV}{\hbox{Si\,{\sc iv}}}
\newcommand{\SiV}{\hbox{Si\,{\sc v}}}
\newcommand{\OI}{\hbox{O\,{\sc i}}}

\newcommand{\OVI}{\hbox{O\,{\sc vi}}}

\newcommand{\HI}{{\hbox{H\,{\sc i}}}}
\newcommand{\HII}{{\hbox{H\,{\sc ii}}}}
\newcommand{\HeII}{{\hbox{He\,{\sc ii}}}}

\title[Tracing the Reionization Era with Metal Lines ]{Tracing the Reionization-Epoch Intergalactic Medium with Metal Absorption Lines}

\author[B. D. Oppenheimer, R. Dav\'e, \& K. Finlator]{Benjamin
D. Oppenheimer$^1$, Romeel Dav\'e$^1$, Kristian Finlator$^1$
\\$^{1}$Astronomy Department, University of Arizona, Tucson, AZ 85721}

\begin{document}

\pagerange{\pageref{firstpage}--\pageref{lastpage}} \pubyear{2008}

\maketitle

\label{firstpage}

\begin{abstract}

  Intergalactic medium (IGM) metal absorption lines observed in $z\ga
  6$ spectra offer the opportunity to probe early feedback processes,
  the nature of enriching sources, and the topology of reionization.
  We run high-resolution cosmological simulations including galactic
  outflows to study the observability and physical properties of 5
  ions ($\CII$, $\CIV$, $\OI$, $\SiII$, $\SiIV$) in absorption between
  $z=8\rightarrow 5$.  We apply three cases for ionization conditions:
  Fully neutral, fully reionized, and a patchy model based on the flux
  from the nearest galaxy.  We find that our simulations can broadly
  fit available $z\sim 5-6$ IGM metal-line data, although all
  observations cannot be accommodated with a single ionization
  condition.  Variations in $\OI$ absorbers among sight lines seen by
  Becker et al. (2006) suggest significant neutral IGM patches down to
  $z\sim 6$.  Strong $\CIV$ absorbers at $z\sim6$ may be the result of
  ionization by the galaxy responsible for that enrichment, although
  the identification of the neighboring galaxy will have to wait to
  confirm this.  Our outflows have typical speeds of $\sim 200$~km/s
  and mass loading factors of $\sim 6$.  Such high mass loading is
  critical for enriching the IGM to the observed levels while
  sufficiently curtailing early star formation to match the observed
  rest-frame UV luminosity function.  The volume filling factor of
  metals increases during this epoch, but only reaches $\sim 1\%$ for
  $Z> 10^{-3} \Zsolar$ by $z=5$.  Detectable absorbers generally trace
  inhomogeneously-distributed metals residing outside of galactic
  halos.  $\CIV$ is an ideal tracer of IGM metals at $z\sim 5-6$, with
  dropping global ionization fractions to either higher or lower
  redshifts.  This results in a strongly increasing global $\CIV$ mass
  density ($\Omega(\CIV)$) from $z=8\rightarrow 5$, in contrast to its
  relative constancy from $z=5\rightarrow 2$.  Our simulations do not
  support widespread early IGM enrichment from e.g. Population III
  stars, as this would overpredict the numbers of weak $\CIV$
  absorbers in the latest data.  High-$z$ absorbers arise from metals
  mostly on their first outward journey, at distances 5-50 physical
  kpc, and often exhibit broad profiles ($\delta v>200 \kms$) as a
  result of outflowing peculiar velocities in the strongest systems.
  The galaxies responsible for early IGM enrichment have typical
  stellar masses of $10^{7.0-8.5}~\msolar$, and star formation rates
  $\la 1~\msolar/$yr.  Future facilities will be able to study the
  high-$z$ galaxy-absorber connection in detail, revealing a wealth of
  information about feedback processes in the reionization epoch.

\end{abstract}


\section{Introduction}  

The Universe undergoes its last major transition at $z\sim 6$ as the
first stars and early galaxies finish the process of reionizing the
previously neutral intergalactic medium (IGM).  Observations have only
recently cracked the $<1$ Gyr Universe with the discovery of galaxies
tracing early star formation \citep{bun04,dic04,yan04,bouw06}, quasars
providing ionizing photons \citep{fan01,fan03}, and the complete
absorption by the IGM of all Lyman-$\alpha$ photons, i.e.  the
\citet{gun65} trough, in $z>6$ quasar spectra \citep{fan02}.

A new type of high-$z$ observation is the recent detection of
metal-line absorbers in the most distant quasar spectra, probing the
nucleosynthetic products of early star formation.  \citet[][hereafter
BSRS]{bec06} surveyed 9 $z>5.5$ quasars and found 4 $\OI$ systems in
the most distant object in their sample, J1148+5251.  \citet{rya06},
\citet{sim06b}, \& \citet{rya09} discovered multiple $\CIV$ absorbers
at similar redshifts ($z\sim6$).  Despite the metals presumably being
formed within galaxies, these absorbers are most readily explained as
arising in the diffuse IGM.  But such absorbers are rare and perhaps
highly inhomogeneous: For instance, \citet{bec09} finds no $\CIV$
along four lines of sight (LOSs) observed at higher signal-to-noise
($S/N$) between $z=5.3-6.0$.

The simple fact that metals exist in the early IGM begs the question,
how did they get there?  Population~III (i.e. metal-free) and early
stars have been proposed to enrich a significant volume of the early
Universe, as early haloes harboring primordial galaxies have small
physical scales and shallow potential wells \citep[e.g.][]{mad01,
  gre07, wis08}.  The relatively invariant amount of $\CIV$ absorption
between $z\approx 2-5$ \citep{son01} and extended to $z\sim6$
\citep{rya06, sim06b} has been interpreted to mean that the majority
of IGM metals were injected at $z>6$~\citep[e.g.][]{sca02}.  But the
most recent data set from \citet{bec09} and \citet{rya09} shows a
decline of a factor of $\sim 4\times$ at $z>5.3$ compared to $z<4.7$,
suggesting that IGM metallicity is increasing at early epochs,
assuming typical ionization conditions do not evolve much.
Furthermore, $z\sim3$ $\CIV$ absorbers are spatially correlated with
Lyman Break Galaxies (LBGs), suggesting ongoing enrichment
\citep{ade03,ade05}, although these measurements are also consistent
with metals being injected by dwarf galaxies at $z=6-12$ coupled with
subsequent growth of clustering~\citep{por05}.

\citet[][hereafter OD06]{opp06} explored $\CIV$ absorbers in
cosmological hydrodynamic simulations, employing various heuristic
prescriptions for galactic outflows to enrich the $z\sim 2-6$ IGM.
They determined that prescriptions based on momentum-driven winds
were best able to reproduce $\CIV$ observations.  OD06 also showed
that an increasing ionization correction for $\CIV$, from both
cosmic evolution and energy injection from outflows, could mask a
comparable increase of $\sim \times 10$ in the true IGM metallicity
from $z\sim 6\rightarrow 2$.  Hence widespread early enrichment
from some exotic stellar population is not required to explain the
relatively invariant $\CIV$ absorption over this redshift range.

Another important question is, how do early galaxies relate to the IGM
metal-line absorbers at $z>6$?  \citet[][hereafter DFO06]{dav06}
explored reionization-epoch galaxies in the simulations of OD06,
finding that strong outflows are required to match the observed
luminosity function (LF) of $z\sim6$ $i$-band dropouts \citep{bouw06}.
\citet{fin07} further found that the star formation histories from
outflow simulations can reproduce broad-band spectral energy
distributions (SEDs) of galaxies such as the lensed $z\sim 6.7$ galaxy
Abell 2218 KESR \citep{kne04}.  The success of these models lies in
the high mass loading of early galactic superwinds, which pushes the
majority of high-$z$ metals nucleosynthesized in massive stars into
the IGM \citep[Oppenheimer \& Dav\'e 2008, hereafter OD08]{dav07}.
Typically, many times more mass is driven out of galaxies than forms
into stars.  Hence if our models are correct, observations of IGM
metal-line absorbers may be the best way to account for the majority
of metals in the $z\ga 6$ Universe.

Finally, what can IGM metal-line absorbers reveal about the topology
of reionization?  Fifth year data from the {\it Wilkinson Microwave
Anisotropy Probe} \citep[WMAP, ][]{hin08} indicate the Universe was
reionized at $z\sim 11\pm 1.4$ if it proceeded instantaneously.
Numerical simulations suggest that reionization is a protracted
process whereby individual sources create $\HII$ bubbles at $z>10$
that propagate outward until overlap by $z\ge6$ \citep[e.g.][]{gne00,
fur05b, ili06, lid07, fin09}.  Observations of $\HI$ absorption
are divided over when reionization ends, with \citet{fan06} showing
a discontinuity in the IGM ionization rate at $z\sim 6$ suggesting
individual ionization bubbles have finally finished overlapping,
while \citet{bec07} finds that the $\lya$ optical depth distribution
shows no evidence of a sudden end to reionization at this epoch.
Metal lines provide an additional handle on how reionization
progresses, allowing the bulk of cosmic volume to be probed in
absorption despite a fully saturated Gunn-Peterson trough.


In this paper we consider the above questions by studying $z=5-8$
metal-line absorbers in cosmological hydrodynamic simulations that
incorporate our favored momentum-driven wind model.  In this wind
model, the outflow speed and mass loss rate are tied to galaxy mass
using a heuristic prescription as expected for momentum-driven
winds~\citep{mur05}.  This wind model has enjoyed a number of
successes when compared to a wide range of observations at $z\la 6$;
most relevant is that it fairly uniquely matches observations of
$z\sim 2-5$ $\CIV$ absorbers~(OD06).  The scaling of wind speed with
galaxy velocity in this model follows the relation observed in local
starburst outflows \citep[e.g.][]{mar05a,rup05}.  We note that such
scalings may arise from other physical scenarios besides momentum (or
radiation) driven outflows~\citep[e.g.][]{dal08}; here we merely
employ the scalings, without reference to the underlying physical
driving mechanism.

We employ state-of-the-art \gad~cosmological hydrodynamic simulations
with a quarter billion particles run to $z=5$; we describe details in \S2.
The salient predicted galaxy and IGM properties are examined in \S3,
and compared to our findings in DFO06.  In \S4 we consider various cases
for the ionization background, including the extreme cases of fully
reionized and fully neutral, as well as a background with a variable
intensity intended to represent ionized bubbles around individual star
forming sources.  \S5 presents a plethora of simulated observations
which are compared to available $z>5$ metal-line observations, as well as
predictions for comparison with future data.  We then study the physical
and environmental parameters of simulated absorbers in \S6, with a
focus on how absorbers relate to galaxies and when they were injected
into the IGM.  \S7 provides a summary.  Throughout we use \citet{asp05}
abundances when calibrating metallicities relative to solar.

\section{Simulations}  

We employ our modified version of the N-body + Smoothed Particle
Hydrodynamics code \gad~\citep{spr05} to run three high-resolution
$2\times 512^3$ particle simulation to explore the high redshift
Universe.  Our simulations self-consistently enrich the IGM via
galactic outflows.  A complete description of our code, including our
wind model, can be found in \S2 of OD06 with further modifications
described in \S2 of OD08; however we also provide some of the
fundamental details involving star formation, enrichment, and feedback
here.

\subsection{Star Formation \& Chemical Production}

Star formation (SF) is modeled using a subgrid recipe where each gas
particle above a critical density is treated as a set of cold clouds
embedded in a warm ionized medium, similar to the interstellar medium
(ISM) of our own Galaxy \citet{spr03a}.  Gas particles eligible for SF
undergo instantaneous self-enrichment from Type II supernovae (SNe).
The instantaneous recycling approximation of Type II SNe energy to the
warm ISM phase self-regulates SF resulting in convergence in star
formation rates (SFRs) when looking at higher resolutions.  The SFR is
scaled to fit the disk-surface density-SFR observed in the local
Universe by \citet{ken98}\footnote{Even if this scaling is not
  applicable at $z>5$, the accretion rate determines the SF as we
  argue in \S\ref{sec:lumfunc}.}.

Star formation below 10 $M_{\odot}$, assumed not to result in Type II
SNe, is decoupled from their high mass counterparts using a Monte
Carlo algorithm that spawns star particles.  The metallicity of a star
particle remains fixed once formed; however, since Type II SNe
enrichment is continuous while stars are formed stochastically, every
star particle invariably has a non-zero metallicity.

\subsection{Chemical Yields}

The chemical yields at high redshift are dominated by Type II SNe, for
which we use the metallicity-dependent yields from \citet{chi04} as
described in OD08.  These yields vary less than 15\% for the 4 species
we track (carbon, oxygen, silicon, \& iron) as long as $Z>10^{-6}$,
which applies to effectively all SF in our cosmological simulation.
We do not consider exotic enrichment yields from zero metallicity, Pop
III, or very massive stars \citep[VMSs, e.g.][]{heg02, heg08}, and
assume a \citet{cha03} initial mass function (IMF) throughout.
Although it may be that such exotic stars have an effect on the IGM
particularly in terms of metal-line
ratios~\citep[e.g.][]{agu04,agu08}, the yields used here provide a
standard baseline for comparison.  The baseline yields are [O/C]=0.27
and [Si/O]=-0.16 if we consider the \citet{chi04} $Z=10^{-3}$ yields
assuming \citet{asp05} abundances; the carbon mass yield ejected by
SNe is $3.0\times10^{-3}$ the total stellar mass in a Chabrier IMF.
The integrated gas-phase abundance ratios of the $z=6$ d16n256vzw
simulation box are [O/C]=0.23 and [Si/O]=-0.13 reflecting the
\citet{chi04} yields inputted in our simulations.

To allow the reader to scale our results to a different set of
theoretical Type II SNe yields we provide Table \ref{table:yields}
listing the integrated yields assuming a \citet{cha03} IMF slope with
$\beta=0.18$, where $\beta$ is the stellar IMF mass fraction in the
range indicated in the second column\footnote{$\beta=0.18$ for a
  Chabrier IMF for stars $10-100 \msolar$, which we assume are the
  stars that undergo Type II SNe.}.  We have interpolated the
published yields to $Z=10^{-3}$ (i.e. $0.08 \Zsolar$) using their
various metallicity models, because this is near the median stellar
metallicity at $z=6$ ($Z_{stellar}(z=6)=1.48\times 10^{-3}$).  There
exists little spread for the oxygen and silicon yields between
\citet{chi04}, \citet{woo95}, and \citet{por98} despite different SNe
mass progenitor ranges.  38\% less carbon is produced from the
\citet{woo95} yield models, which is significant when we consider a
measurement such as $\Omega(\CIV)$ (\S\ref{sec:omegac4}).

\begin{table*}
\caption{Type II Supernovae Yields$^a$}
\begin{tabular}{lccccccc}
\hline
Reference &
Mass Range &
$Z$ & 
[C/H] &
[O/H] &
[Si/H] &
[O/C] &
[Si/O]
\\ 
\hline
\multicolumn {7}{c}{} \\
{\bf Chieffi \& Limongi (2004)}$^b$ & 13-35 $M_{\odot}$  & $10^{-3}$ &  0.13 &  0.40 &  0.24 &  0.27 &  -0.16 \\
Woosley \& Weaver (1995)  & 12-40 $M_{\odot}$  & $10^{-3}$ & -0.08 &  0.39 &  0.25 &  0.47 &  -0.14 \\
Portinari et al. (1998)   & 10-120 $M_{\odot}$ & $10^{-3}$ & -0.05 &  0.43 &  0.17 &  0.48 &  -0.26 \\ 
Portinari et al. (1998)   & 30-120 $M_{\odot}$ & $10^{-3}$ & -0.21 &  0.35 & -1.54 &  0.56 &  -1.89 \\ 
Heger \& Woosley (2008)   & 10-100 $M_{\odot}$ & $0$       &  0.02 &  0.20 & -0.29 &  0.18 &  -0.49 \\
d16n256vzw $z=6$ SPH      & --                 & --        &  --   &  --   &  --   &  0.23 &  -0.13 \\
\hline
\end{tabular}
\\ 
$^a$ Assuming a Chabrier IMF slope and $\beta=0.18$ (i.e. 18\% of
stellar IMF is covered by the mass range in column 2 and only this
mass range contributes to Type II SNe yields.). \\
$^b$ We apply this set of metallicity-dependent yields in our simulations.
\label{table:yields}
\end{table*}

We note that there may be reasons to favor a more top-heavy or
bottom-light IMF at high redshifts \citep[e.g.][i.e. $\beta$
increases]{dav08a}.  However, since our simulations are well-matched
to observed $z\sim 6$ rest-UV LFs (as we will show in \S3), and since metals
mostly come from the same massive stars responsible for UV emission,
our total metal production should be reasonably well-constrained.
Along these lines we show the yields using the $30-120 \msolar$ from
\citet{por98} in Table \ref{table:yields}; silicon appears to be
significantly under-produced in SNe ejecta.  For comparison we include
yields from \citet{heg08} from zero metallicity stars.  These yields
show less difference relative to \citet{chi04} than the top-heavy
\citet{por98} (except for oxygen), perhaps indicating just how hard it
is to clearly identify the nucleosynthetic signatures of Pop III
stars.

\subsection{Feedback Model}

The implementation of winds follows \citet{spr03a}, using a Monte
Carlo ejection mechanism governed by two parameters, the outflow
velocity $\vw$ and the mass loading factor $\eta$, where the mass
loading factor is defined as the outflow mass rate ($\dot{M}_{wind}$)
in units of the galaxy's SFR, i.e. $\dot{M}_{wind} = \eta \times$
SFR.  The wind model we employ here follows scalings of $\vw$ and
$\eta$ expected for momentum-driven winds; this fairly uniquely
matches a broad range of observations including IGM enrichment between
$z=6\rightarrow 1.5$ traced by $\CIV$ (OD06) and at $z<0.5$ traced by
$\OVI$ \citep{opp08b}, the LFs of $z=6$ galaxies (DFO06), the galaxy
mass-metallicity relations \citep{fin08a}, and the enrichment and
entropy levels in intragroup gas \citep{dav08b}.  The scalings depend
on galaxy velocity dispersion ($\sigma$):
 \begin{eqnarray}
  \vw &=& 3\sigma \sqrt{f_L-1}, \label{eqn: windspeed} \\
  \eta &=& {\sigma_0\over \sigma} \label{eqn: massload},
 \end{eqnarray} 
where $f_L$ is the luminosity factor, which is the luminosity of the
galaxy in units of the critical (or sometimes called Eddington)
luminosity of the galaxy, and $\sigma_0$ provides a normalization for
the mass loading factor.  Here we randomly select $f_L=[1.05,2]$ for
each SPH particle we kick, following observations of local
starbursts by \citet{rup05},
and we take $\sigma_0=150 \kms$.  $\sigma$ is estimated from the
galaxy's mass following \citet{mo98}, where the mass is obtained from an
on-the-fly friends-of-friends galaxy finder.

\subsection{Runs}

The simulations adopt the cosmological parameters based on the 5-year
WMAP results~\citep{hin08}, namely $\Omega_{0} = 0.25$,
$\Omega_{\Lambda} = 0.75$, $\Omega_b = 0.044$, $H_0 = 70$ km s$^{-1}$
Mpc$^{-1}$, $\sigma_8 = 0.83$, and $n = 0.95$.  The value of
$\sigma_8$ slightly exceeds the favored WMAP data alone, but agrees
better with the combined results including Type Ia SNe and baryonic
acoustic oscillation data.  A greater $\sigma_8$ results in
appreciably more SF at high-$z$, which is important since SF drives
the enriching outflows.  Our runs with this cosmology are named the
{\it d-series}; the primary run is named d16n512vzw using our standard
naming convention, indicating the box spans 16 comoving $\hmpc$s with
512$^3$ each of gas and dark matter particles.  The {\it vzw} suffix
indicates our implementation of the momentum-driven wind model as
described in OD08 (with one inconsequential exception: we no longer
impose an upper limit due to SN energy limitations on $\vw$).  The gas
particle mass is $5.4\times10^{5} \msolar$, which translates to a
minimum galaxy baryonic mass of $1.7\times10^{7} \msolar$ assuming a
galaxy is resolved with 32 particles (which we will demonstrate
later).  The softening length is set to 0.6 comoving $\hkpc$.

Two more simulations, d8n512vzw and a d32n512vzw, are also run to explore
numerical resolution convergence.  The respective softening lengths are
0.3 and 1.2 $\hkpc$, and the minimum galaxy baryonic mass resolutions
are $2.1\times10^{6}$ and $1.4\times10^{8} \msolar$.  We will show
that the primary enrichers of the IGM are galaxies with masses $\sim
10^{7-8.5}M_\odot$, which are statistically sampled best in the $16
\hmpc$ box.

The d16n512vzw simulation was run to $z=3$, however we only consider
outputs down to $z=5$ here.  This simulation was run on the Intel
64-bit Abe Cluster at the National Center for Supercomputing
Applications on 128 processors, requiring 32,500 processor hours.  The
d8n512vzw and d32n512vzw simulations were run only to $z=5.5$ and
$4.5$ respectively, each consuming approximately half the above
processor hours.  We run the Spline Kernel Interpolative DENMAX
(SKID)\footnote{http://www-hpcc.astro.washington.edu/tools/skid.html}
group finder on the $z=$8, 7, 6, and 5 outputs to identify galaxies.

\section{Physical Properties}

In this section we explore key physical and observational predictions
of our new simulations and relate them to findings from our previous
lower-resolution simulations.  This is intended to demonstrate
continuity with our previous works, as well as to provide a more
global context for our exploration of high-$z$ metal-line absorbers.
We begin by presenting the $z=6$ galaxy stellar mass ($M_*$) function
in our two simulation boxes, and relating this to earlier papers where
we fit observables best with a momentum-driven wind model.  We show
that the 16 $\hmpc$ box is the best one for studying high-$z$ IGM
metals, despite the fact that it is not large enough to sample the
space density of the most luminous $z\sim 6$ galaxies.  Finally, we
show global evolutionary trends in star formation, wind properties,
and metal enrichment.

\subsection{Galaxy Mass Function} \label{sec:galmass}

In DFO06 we explored the properties of reionization-epoch galaxies
($z=9\rightarrow 6$), finding that the {\it mzw} wind model most capable
of reproducing observations \citep[specifically][]{bouw06} compared to
the constant wind ({\it cw}) and no wind ({\it nw}) cases.  The {\it mzw}
model is similar to the {\it vzw} model, the only difference being that
the wind speed at a given $\sigma$ is constant in {\it mzw} while it has
a random spread in {\it vzw}.  The key point is that $\eta$ varies with
$\sigma$ the same way in both models, and as we showed in DFO06, $\eta$
governs high-$z$ SF by removing gas that would otherwise form into stars.
The constant wind model is that of \citet{spr03b}, with $\vw=484$~km/s
and $\eta=2$.

\begin{figure}
\includegraphics[scale=0.80]{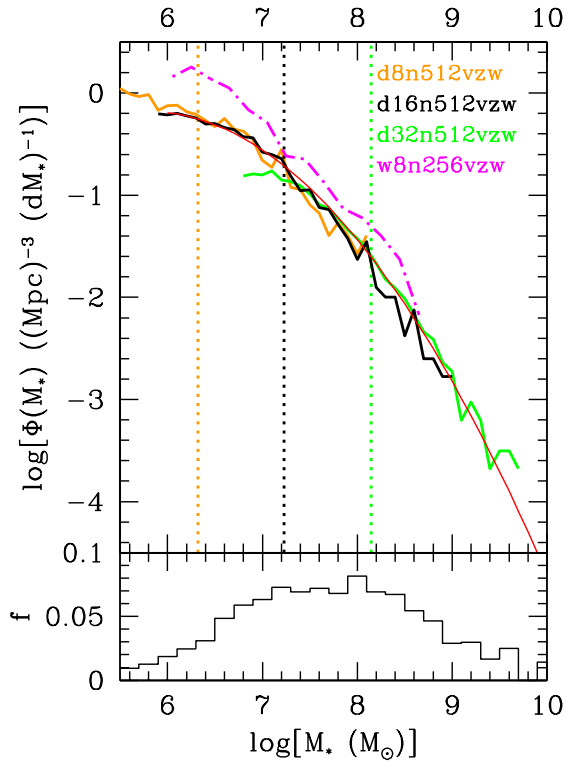}
\caption[]{The $z=6$ stellar mass function of galaxies in the
d8n512vzw (orange line), d16n512vzw (black line), and d32n512vzw
(green line) boxes in units of $\#/\logrm(dM_*)/{\rm Mpc}^{3}$ (comoving).
Corresponding dotted vertical lines show the galaxy mass resolution limit.
For comparison we show the mass function of our older w8n256vzw model
(dot-dashed magenta line, from DFO06), which is higher due to the different
cosmology.  The solid red line is a least-squares fit to our three
simulations (Equation \ref{eqn:M*func}).  We define the characteristic
$M_*$ as where the fit matches the slope of a $\Phi\propto M_*^{-1}$
power law, which is at $M_*=10^{7.7} \msolar$.  The bottom panel shows a
histogram of the stellar mass.  72\% of the mass is in galaxies with
$M_*=10^{6.7-8.7} \msolar$.  }
\label{fig:massfunc}
\end{figure}

We plot the stellar mass function of $z=6$ galaxies ($\Phi(M_*)$) in
Figure \ref{fig:massfunc} for the d8n512vzw (orange), d16n512vzw
(black), and d32n512vzw (green) simulations.  We compare it to the
w8n256vzw simulation (dot-dashed magenta), which is part of the
w-series simulations explored in DFO06.  The w8n256vzw simulation
has nearly the same resolution as the d16n512vzw simulation, but uses
the first-year WMAP results \citep{teg04}.  This cosmology
($\Omega_{0} = 0.3$, $\Omega_{\Lambda} = 0.7$, $\Omega_b = 0.04$, $H_0
= 70$ km s$^{-1}$ Mpc$^{-1}$, $\sigma_8 = 0.90$) produces almost
double the stars over $M_*=10^{7.0-8.5} \msolar$ owing to its higher
$\sigma_8$.  In the d16n512vzw simulation, we find 1.83\% of baryons
in galaxies and 0.13\% in stars; the respective numbers in the
w8n256vzw simulation are 3.10\% and 0.26\%.

We use $\Phi(M_*)$ to argue that we are resolving the galaxies
responsible for metal enrichment.  Stellar mass can be used as a
direct proxy for the integrated metal production if two conditions are
met: (i) All metals (those not trapped in stellar remnants) are
produced in short-lived, massive stars (i.e. Type II SNe), which is a
fair assumption given that longer lived stars (AGB stars) have not had
time to enrich much by $z\sim 5$; and (ii) stellar mass is a direct
proxy for the integrated number of Type II~SNe.  The stellar mass
contained in \gad~star particles consists only of long-lived, non-SN
stars, so the mass in star particles is a good record of the
integrated Type II SNe contribution.

We find that a two-dimensional polynomial fit to the logarithmic stellar
mass function provides a good, simple fit:
\begin{equation}\label{eqn:M*func}
\logrm[\Phi(M_*)] = -8.70 + 2.95 \logrm M_*  - 0.255 (\logrm M_*)^2.
\end{equation}
Shown as the red line in Figure \ref{fig:massfunc}, this fit obtains a
slope of $-1$ at $M_*=10^{7.7} \msolar$ , making this the
characteristic $M_*$ at $z\sim 6$ where the most mass per logarithmic
$M_*$ bin is held.  We cannot fit the mass function over 4 dex
effectively with a Schechter function, because the high mass end does
not display an exponential cutoff and the low-mass end shows a gradual
turnover owing to filtering by the cosmic UV
background~\citep[e.g.][]{tho96, ili07}.  Integrating Equation
\ref{eqn:M*func}, we find that 72\% of mass is in galaxies with
$M_*=10^{6.7-8.7} \msolar$.  To illustrate the stellar mass
distribution, the bottom panel of Figure \ref{fig:massfunc} shows the
fractional mass in each bin.  The median $M_*$ in terms of integrated
mass is $M_*=10^{7.6} \msolar$ for the d16n512vzw simulation, and
$M_*=10^{7.7} \msolar$ when including all three simulations.  The
d32n512vzw box is better able to account for the cosmic variance of
more massive galaxies, but its mass resolution is comparable to the
median stellar mass, so it cannot account for early enrichment from
small galaxies.  The d8n512vzw run, conversely, does not produce many
galaxies above $10^8 M_\odot$, so it does not sample the high-mass end
sufficiently.  The d16n512vzw run is well-suited to resolve the
relevant galaxy mass range, so we employ this run from here on in.



\subsection{Star Formation Rate Function} \label{sec:lumfunc}

A high-$z$ galaxy's SFR is more easily measured than its stellar mass,
since the former can be estimated from optical data while the latter
requires near-infrared data.  Our simulations predict a tight relation
between SFR and $M_*$ at $z=6$, where
\begin{equation}\label{eqn:SFR_M*}
{\rm SFR} = 9.2\; M_*^{0.94}~\msolar {\rm Gyr}^{-1}.
\end{equation}
This nearly linear relation is similarly tight in our simulations at
all redshifts \citep[DFO06, see their Figure~4; ][]{fin06, dav08a},
with a low scatter, and is mostly insensitive to outflow
model~\citep{dav08a}.  This relation arises from the dominance of
smooth, filamentary cold mode gas accretion, where the SFR is
regulated by the dynamical timescale for gas infall in slowly-growing
halo potentials \citep[e.g.][]{ker05}; therefore the SF law we scale
to \citep{ken98} has little effect on the SFR, chemical production,
and feedback.  Mergers are a sub-dominant process for fueling
galaxies, particularly at early times and in smaller
galaxies~\citep{ker08}, which is why the scatter is small.

\begin{figure}
\includegraphics[scale=0.80]{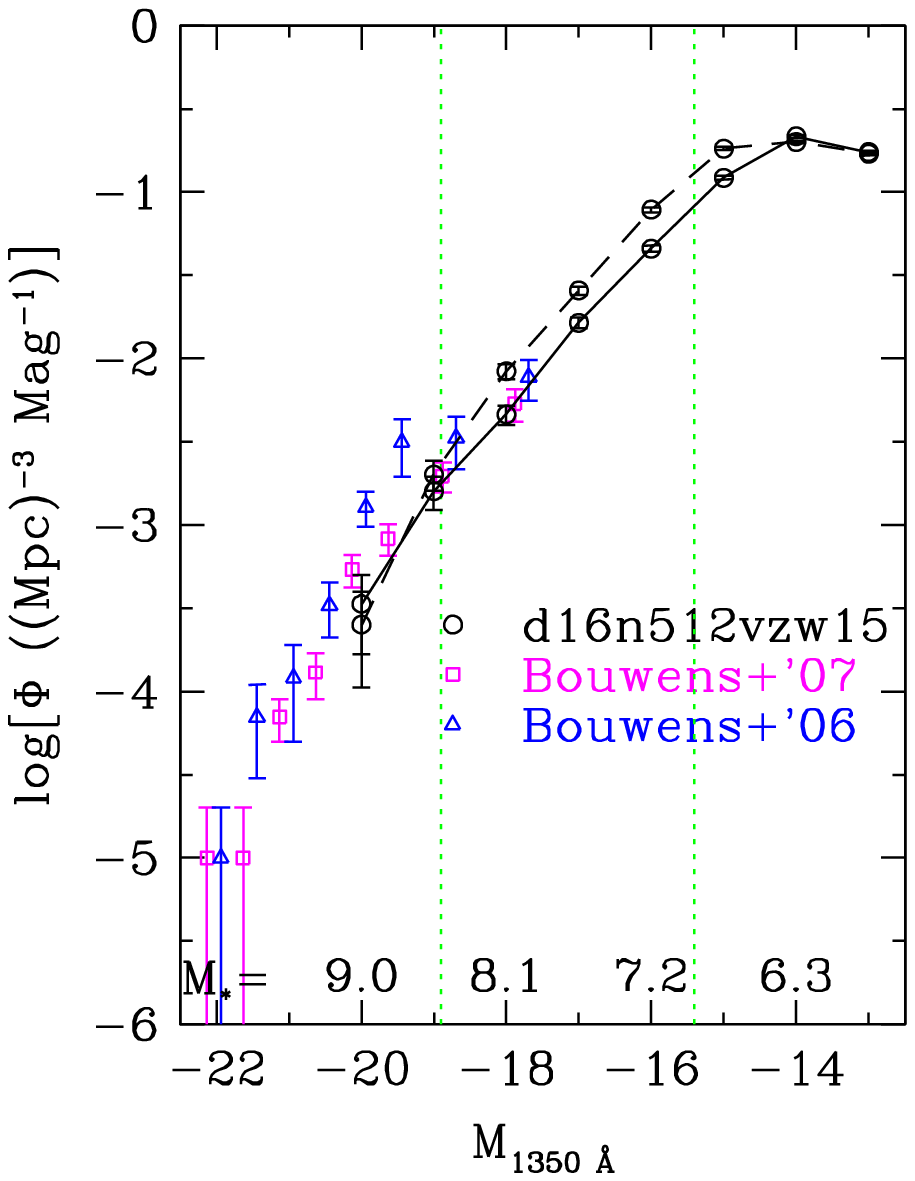}
\caption[]{The simulated $z=6$ rest-UV luminosity function
  (1350\AA~magnitudes AB) from our d16n512vzw run.  Solid line assumes
  uniform dust extinction of $E(B-V)=0.069$, dashed line uses the
  \citet{fin06} metallicity-dependent extinction.  These are compared
  to observations from \citet{bouw06} (blue triangles) and
  \citet{bouw07} (magenta squares).  The agreement is good in the
  overlap region.  The majority of SF remains below current detection
  limits, which implies that the sources of high-$z$ IGM metals have
  yet to be detected.  The range of $M_{1350 {\rm{\AA}}}$ for parent
  galaxies of $z=6$ $\CIV$ absorbers are demarcated by the dotted
  green lines.  Median logarithmic stellar masses at a given magnitude
  are listed along the bottom.}
\label{fig:lumfunc}
\end{figure}

In Figure \ref{fig:lumfunc} we show our d16n512vzw simulated
rest-frame 1350 \AA~UV luminosity function at $z=6$, which is
effectively a SFR function (including extinction).  We compare to data
from \citet{bouw06} and \citet{bouw07}.  We compute simulated galaxy
luminosities by passing their SF histories through the \citet{bru03}
population synthesis model, and applying a tophat filter around
1350\AA.  Dust extinction is critical in this comparison; we apply two
extinction laws to our simulated galaxies: (i) the solid line
represents a constant dust extinction, where $E(B-V)=0.069$, which
leads to 50\% flux attenuation at 1600 \AA~using the \citet{cal00} law
(we will use this value for our local ionizing field); and (ii) the
dashed line uses the metallicity-dependent extinction law calibrated
from local galaxies introduced in \citet{fin06} and employed in DFO06,
which predicts less attenuation for the lower mass galaxies.  In both
cases the agreement with data is reasonable.  This shows that the star
formation rates, and hence metal production rates, in our simulated
galaxies are plausible.

Numbers across the bottom of this figure show the median $\logrm[M_*]$
corresponding to a given rest-UV magnitude.  The characteristic $M_*$
from Equation~\ref{eqn:M*func} corresponds to $M_{1350
\rm{\AA}}=-17.1$, which is roughly one magnitude below current
detection limits.  The green dotted lines indicate the magnitude range
of the parent galaxies of $z=6$ $\CIV$ absorbers, which we identify in
\S\ref{sec:galabs}.  Hence the originating galaxies of currently
observed high-$z$ $\CIV$ systems are mostly below today's detection
thresholds.

\subsection{Cosmic Star Formation}

\begin{figure}
\includegraphics[scale=0.80]{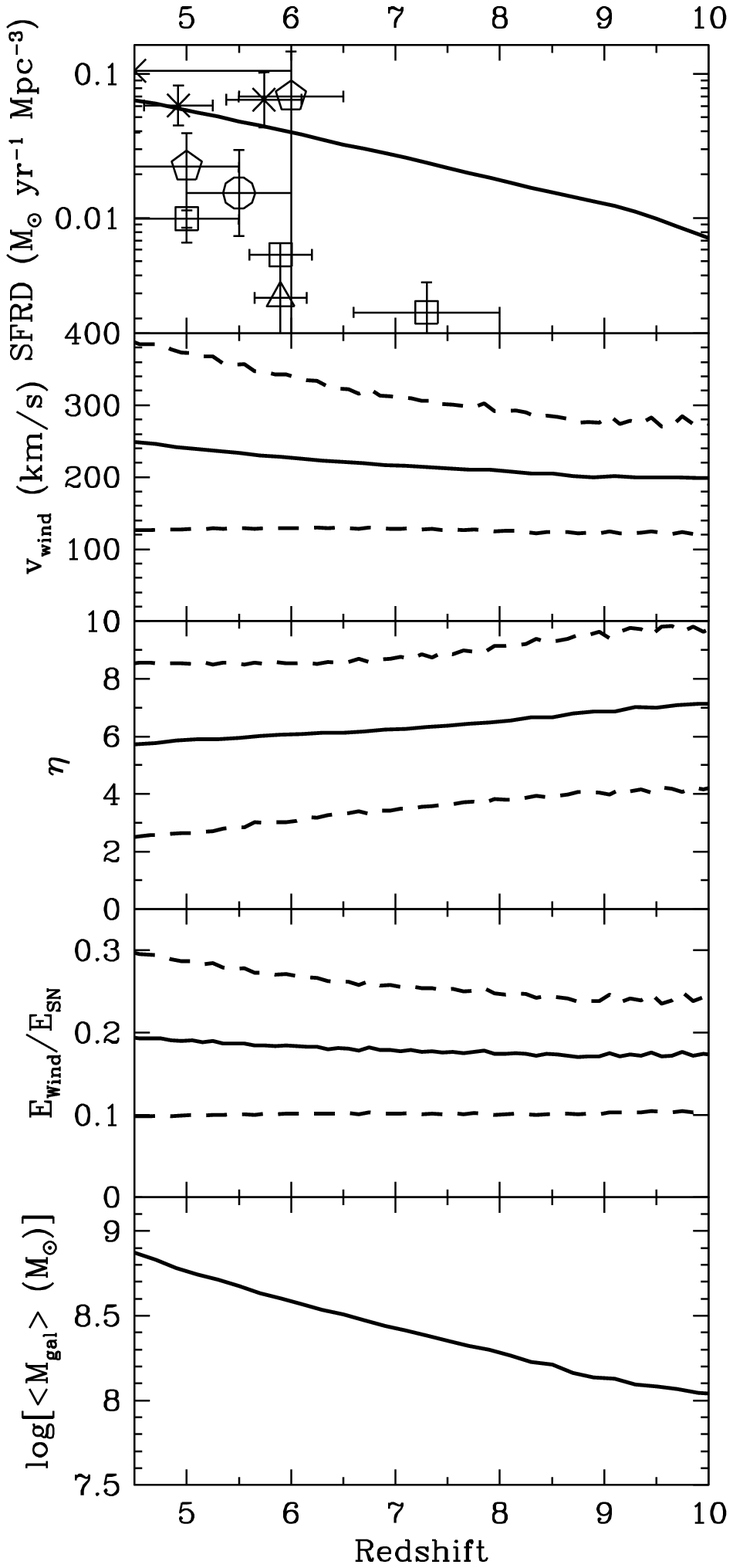}
\vskip -0.13in
\caption[]{{\it Top:} The cosmic SFRD from all galaxies in the
d16n512vzw simulation (solid line), compared to data assuming a
Chabrier IMF at high redshift.  Data are from the
\citet{hop04} compilation (X's), \citet{fon03} (circle), \citet{bun04}
(triangle), \citet{tho06} (pentagons), and \citet{bouw08} (squares).
Average wind properties (solid lines) in the next three panels show
typical $\vw=200-250 \kms$ with mass loading factors $\eta\sim
6-7$.  The energy in outflows is $\approx 20\%$ of the available
supernova energy at these redshifts, assuming a Chabrier IMF.  One
sigma dispersions are indicated by dashed lines.  The baryonic mass
of galaxies responsible for winds (bottom panel) increase 6$\times$
between $z=10\rightarrow 4.5$ (bottom panel).  }
\label{fig:sfr_wind}
\end{figure}

Figure \ref{fig:sfr_wind} (top panel) shows evolution of the cosmic
SFR density (SFRD) per comoving volume, indicating a rise of a factor
of three over the range $z=8\rightarrow 5$.  The values are in the
range of data, though the observations span a large range, in part
owing to differences in the adopted limiting luminosity in each case.
For example, \citet{bouw07} finds $6.7\times$ more SFRD when they
reduce the limiting luminosity by a factor $7.5\times$ for the z-band
dropouts ($\langle z \rangle = 7.4$).  Most of the $z=6$ stellar mass
and SF in our simulations remain below the detection limits of current
surveys.  At $z=6$, we find 31\% of the SFRD arises from galaxies with
$M_*>10^{8} \msolar$ , 39\% from $M_*=10^{7-8} \msolar$, and the
remaining 30\% from masses below this.  At $z=8$, 61\% of the SFRD
arises in galaxies with $M_*<10^{7} \msolar$.  We note that
\citet{bouw07} examined the evolution of the cosmic SFRD and found
that our momentum-driven wind model provided the best match to the
observed evolution from $z\sim 7\rightarrow 4$, of the models
considered there.


\subsection{Winds}

The adopted dependence of momentum-driven wind scalings from galaxy
mass results in the median wind speed growing and median mass loading
factor declining towards lower redshifts (OD08).  This has direct
ramifications for reionization-era galaxies, since SFRs are attenuated
by a factor of $(1+\eta)$ relative to the no-wind case.  This is
because $\frac{\eta}{1+\eta}$ of infalling gas that would otherwise be
converted into stars, instead is ejected in outflows.  This factor
assumes that outflows do not return to galaxies, which is a fair
assumption at high redshift because outflow material generally does
not recycle back into galaxies within a Hubble time at these epochs
(OD08).

The lower panels of Figure \ref{fig:sfr_wind} show that high-$z$ wind
properties have a relatively weak redshift dependence from
$z=10\rightarrow 5$: $\vw$ increases by 20\% (second panel) and $\eta$
declines by 17\% (third panel), despite an increase of 6$\times$ in
the average baryonic mass of a galaxy driving the winds (bottom panel,
i.e. weighted by $\dot{M}_{wind}$).  The reason is that $\sigma$ is
proportional to $M_{gal} \times H(z)$, and the declining $H(z)$ turns
out to roughly cancel the increase in $\langle M_{gal} \rangle$ during
this epoch, resulting in only modest evolution of wind properties.
Fiducial values during reionization in our model are therefore
$\vw\sim 200 \kms$ and $\eta\sim 6$.  The energy budget of momentum
driven winds remains roughly 20\% of the total SN energy assuming a
Chabrier IMF (fourth panel).  High-$z$ winds require substantial
energy input, but the wind speeds are relatively mild compared to
low-$z$ outflows from LIRGs and ULIRGs \citep[e.g.][]{mar05a, rup05}.
As before, we find that substantial mass ejection is a key feature of
our wind model: the amount of material ejected at these epochs is many
times that which forms into stars.

\subsection{Metallicity Distribution} \label{sec:Zdist}

\begin{figure}
\includegraphics[scale=0.80]{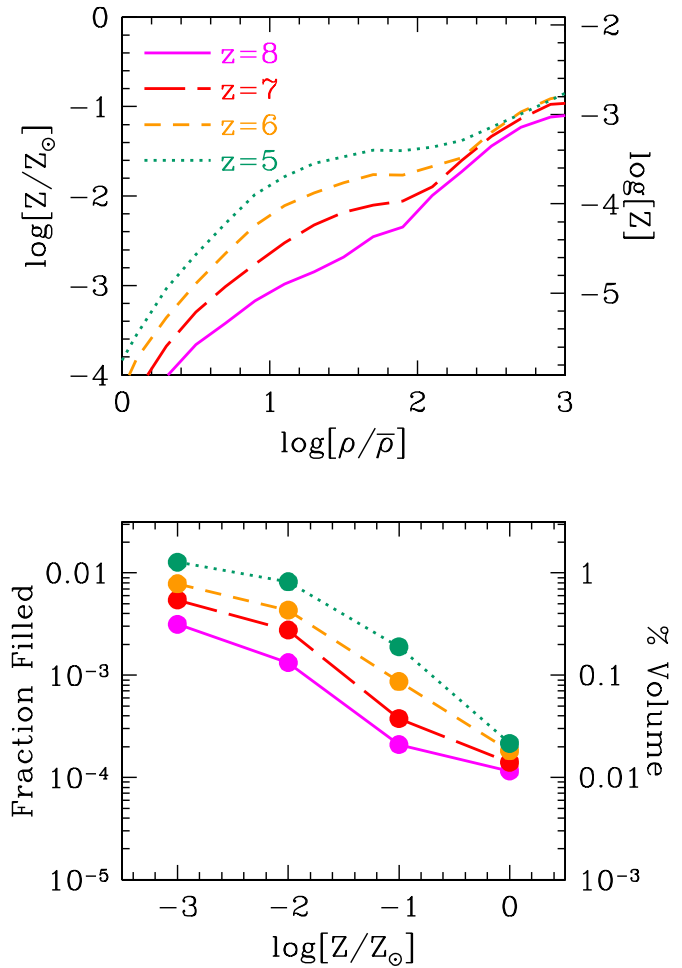}
\caption[]{{\it Top:} The average metallicity-density relationship at
$z=8,7,6,5$ from the d16n512vzw simulation.  $Z$ is the total mass
fraction of metals; $\Zsolar=0.0122$.  A clear gradient is established
early on, and as time passes the diffuse IGM is enriched more quickly
than high-density regions.  {\it Bottom:} The cumulative volume
filling factor of metals versus metallicity, indicating an
inhomogeneous distribution of metals with only 1\% of the Universe
enriched to $10^{-3} \Zsolar$ by $z=6$, despite some metals spreading
to near the cosmic mean density.  }
\label{fig:rho_Z}
\end{figure}

One reason why enriching the IGM is easier at high-$z$ is that the
physical scale of the Universe is small.  OD08 showed that in our wind
model, winds propagate a similar physical distance ($\sim 60-100$~kpc) for
all redshifts and galaxy masses; the comoving volume a galaxy can enrich
therefore goes as $(1+z)^3$.  Figure \ref{fig:rho_Z} illustrates the
rapid growth of metallicity at IGM overdensities from $z=8\rightarrow 5$.
The gradually rising $Z-\rho$ trend in place at all four redshifts is
one of the defining signatures of the momentum-driven wind enrichment.
In contrast, a wind model with constant $\vw$~\citep[e.g.][]{spr03b}
will create a flatter metallicity-density relationship with either
a sharp drop-off at lower overdensity (cf. Figure 10 of OD06) or a
rise in metallicity if superwinds are even stronger (cf. Figure 8 of
\citet{cen05}) signifying efficient enrichment of voids.  Under our
model the IGM metallicity grows between $z=8\rightarrow 5$ by a factor
of $7\times$ at an overdensity ($\delta\equiv\rho/\bar{\rho}-1$) of 80
and $15\times$ at $\delta=8$.

Despite some metals reaching cosmic mean densities, most of the
volume remains unenriched, especially the voids.  The cumulative
volume filling factors of metals in the bottom panel of Figure
\ref{fig:rho_Z} show this, as even regions with $Z>10^{-3} \Zsolar$
fill only about 1\% of the simulation volume at $z=5$.  Filling
factors increase only modestly to lower $z$ (OD08), which is somewhat
remarkable given that these models can match most available IGM
metal-line data.  Hence the presence of diffuse IGM metal absorbers
does not necessarily indicate widespread metal distribution.  We
will return to this point, and the inhomogeneous enrichment it
implies, later.




Regions of galactic overdensities ($\rho/\bar\rho\ga 10^3$) are
generally enriched rapidly to $Z\sim 0.1 \Zsolar$ or higher.  The
metallicity does not grow much at such overdensities, which is
reflected in the lack of predicted evolution in the mass-metallicity
relation from $z=8\rightarrow6$ in DFO08 (see their Figure 5).  The
fraction of cosmic volume enriched to $Z=\Zsolar$ is 0.01-0.02\%,
which is actually a factor of 1.5-3$\times$ {\it higher} than that
at $z=0.25$~\citep{opp08b}.  It is astounding to consider that the
volume fraction of the Universe enriched to around solar metallicity
may change very little between $z=8\rightarrow 0$; the key difference
is the corresponding regions at low redshift contain orders of
magnitude more baryons.  This is a result of our momentum-driven
wind model based on locally observed outflows.  \citet{dav07} showed
these winds efficiently enrich IGM overdensities at high-$z$, some
to $\ge \Zsolar$, while by $z=0$ most of these metals recycle back
into galactic environments and later winds often cannot escape out
of galactic haloes (OD08).

\begin{figure*}
\includegraphics[scale=0.66]{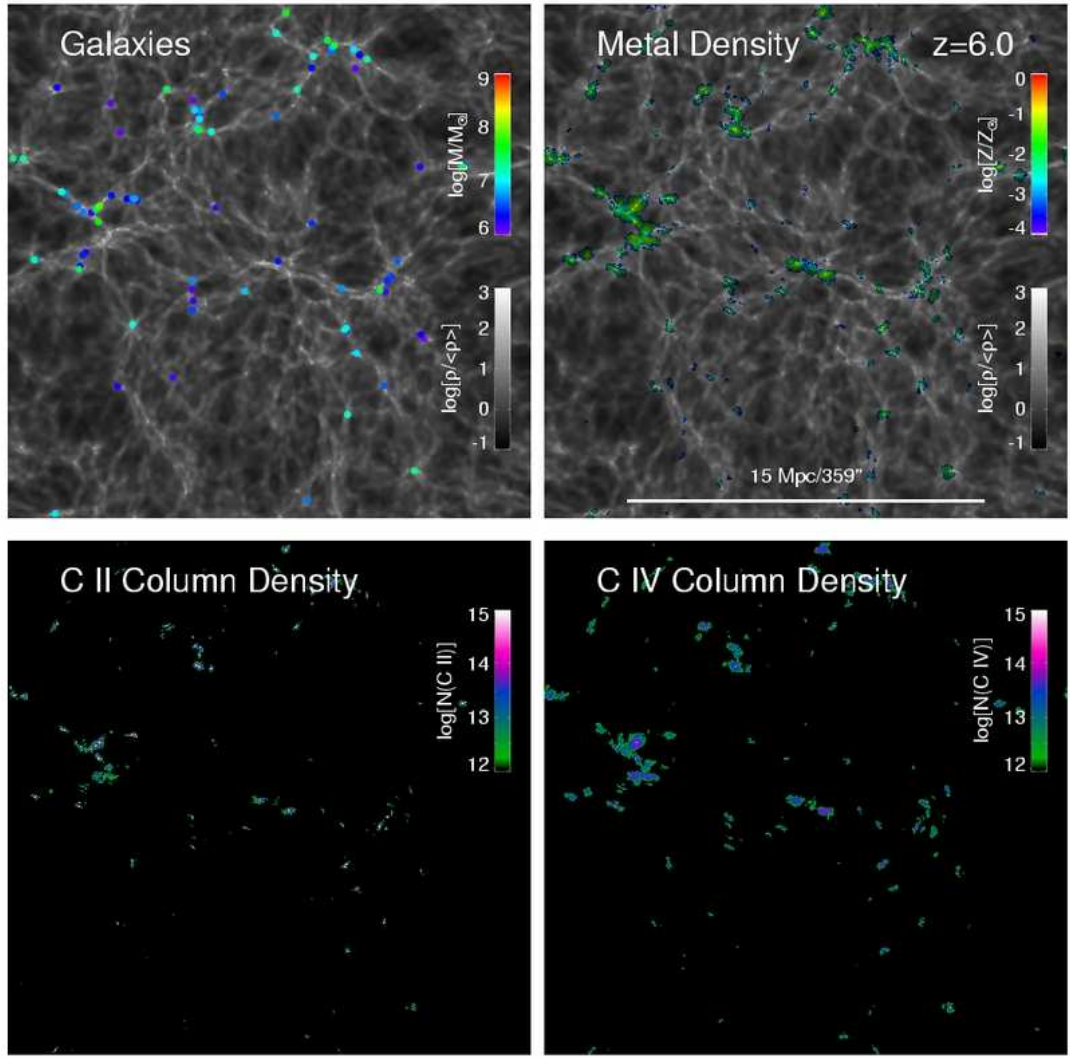}
\caption[]{A 25 $\kms$ slice of IGM spanning the 16 comoving $\hmpc$s
  of the d16n512vzw simulation at $z=6$.  {\it Upper left:} The
  locations of galaxies are shown as colored points, where the color
  corresponds to stellar mass, overlaid on the gas overdensity
  represented in greyscale.  {\it Upper right:} Color indicates the
  enrichment level; the vast majority of the IGM remains unenriched,
  and metals are close to galaxies.  The scale bar indicates the
  arcsecond scale on the sky using the \citet{wri06} on-line
  calculator.  {\it Bottom panels:} $\CII$ (left) and $\CIV$ (right)
  column densities in this 25~km/s slice, assuming a uniform
  \citet{haa01} ionization background. $\CII$ traces denser regions
  compared to $\CIV$, which traces the volume distribution of metals
  extremely well.}
\label{fig:snaps60}
\end{figure*}

Many of the trends noted above can be seen in Figure~\ref{fig:snaps60},
which shows the locations of galaxies in the upper left and the
metallicity-density distribution in the upper right in a 25
$\kms$-thick slice extending across the d16n512vzw simulation box.
Gray indicates zero metallicity in these frames, which corresponds
to the vast majority of the IGM, and shows the nascent cosmic web.
Most of the volume remains unenriched at these redshifts, while
regions within early galaxies become enriched up to a tenth solar
or more.  Clustering between early galaxies and enriched portions
of the IGM is strong.  This illustrates that the galaxy-absorber
connection at high-$z$ is extremely tight, as we investigate further
in \S\ref{sec:galabs}.

\section{Ionization Backgrounds} 

To make predictions for observable metal-line absorbers, we must
obtain ionization corrections for relevant metal ions.  This requires
that we have a model for the photo-ionization rate at each point in
the simulation volume.  Ideally, we would like to self-consistently
track the local ionization spectrum in these simulations.  But this
must await a full time-dependent, non-equilibrium radiative transfer
implementation, which we are in the process of developing
\citep{fin08b, fin09} but do not yet have.  Hence for this work, we
will consider three ad hoc cases for the ionization background (dubbed
{\it No Field, HM2001,} and {\it Bubble}) that span a plausible range.

Note that our simulations are originally run with a uniform
\citet{haa01} ionizing background, and we are only imposing these
backgrounds in post-processing.  This obviously does not properly
capture the dynamical effects of different pressure forces in our
three cases.  However, since dark matter is dynamically dominant
in the relatively low-density regions giving rise to IGM absorbers,
post-processing the background to obtain ionization corrections is
a good approximation.

\subsection{Ionization Cases} \label{sec:ioncase}

\noindent {\bf No Field:} Our first case assumes a fully neutral IGM,
where no ionizing background exists at wavelengths below the Lyman
limit, while lower energy photons stream freely.  This extreme case
may be applicable if metal injection precedes ionization fronts, or
more realistically if the IGM recombines in regions enriched and
ionized at earlier times.  Oxygen remains in its ground state due to
the $\OI$ ionization potential (13.618 eV) being barely higher than
the $\HI$ ionization potential, while $\CI$ (11.260 eV) and $\SiI$
(8.151 eV) are completely ionized to the second level.\\


\noindent {\bf HM2001:} Our second case assumes reionization is
complete and a spatially-uniform background applies everywhere.  We
use the \citet{haa01} background assuming a contribution from quasars
and 10\% of photons above the Lyman limit from star forming galaxies
generated by their Cosmic Ultraviolet Background (CUBA) program;
at $z>6$, star-forming galaxies dominate the ionizing flux.  We
divide the total flux by a factor of 1.6 as we did in OD06 in order
to match the summed flux decrement of the $\lya$ forest down to
$z=2$.  An assumption of a spatially-uniform ionization background
is probably far from correct at $z>6$, but this is intended to
represent another extreme case like {\it No Field}.  Ionization
fractions for metal lines are calculated using CLOUDY-generated
\citep{fer98} ionization tables as a function of density and
temperature, as described in OD06.\\

\noindent {\bf Bubble:} The last case assumes that the ionizing flux of
the nearest galaxy dominates, which may be the most realistic scenario
during reionization.  This is applicable if metals remain near enough
to their parent galaxies to be dominated by the local ionizing intensity,
which we will show is often true since metals typically reside within
30~kpc of the originating galaxy.  Of course, detailed radiative transfer
effects of escaping ionization and subsequent recombinations will play
a critical role, so this scenario should be considered a crude proxy for
a patchy reionization model.

To implement the {\it Bubble} field, we consider the ionizing field
from only the nearest galaxy, weighted by the inverse square law.  We
assume an average SED per stellar mass by taking the average
\citet{bru03} stellar synthesis spectrum of {\it all} the stars in the
simulation box for a given redshift, accounting for their ages and
metallicities.  We then calculate the impinging flux using the stellar
mass and distance of the nearest galaxy, which is identified for each
SPH particle.  Note that the same ionizing flux per unit stellar mass
is assumed to emanate from each galaxy at a given epoch; we do not
account for variations in nearest galaxy SEDs, which is another
approximation required to make the problem computationally tractable.


We attenuate the emission from galaxies by dust extinction.  The
assumed extinction has a fixed value of $E(B-V)=0.069$, as shown in
Figure~\ref{fig:lumfunc} to broadly match the observed rest-UV LF.  To
calculate how the emergent SED is attenuated, we use the theoretical
dust absorption curve from the X-ray to the far-IR of \citet{wei01},
using their grain size distribution of silicate and carbonaceous
grains found to fit the Small Magellanic Cloud (SMC) bar.  We use this
curve because $z\sim 3$ LBGs exhibit the smaller grain size
distributions of SMC-like dust \citep{vij03}.  Note that we cannot
directly apply the empirical \citet{cal00} extinction law because it
extends only down to 1200 \AA, while we are especially interested in
the attenuation at energies where $\CIV$ and $\SiIV$ are ionized
(i.e. $<400$ \AA).  For each SPH particle, the impinging flux is then
the dust-attenuated emission divided by the distance from the nearest
galaxy squared.  We then use CLOUDY to generate the resulting metal
ionization fractions, which is then used to obtain the optical depth
for any ion along the LOS as described in OD06.

\begin{figure}
\includegraphics[scale=0.90]{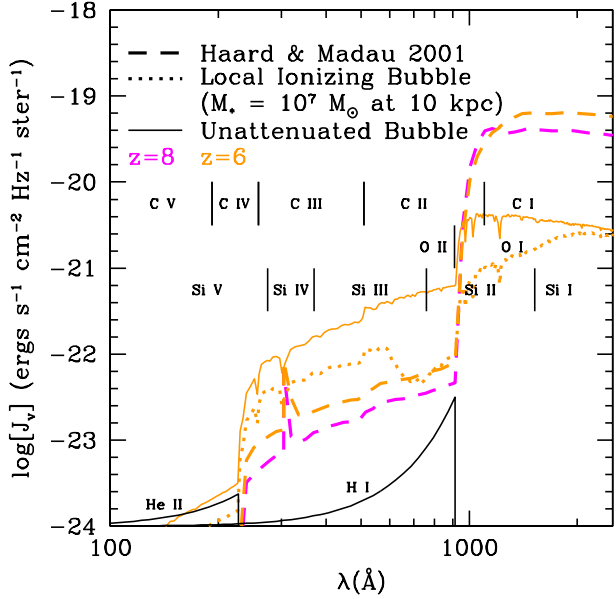}
\caption[]{Ionizing background spectra: The {\it HM2001} (dashed)
ionization background at $z=8$ (magenta) and $z=6$ (orange) are
compared to the {\it Bubble} field arising from a $z=6$ $M_* = 10^7
\msolar$ galaxy at 10 kpc.  The thin solid orange line represents the
unattenuated \citet{bru03}-computed SED at $z=8$ to compare with the
dust-attenuated (thick dotted orange line) SED.  For reference,
ionization potentials for carbon, oxygen, and silicon are indicated,
and the relative cross-sections from $\HI$ and $\HeII$ as a function
of wavelength are shown along the bottom.  }
\label{fig:ionbkgd}
\end{figure}

We plot the {\it HM2001} field at $z=8$ and 6 in Figure
\ref{fig:ionbkgd}, along with the {\it Bubble} field, where we assume
a $M_* = 10^7 M_{\odot}$ galaxy at 10 kpc (typical values) for
concreteness.  The difference between the solid and dotted orange
lines for the {\it Bubble} case at $z=6$ illustrates the effects of
dust attenuation.  Dust absorption in all of the \citet{wei01} models
sharply peaks at 600-900\AA, causing a dip.  There is a clear Lyman
break, which in our cases arises mostly from stellar atmospheres, and
secondarily from dust.  We note that this break is not nearly as
strong as that observed by \citet{sha06} in Lyman break galaxies at
$z\sim 3$, perhaps indicating that there is a fairly small escape
fraction of ionizing photons; for instance, \citet{haa01} assume a
global 10\% escape fraction.  However, because these data are from
lower-$z$ and higher-$M_*$ galaxies, it may not be applicable here.
For simplicity, we do not assume an overall escape fraction.


Comparing the {\it Bubble} and {\it HM2001} fields at $z=6$, the
strength of the Lyman break is much less in the {\it Bubble} case.
This is mostly because our typical galaxy is younger and more
metal-poor than that assumed by \citet{haa01} ($0.2 \Zsolar$, 0.5
Gyrs), which leads to a harder ionizing source.  Two secondary reasons
for greater hardness are: (i) the {\it Bubble} Lyman-break attenuation
due to dust is less than the {\it HM2001} attenuation applied ($\sim
3$ vs. an overall 10\% escape fraction); and (ii) the dust attenuation
declines at the ionization potentials of $\CIV$ and $\SiIV$ (shown).
This leads to greater ionization fractions at these higher states in
the {\it Bubble} field, which has observable implications as we will
show throughout \S\ref{sec:observe}.


\begin{figure*}
\includegraphics[scale=0.90]{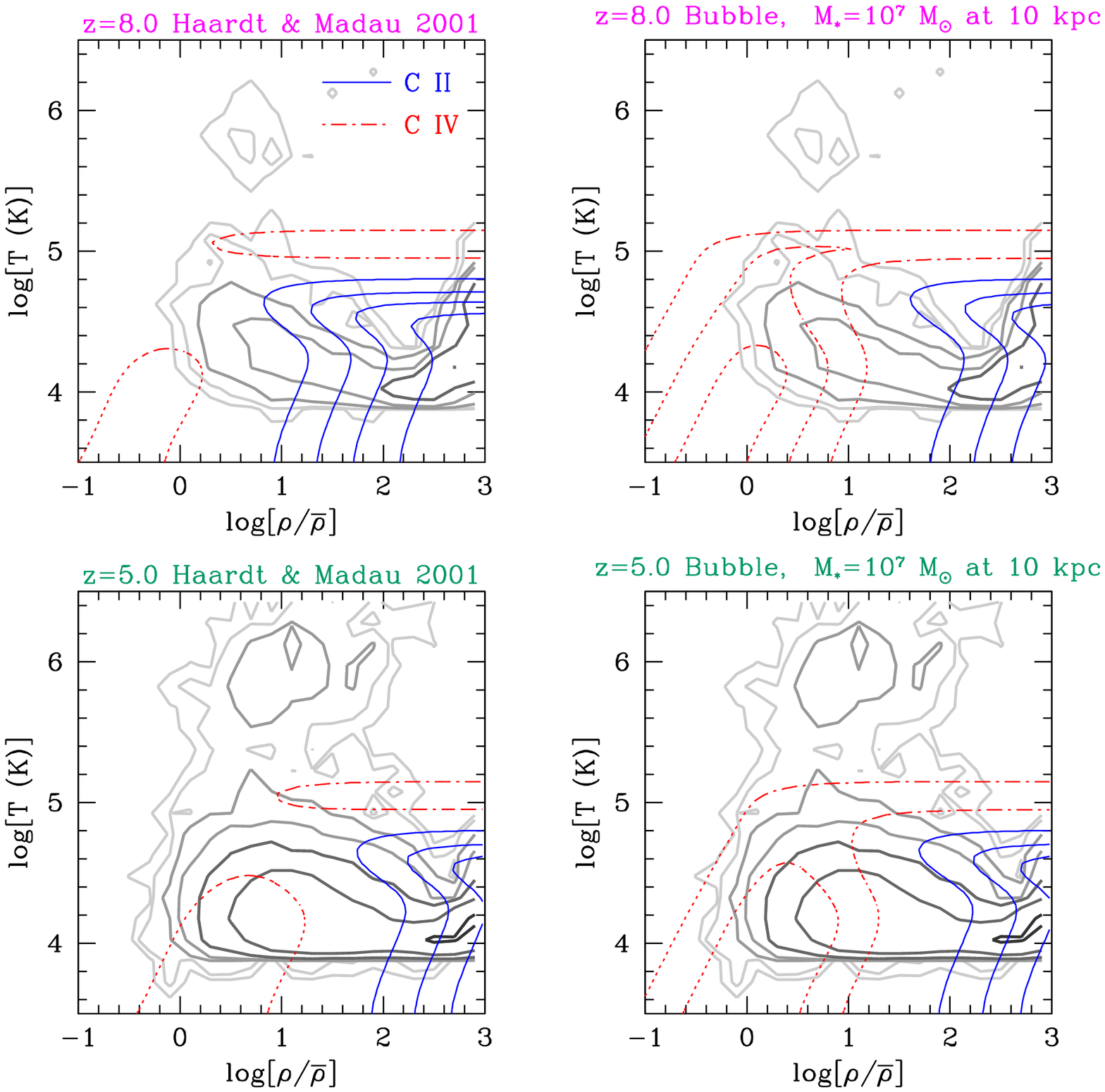}
\caption[]{Metal phase space plots: Colored contours (solid blue for
  $\CII$ and red dot-dashed for $\CIV$) correspond to the ionization
  fractions, while thick gray contours indicate the metal distribution
  in phase space.  Overlapping regions give rise to absorption in that
  ion.  The left panels assume the uniform \citet{haa01} background,
  and the right panels use the local ionizing {\it Bubble} field
  arising 10 kpc from a $10^7 \msolar$ galaxy; the contour spacings
  correspond to ionization fraction steps of 20\% (i.e. $f(\CIV)=$0.2,
  0.4, etc.).  The gray metallicity contours are in logarithmic steps
  of 0.5 dex.  At $z=8$ (top panels), $\CII$ should trace metals
  better than $\CIV$ for the {\it HM2001} field, while by $z=5$
  (bottom panels) $\CIV$ should become a better tracer of metals in
  the diffuse IGM.  The {\it Bubble} contours depends on galaxy size
  and distance: Contours move right as $M_*$ increases and left as
  $d_{gal}$ increases (i.e. $U\propto M_* d_{gal}^{-2}$).  $\CIV$ is
  always a good tracer of IGM metals from $z=8\rightarrow 5$ in the
  {\it Bubble} case.  }
\label{fig:ioncomp}
\end{figure*}

\subsection{Ionization Fraction Behavior}\label{sec:ionbehave}

Before delving into the observables, we present here a primer for
understanding metal-line observations resulting from our two assumed
ionization backgrounds, the {\it HM2001} and the {\it Bubble} fields.
The {\it No Field} case is straightforward (i.e. ionization fractions
of unity for $\CII$, $\OI$, and $\SiII$), but the often nuanced
behavior of the other two fields has some perhaps unexpected
consequences.  We consider metals in $\rho-T$ phase space exploring
both low and high-ionization species at two redshifts that span the
redshift range we consider.

Figure \ref{fig:ioncomp}~plots colored contours corresponding to the
{\it HM2001} (left panels) and the {\it Bubble} (right panels)
backgrounds in $\rho-T$ phase space for $\CII$ (solid blue) and $\CIV$
(dot-dashed red) at $z=8$ and $z=5$ (upper and lower panels,
respectively).  Each contour step represents an increase of 20\% in
the ionization fraction for the corresponding ion.  Overlaid are thick
gray logarithmic metal density contours (darker contours indicate
higher metal densities) at the respective redshifts.  Overlapping
color and gray contours represent regions of phase space where
absorption in that ion will arise for a given ionization field.

The {\it HM2001} contours are the same at all spatial locations within
the simulation box at a given redshift, so the resulting behavior is
simpler to understand.  The corresponding contours for both $\CII$ and
$\CIV$ move to higher overdensity by almost a factor of 10 from
$z=8\rightarrow 5$, owing to the ionization parameter $U$ increasing.
$U$ is defined in this paper as the number density of photons capable
of ionizing to a particular state (e.g. ionizing $\CIII$ to $\CIV$)
divided by the number density of a particular atomic species (carbon
in this case).  $U$ increases $\sim\times 10$ because the physical
densities decline by a factor of 3.4 from Hubble expansion, and the
ionization intensity increases by $\sim\times 3$ between
$z=8\rightarrow 5$.  $\CII$ evolves to trace a smaller fraction of
phase space at higher densities; however the metal density is greater
at these higher densities and at later times, so the total amount of
$\CII$ is expected to be fairly stable from $z=8\rightarrow 5$.  The
contours extending to higher densities at $T\sim10^{5}$ K in Figure
\ref{fig:ioncomp} correspond to $\CIV$ in collisional ionized
equilibrium (CIE).  This overlaps few metals, because $\CIV$ is an
efficient coolant and SPH particles evolve rapidly through this
region, much like we find for $\OVI$ at $z<0.5$ \citep{opp08b}.

\begin{figure}
\includegraphics[scale=0.90]{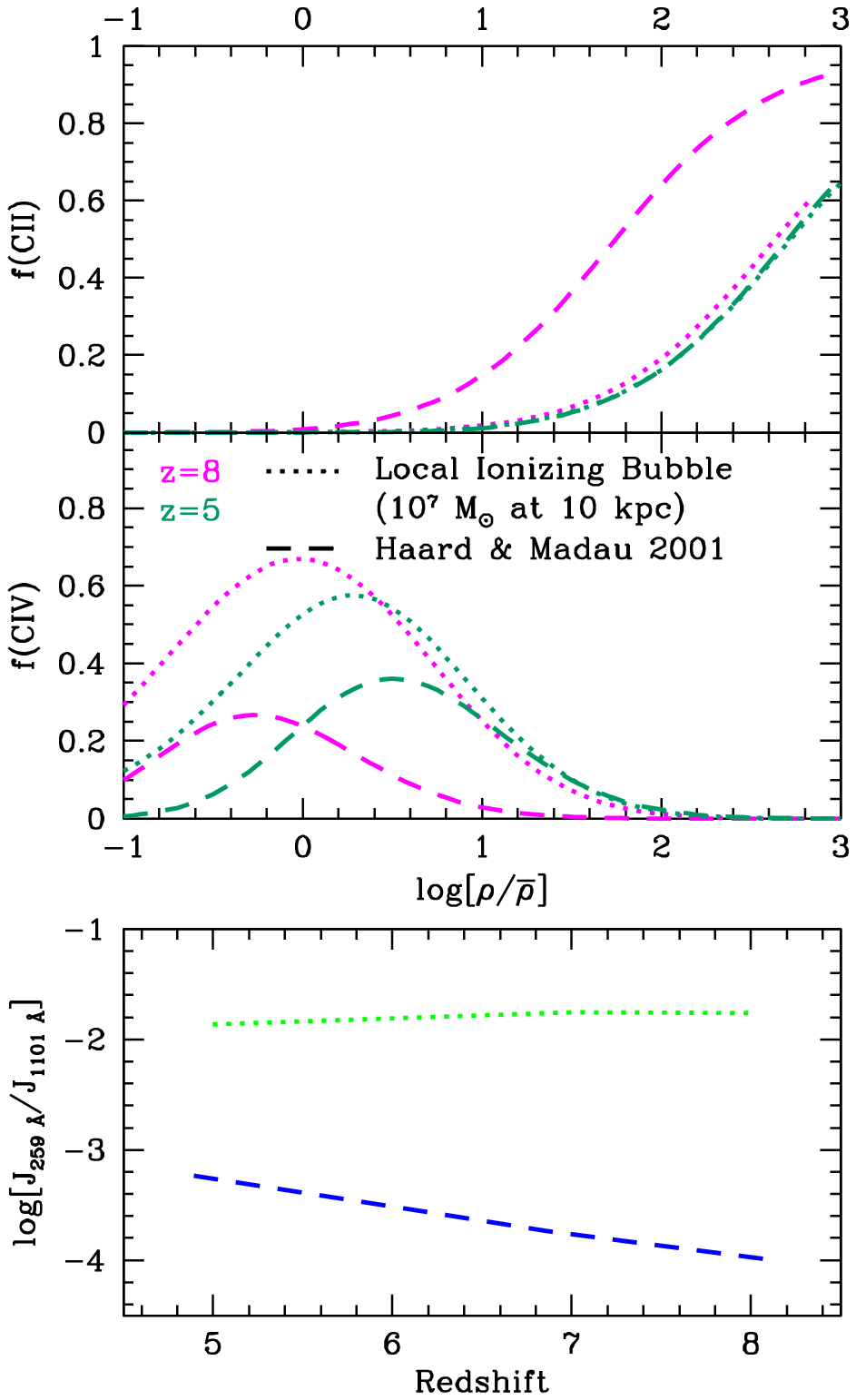}
\caption[]{{\it Top panels:} $\CII$ (upper) and $\CIV$ (lower)
ionization fractions as a function of overdensity at $T=10^4$ K for
the {\it HM2001} (dashed lines) and {\it Bubble} (dotted lines) at
$z=8$ (magenta) and $z=5$ (green).  The {\it HM2001} field evolves
more as quasars provide increasing intensity above 47.9 eV at later
times.  The {\it Bubble} field shows less evolution, and generally
$\CIV$ traces lower overdensities.  {\it Lower panel:} The mean
carbon spectral ratio $\equiv J_{259 {\rm \AA}}/J_{1101 {\rm \AA}}$
for our two ionization fields as a function of redshift.  The star
forming galaxy-dominated {\it Bubble} field is significantly harder
than the {\it HM2001} field with quasars, because {\it HM2001} is
dominated by galaxies at high-$z$ and the attenuation above 47.9
eV is greater than the {\it Bubble} attenuation due to dust.  This
results in overall lower values for $f(\CIV)$ in the {\it HM2001}
case.
}
\label{fig:rho_ionf}
\end{figure}

As most IGM metals remain near $T=10^{4}$ K, we plot the $\CII$/C
and $\CIV$/C ionization fractions ($f(\CII)$, $f(\CIV)$) at this
temperature in Figure \ref{fig:rho_ionf} (top panels), showing the
{\it HM2001} curves moving upward in overdensity by the factor of
10 from $z=8\rightarrow 5$.  $\CIV$ should dramatically increase
as the $z=8$ field is too weak to ionize $\CIV$, while by $z=5$ the
gray and $\CIV$ contours move to overlap as more metals enrich lower
overdensities while $U$ increases.  A second effect illustrated in
Figure \ref{fig:rho_ionf} is that the maximum $f(\CIV)$ grows as a
result of the harder background with more quasar contribution at
$z=5$ relative to $z=8$; more photons exist at the ionization
potential of $\CIII$ (47.9 eV or 259 \AA) to create $\CIV$ relative
to the $\CI$ potential (11.3 eV or 1101 \AA) to make $\CII$, resulting
in a larger density range where $\CIV$ resides.  We define and plot
in Figure \ref{fig:rho_ionf} (bottom panel) the ratio $J_{259 {\rm
\AA}}/J_{1101 {\rm \AA}}$, which we call the ``carbon spectral
ratio''.  This has significant implications for $\CII$ and $\CIV$
observations.  The carbon spectral ratio is $200\times$ greater for
the {\it Bubble} versus the {\it HM2001} field at $z=8$.

We now focus on the {\it Bubble} field behavior.  The example fields
signified by the right panels in Figure \ref{fig:ioncomp} demonstrate
the expected ionization in $\rho-T$ phase space at 10 physical kpc
from a $10^7 \msolar$ galaxy.  These contours shift right as $M_*$
increases and left as $d_{gal}$ increases due to the inverse square
law; e.g. the contours are equivalent to the ionization field from a
$10^9 \msolar$ galaxy at 100 kpc.  We choose this field strength
because this is the representative of where our simulated metals
typically reside.  Like the {\it HM2001} field, our implementation of
the {\it Bubble} field depends on only two variables: $U$ and $T$.
The difference is that $U$ depends on a combination of density,
stellar mass, and galaxy distance.

If we consider the {\it Bubble} case shown in Figure \ref{fig:ioncomp}
at $z=8$, there exists moderate overlap with the metal contours
between $1-10\times$ mean overdensity, which are reasonable densities
at 10 physical kpc from our simulated $10^7 \msolar$ galaxies.  But
note that the calculated incidence of intersecting at least this close
to one of 332 $M_*\ge 10^{7} \msolar$ galaxies in the d16n256vzw box
is $0.22 \Delta z^{-1}$, so a high-$z$ quasar spectrum would be lucky
to intersect one such region.  The contours move to higher overdensity
for larger galaxies (or at closer distances), but these are even less
likely.  Meanwhile, by $z=5$, the space density of $M_* \ge 10^7
\msolar$ galaxies is $\times 8$ higher and more metals have enriched
IGM overdensities while the ionization contours have not evolved too
much.  This is why we obtain an increase in $\CIV$ absorption from
$z=8\rightarrow 5$, as we show in \S\ref{sec:omega_hiion}.

The greater carbon spectral ratio in the {\it Bubble} field leads to
greater $f(\CIV)$ at photo-ionized temperatures, as shown in Figure
\ref{fig:rho_ionf}.  At $T=10^4$ K, $f(\CIV)$ maximizes at 0.67 at
$z=8$ and extends over 3.2 decades of density where $f(\CIV)\ge 0.1$,
compared to $f(\CIV)=0.27$ and 1.6 decades with $f(\CIV)\ge 0.1$ for
the {\it HM2001} field.  The {\it Bubble} hardness declines as stars
age and grow more metal rich by $z=5$, (maximum $f(\CIV)=0.57$), and
rises for {\it HM2001} as the quasar density increases
($f(\CIV)=0.37$, cf. the carbon spectral ratio evolution in the bottom
of Figure \ref{fig:rho_ionf}).  The greater {\it Bubble} hardness
allows significantly more $\CIV$, especially at the highest redshifts
we explore.  \citet{sim06b} also finds $f(\CIV)$ increases at $z>4.5$
at $\rho/\bar{\rho}=10$ when the dust attenuated flux from a $10^{8}
\msolar$ galaxy at 100 kpc is included.  In the end, this results in
less evolution for $\CIV$ absorption in the {\it Bubble} model
relative to {\it HM2001}, as we show in \S\ref{sec:omega_hiion}.  Our
{\it Bubble} evolution is greater than the \citet{sim06b} example,
because the metals are 5-50 kpc from ionizing galaxies.

A key point applicable to all metal-line IGM observations is that
while it is well known that taking ratios of different photo-ionized
species can determine spectral hardness
\citep[e.g.][]{sch03,agu04,agu08}, a single species' ionization
correction alone can vary dramatically due to the spectral shape.  It
is risky therefore to assign a fiducial ionization correction,
especially considering we know so little about the ionization
background for high-ionization species; as a result, an observable
such as $\Omega(\CIV)$ could easily change by a factor of a few, as we
stress in \S\ref{sec:omegac4}.

To visualize the spatial distribution of metal absorbers, we
illustrate the absorption column densities in two-dimensional maps of
the 25 $\kms$ slices in the bottom panels of Figures
\ref{fig:snaps60}~assuming a uniform {\it HM2001} background.  $\CII$
is confined to overdense regions close to galaxies, while $\CIV$ is a
more capable tracer of the IGM volume enriched, albeit mostly at
column densities that are below current observational limits.  The
scarcity of pink in the bottom right frame indicating column densities
similar to those observed by \citet{rya06} and \citet{sim06b}
indicates just how rare detectable $\CIV$ is.

\begin{figure*} \includegraphics[scale=0.66]{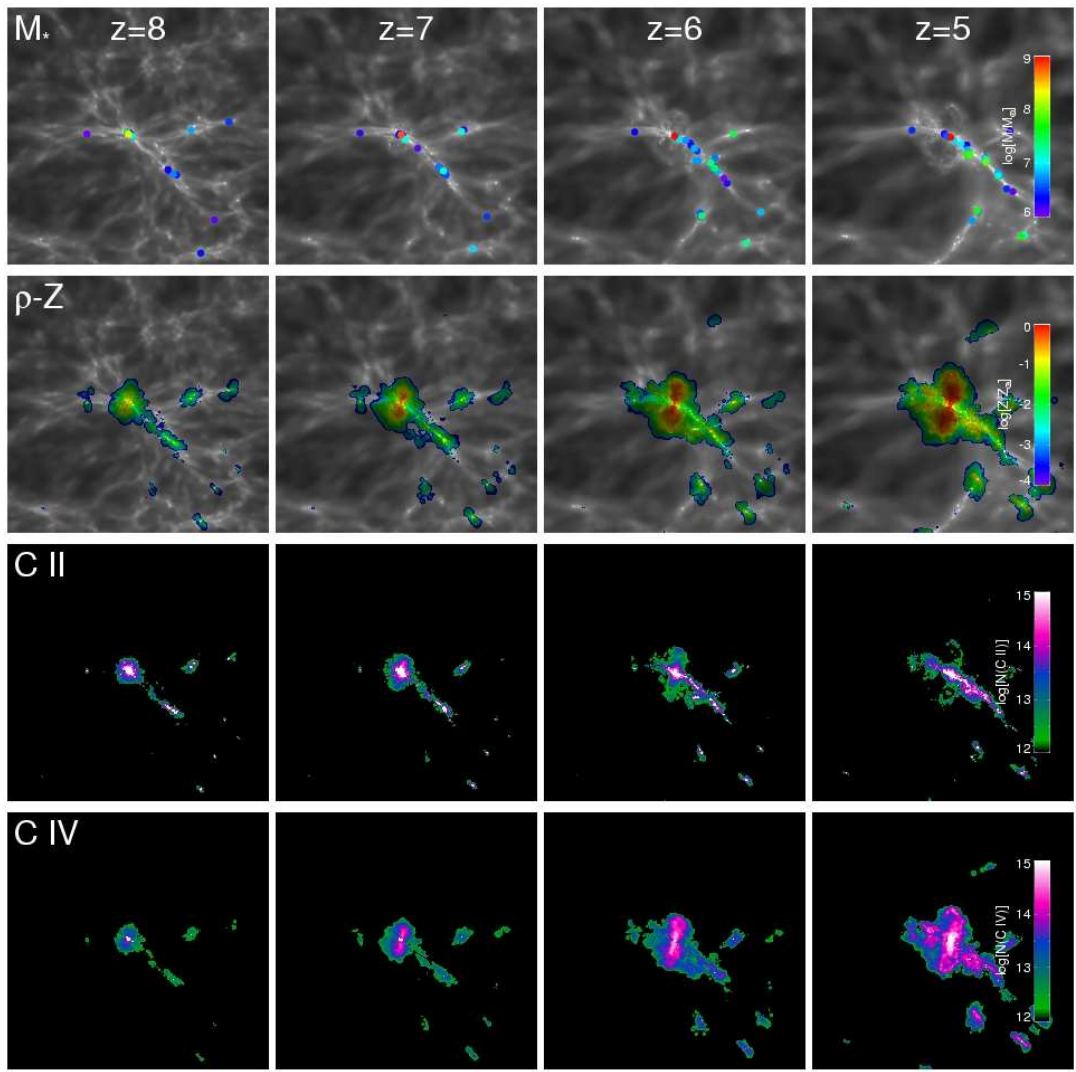}
  \caption[]{The evolution of a $4\times 4 \hmpc \times 25 \kms$
    slice, centered on a galaxy that grows from $2.4\times10^8
    \msolar$ to $1.8\times10^9 \msolar$ from $z=8\rightarrow 5$.  The
    global metal volume filling factor increases by $\sim \times 10$
    from $z=8\rightarrow 5$, but this slice shows less growth owing to
    early enrichment from the massive galaxy at the center.  $\CII$
    traces higher overdensities in the proximity of galaxies, while
    $\CIV$ evolves to trace the more diffuse IGM by $z=5$, including
    the bipolar outflow.  $M_*=10^{7-8} \msolar$ galaxies such as
    those in the bottom right of the frame represent the most prolific
    enrichers of the $z=6$ IGM, since they exist in greater numbers
    and have higher mass loading factors.  Visit {\tt
      http://luca.as.arizona.edu/\~{}oppen/IGM/hiz.html} for
    simulation movies.}
\label{fig:snapsevol}
\end{figure*}

We demonstrate evolution using a zoom-in to the second most massive
galaxy in our box in Figure \ref{fig:snapsevol}.  This visualization
shows the evolution of the surrounding $4\times 4 \hmpc \times 25
\kms$ region using the same scales as Figure \ref{fig:snaps60}.  A
clear bipolar outflow emanates perpendicularly to a filament feeding
the growth of a $M_*=10^9 \msolar$ galaxy (at $z=6$).  Strong $\CII$
absorption traces the overdense regions around the galaxy at all
epochs, while $\CIV$ evolves to trace the extended IGM enrichment;
observed $\CIV$ evolves faster than $\CII$ between $z=8\rightarrow 5$.
The smaller galaxies in the bottom right of this frame, $M_*\sim
10^{7-8} \msolar$, are the most prolific IGM enrichers at high-$z$ due
to their large numbers and high mass loading factors.

\section{Observational Predictions}\label{sec:observe}

Observations of the high-$z$ IGM are leaping forward with the advent
of near-IR spectroscopy, along with the discovery of quasars at $z>6$,
with an eye to the tantalizing possibility of using gamma-ray bursts
\citep[GRBs;][]{kaw06} to briefly illuminate IGM absorbers at very
high redshifts.  We present here predictions for high-$z$ metals line
transitions occurring redward of $\lya$, where the quasar continuum is
free of $\HI$ absorption.  We consider five species already observed
at $z\ge 5$ with rest wavelengths ($\lambda_0$) redward of $\lya$:
$\CII$ (1334.5\AA), $\CIV$ (1548.2, 1550.8 \AA), $\OI$ (1302.2 \AA),
$\SiII$ (1260.4 \AA), and $\SiIV$ (1393.8, 1402,8 \AA) at our four
chosen redshifts: $z=8$, 7, 6, and 5.  Three ionization fields
introduced in \S\ref{sec:ioncase} are considered: {\it No Field}, {\it
  HM2001}, and {\it Bubble}.

We use our quasar absorption line spectral generator {\tt
  specexbin}~(see \S2.5 of OD06 for details) to produce continuous
angled LOSs through our simulation volume, each covering $\Delta z =
0.9$ for each of the four redshifts.  We make an ``ideal''-quality
sample of 70 LOSs, assuming a high $S/N$ ratio of 50 for each
$R=50,000$ resolution element, and a moderate instrumental resolution
of 15 $\kms$.  This is comparable to what is achievable on current
optical spectrographs on $8-10$ meter class telescopes, which can trace
$\CII$ out to $z\sim 6.5$ (e.g. BSRS).  At higher redshifts, near-IR
data will be required.  The current generation of near-IR
spectrometers have resolutions of $R=5000-10,000$ and a $S/N=10-20$
per resolution element for $z\sim6$ quasars (R.  Simcoe, private
communication).  The next generation should achieve $R=20,000$ with
much higher S/N using 20-30 meter class telescopes, e.g.  the
near-Infrared Multi-Object Spectrograph (IRMOS) on the Thirty Meter
Telescope (TMT) \citep{eik06}.  Our simulated spectra can be used both
to compare to current optical spectra, as well as make predictions for
future instruments.

We first consider column density distributions for three of the five
ionization species.  We fit Voigt profiles using the automated fitter
AutoVP~\citep{dav97}, generally with little uncertainty since metal
lines are typically unsaturated and unblended.  Then we sum column
densities into a cosmic mass density as a function of redshift for
each species.  We present predictions for equivalent widths next, for
comparisons with large samples of $z>6$ metal absorbers that may
obtained in the near future.  We then consider some simple absorber
ratios, emphasizing what can be inferred about the ionizing
conditions.  Lastly, we show some sample absorption line profiles and
explain the velocity profiles of systems.  Along the way, we compare
to published $z\ga 5$ metal absorber data, and offer some
interpretations based on our models.  We consider the observed
completeness limits assuming $S/N$=10 and 20 per resolution element at
$R=5000$, and we confirm that the results we show in our high quality
spectra hold at lower resolution and $S/N$.

\subsection{Column Density Distributions}

The column density distribution (CDD), $d^2n/dXdN({\rm ion})$, is the
most basic metal absorber counting statistic.  Here $n$ is the number
of lines with column density between $N({\rm ion})$ and $N({\rm
  ion})+dN({\rm ion})$, and $dX$ is absorption pathlength.  Following
general practice among observers, we sum together all individual Voigt
profile {\it components} within 100 $\kms$ of each other into {\it
  systems}.  This minimizes differences due to vagaries of Voigt
profile fitting, and allows more straightforward comparisons among
data sets with varying resolution and signal-to-noise.  For $dX$ we
use
\begin{equation}
X(z) = \frac{2}{3\Omega_M}\sqrt{\Omega_M(1+z)^3+\Omega_\Lambda}.
\end{equation}
$dX$ for our adopted $\Omega_M=0.25$ cosmology is almost double that
in an Einstein-deSitter universe (the $\Omega_\Lambda$ term is
negligible at these redshifts).  We scale other observations to our
assumed cosmology where necessary.

\begin{figure*}
\includegraphics[scale=0.80]{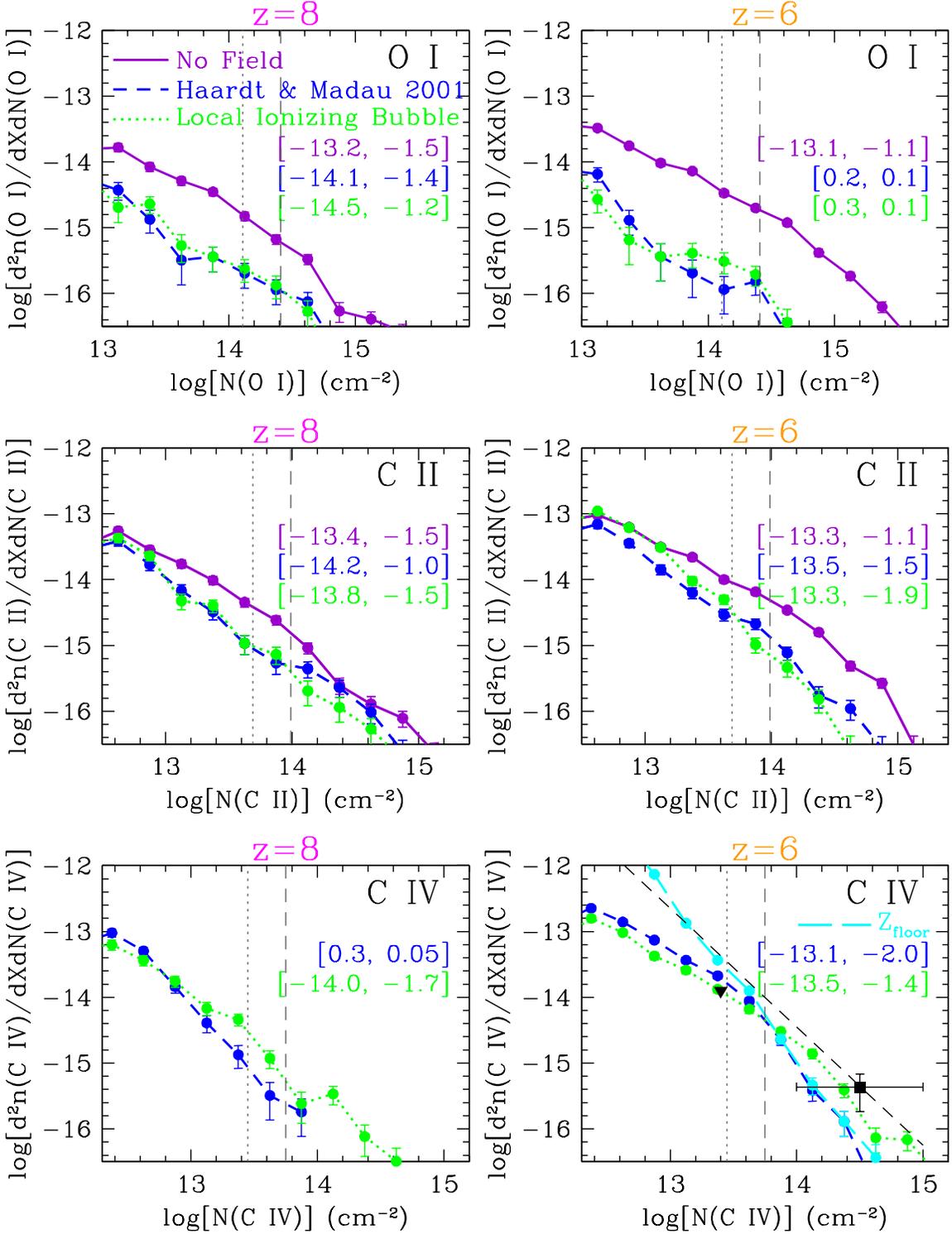}
\caption[]{Predicted differential column density distributions for
  $\OI$, $\CII$, and $\CIV$ systems at $z=8$ (left) and $z=6$ (right),
  for our three ionization field cases.  50\% detection thresholds are
  indicated by gray lines for $S/N=10$ (dashed) and 20 (dotted) per
  resolution element in an $R=5000$ spectrum.  The two values on the
  right side of each panel are log[$B$] and $\alpha$ from Equation
  \ref{eqn:CDD_powerlaw} if both numbers are negative, or if there are
  too few lines for a power law fit the line frequencies per $\Delta
  z$ in the lower and upper column density bins if both numbers are
  positive.  $\OI$ (upper panels) shows frequent strong lines and a
  shallow decline toward high $N(\OI)$ if all oxygen is neutral, while
  significantly less arises in the ionized cases.  $\CII$, middle
  panels, is assumed to have no ionization correction in the {\it No
    Field} case, showing much the same CDDs as neutral $\OI$.  $\CII$
  is more prevalent in the {\it Bubble} than $\OI$ due to this state's
  higher ionization potential.  $\CIV$, bottom panels, shows more
  strong absorbers for the {\it Bubble} field, while the {\it HM2001}
  uniform field predicts much stronger evolution.  The long dashed
  cyan CDD at $z=6$ is the case where a metallicity floor of
  $Z=10^{-3} \Zsolar$ is added, resulting in appreciably more lines
  below $10^{13.5} \cms$.  The black square in the bottom right panel
  correspond to the total number of components (3) with
  $N(\CIV)=10^{14.2-15.0} \cms$ in the combined sample of
  \citet{bec09} \& \citet{rya09} over a pathlength of $\Delta X =
  27.4$.  The black upside-down triangle is the upper limits from the
  \citet{bec09} null detection at $N(\CIV)=10^{13.4} \cms$ assuming
  $\alpha>-2.6$.  The dashed black line is the power law fit from
  \citet{son01} to $\CIV$ lines at $z=2.90-3.54$.}
\label{fig:cdd}
\end{figure*}

Figure \ref{fig:cdd}~shows CDDs for oxygen and carbon using different
choices for the ionizing field at $z=8$ (left) and $z=6$ (right).  The
first ion considered is $\OI$ in the top panels, which has a unique
association with $\HI$ because these two species remain in
charge-exchange equilibrium \citep{ost89} due to their neutral
ionization potentials being only 0.15\% apart in energy.  \citet{oh02}
and \citet{fur03} hypothesize the existence of an ``$\OI$ forest''
tracing the neutral metals that have enriched dense IGM regions via
early enrichment.  Higher ionization states easily recombine at the
high densities of the early Universe. Dozens, even hundreds of lines
with $N(\OI)> 10^{14} \cms$ should be detectable if the IGM is neutral
and enriched with a large filling factor of metals, but BSRS find very
few $\OI$ lines.  No $\OI$ systems above $10^{15} \cms$ are found
either, despite such predictions by \citet{oh02}.

Our {\it No Field} case (upper panels of Figure \ref{fig:cdd}) shows
that $\OI$ lines should be common at $z=8$, despite our metal filling
factors being under 1\%.  We predict line frequencies of 1.9 and 0.6
$\Delta z^{-1}$ for $10^{14}\le N(\OI)<10^{15} \cms$ and $10^{15}\ge
N(\OI)$ respectively at $z=8$.  The corresponding frequencies rise to
5.4 and 1.9 $\Delta z^{-1}$ by $z=6$.  $\OI$ provides the clearest
contrast of any species between a neutral and ionized IGM; the {\it
  Bubble} field yields frequencies of 0.4 and 0.3 $\Delta z^{-1}$
above $N(\OI)=10^{14} \cms$ at $z=8$ and 6 respectively.  This clear
contrast may already have been observed by BSRS, as all of their $\OI$
absorbers detected at $z>6$ are along one of their three LOSs probing
these redshifts.  This could suggest a patchy reionization scenario
\citep[e.g.][]{fur05b, ili06, lid07} where individual LOSs probe
large-scale structures at different ionization states.

We fit a power law of the form
\begin{equation}\label{eqn:CDD_powerlaw}
d^2n/dXdN({\rm ion}) = B\times \left(\frac{N_{\rm ion}}{N_0}\right)^{\alpha} 
\end{equation}
to the $\OI$ distribution by summing systems into two bins
($N(\OI)=10^{13.7-14.45}$ and $N(\OI)=10^{14.45-15.2} \cms$) and
finding the best-fit line.  We list $\log[B]$ and $\alpha$ for each
CDD in Figure \ref{fig:cdd} if there are at least 10 systems in each
bin; otherwise we list system frequencies per $\Delta z$ of the lower
and upper bins (positive numbers).  We set $N_0 = 10^{13.0} \cms$ for
the normalization.  The {\it No Field} $\OI$ $z=6$ CDD has
$\alpha=-1.1$ over these column densities, which translates to 85\% as
many systems in the upper $\OI$ bin compared to the lower bin.  A
power law where $\alpha>-2$ indicates more metals at higher column
densities, meaning this power law must turn over so that the summed
$\OI$ is finite.  Power law slopes can yield information about the
nature of the metallicity distribution and ionizing source, and
provide an easy way to compare datasets to simulations.  Not only does
the {\it Bubble} field create fewer $\OI$ lines at every column
density, it has a much steeper power law, which we cannot calculate
accurately as there are too few absorbers in the higher bin.  Oxygen
and hydrogen are easily ionized out of their neutral states by any
ionizing background above 1 Rydberg.

$\CII$ CDDs shown in the middle panels of Figure
\ref{fig:cdd}~recreate similar behavior to $\OI$ for the {\it No
  Field} case, but has a much higher frequency for the ionization
field cases since the $\CI$ ionization potential lies below 1 Ryd,
unlike $\OI$.  The $z=6$ frequency of the lines with
$N(\CII)>10^{14.0} \cms$ is 4.5, 0.8, and 0.4 $\Delta z^{-1}$ for the
{\it No Field}, {\it HM2001}, and {\it Bubble} fields, respectively.
Furthermore, the slopes of the CDDs between $N(\CII)=10^{13.2-14.7}
\cms$ may help constrain the ionization field, being $\alpha=-1.1$,
$-1.5$ and $-1.9$, respectively.  The {\it Bubble} field produces
fewer strong $\CII$ lines than {\it HM2001}, because the metals at
$z=6$ still reside close enough to the local ionizing source that
carbon is ionized to higher states by the greater intensity above the
Lyman limit.

$\CIV$ has a higher frequency of lines with $N(\CIV)\ge 10^{14.2}
\cms$ (0.9 vs. 0.2 $\Delta z^{-1}$) and a shallower slope ($\alpha=-1.4$
vs. $-2.0$ over $N(\CIV)=10^{13.0-14.5}$) for the {\it Bubble} field
compared to the {\it HM2001} field in the bottom panels of Figure
\ref{fig:cdd}.  There is no $\CIV$ in the {\it No Field} case, as
shocks from structure formation rarely heat the gas to temperatures
where $\CIV$ is collisionally ionized at these early epochs.  We
plot a black square from the three observed $\CIV$ absorbers of
\citet{rya09} where $N(\CIV)=10^{14.0-15.0} \cms$ and include the
additional pathlength from the SDSS J0002+2550 LOS observed only
by \citet{bec09} (total $\Delta X=27.4$).  We find an observed
frequency of $0.5\pm 0.3 \Delta z^{-1}$ for these absorbers averaging
$\langle z \rangle=5.75$.  To compare directly to our models, we
(i) multiply the frequency of our $\CIV$ absorbers by the \citet{rya09}
sensitivity function (their Figure 6), and (ii) estimate the $z=5.75$
$\CIV$ frequency by taking a log-weighted average of our $z=6$ and
$5$ model frequencies, because strong $\CIV$ absorbers rapidly
increase in number from $z=6\rightarrow 5$ (especially for {\it
HM2001}).  Our $N(\CIV)=10^{14.0-15.0} \cms$ frequencies are 1.2
and 0.4 $\Delta z^{-1}$ for the {\it Bubble} and {\it HM2001} models
respectively.  The uniform field is consistent with the observed
frequency, while the local ionizing bubble actually over-predicts
strong $\CIV$ absorbers.  We actually prefer the the {\it Bubble}
model, because we will argue IGM metals are not completely ionized
and that a significant fraction of them are in lower ionization
species, either recombined or shielded from UV flux.

We also plot the maximum allowed frequency of $N(\CIV)=10^{13.4} \cms$
systems from the null \citet{bec09} detection (upside-down triangle);
this is the column density where they calculate they have 50\%
completeness.  We appear to over-predict the amount of these weaker
absorbers in either model, although by how much will have to wait for
future observations.  The long dashed cyan line is the case where a
$10^{-3} \Zsolar$ metallicity floor is added to the simulation; we
discuss this more in \S\ref{sec:ew}. The fact that \citet{bec09} finds
no weak $\CIV$ lines in the largest pathlength yet sampled at $z>5.5$
indicates a large volume of the IGM is not enriched much above
$Z>10^{-3} \Zsolar$.

For comparison with lower $z$, we plot \citet{son01} power law fit to
$\CIV$ absorbers at $z=2.90-3.54$.  There exists less $\CIV$ at all
column densities in our simulations, which agrees with our
reproduction of a declining $\Omega(\CIV)$ at $z>5$ as we discuss
next.

\subsection{Metal Mass and Ion Densities} \label{sec:omega}

We now consider the summed mass densities as a way to illustrate the
evolutionary differences between ionization fields.  The evolution of
$\CIV$ for the two fields differs since the weak {\it HM2001} field at
$z=8$ is incapable of ionizing overdensities containing carbon up to
$\CIV$, creating very few strong absorbers, while the {\it Bubble}
field has a harder spectrum early on (cf. red contours in upper panels
of Figure \ref{fig:ioncomp}).

An ion's global mass density may be obtained by integrating the total
column density of their systems and dividing by pathlength, as
follows:
\begin{equation}\label{eqn:omega}
\Omega({\rm ion}, z) = {H_0 m_{\rm species} \over c \rho_{crit}} {\Sigma N({\rm ion},z) \over \Delta X(z)}.
\end{equation}
where $H_0$ is the Hubble constant, $c$ is the speed of light,
$\rho_{crit}$ is the critical density of the Universe, and $m_{\rm
species}$ is the atomic weight of the given species.  

For comparison to the data, we count systems over two decades of
column density chosen to correspond to the species' detectability.  We
count lines for $\CIV$ over $N(\CIV)=10^{13.0-15.0} \cms$, and scale
other species' column density range using the ratio of the oscillator
strength of a line ($g$, the strongest of a doublet) times its rest
wavelength ($g \times \lambda_0$) for the considered species versus
that of $\CIV$.  For example, the range considered for $\OI$ works out
to be $N(\OI)=10^{13.7-15.7} \cms$, because the resulting $\CIV:\OI$
ratio is $4.6$; $\OI$ has a $\times 3.9$ weaker oscillator strength
and a $\times 1.2$ smaller wavelength.  $\OI$ is harder to observe due
to its weaker oscillator strength, so the column densities considered
are higher.  Conversely, silicon species have high oscillator
strengths, meaning that column densities below $10^{13} \cms$ are
achievable for $\SiII$ ($N(\SiII)=10^{12.4-14.4} \cms$) and $\SiIV$
($N(\SiII)=10^{12.6-14.6} \cms$).  The range for $\CII$ is
$N(\CII)=10^{13.2-15.2} \cms$.  Nature creates a convenient conspiracy
whereby greater oscillator strengths make up for lower Type II SNe
yield abundances, leading to similarly strong lines for our five
considered ionic species.

\begin{figure*}
\includegraphics[scale=0.80]{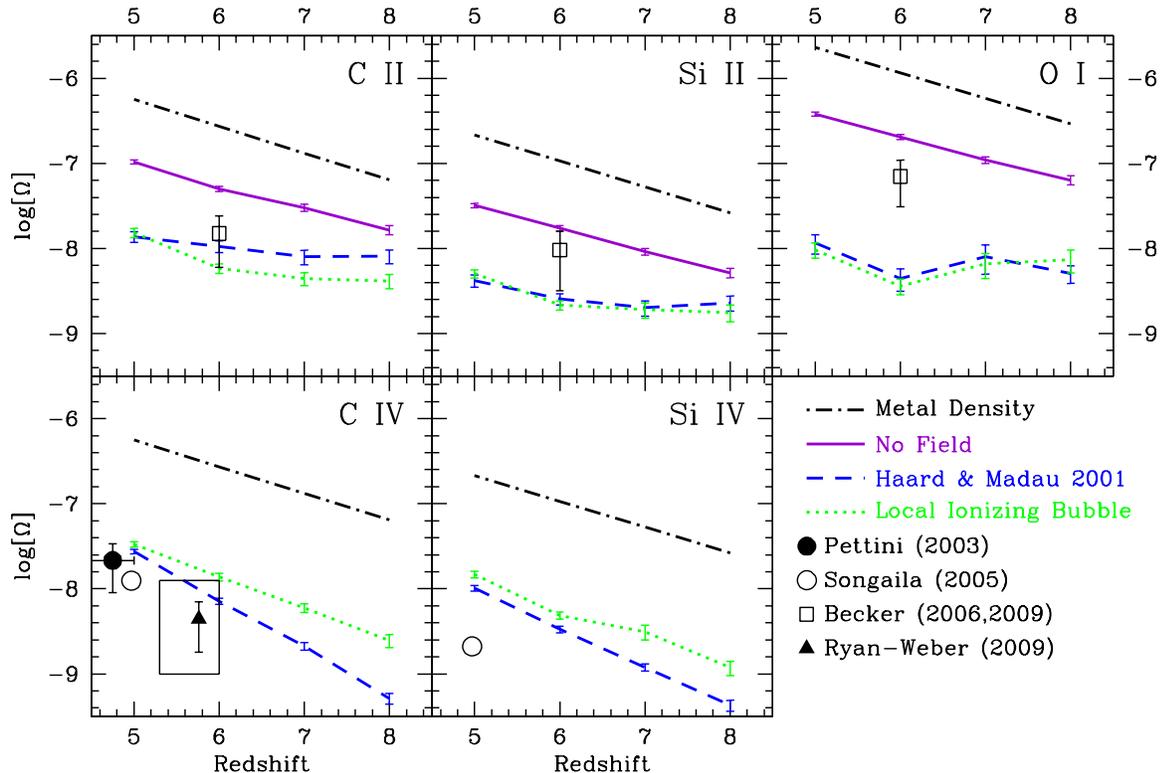}
\caption[]{The evolution of the global ion mass densities, determined
  from simulated quasar LOSs for the three ionization cases, compared
  to the metal density summed for the corresponding atomic species
  calculated by summing over all the SPH particles in the simulation
  box; the difference between the two represents a global ionization
  correction for that ion.  We sum absorbers over 2 decades of column
  density corresponding to the range the given transition is
  observable and unsaturated.  The {\it No Field} case assumes no
  ionization correction, but falls below the total metal density
  because most of the absorption lies in extremely rare saturated
  lines, which are not in the included column density range when
  summing $\Omega$.  The observations of BSRS in quasars at $z>6.2$
  appear to favor a partially neutral IGM.  Most $\CIV$ observations
  cannot distinguish between the {\it HM2001} and {\it Bubble} models,
  although the rapid evolution observed by \citet{bec09} and
  \citet{rya09} at $z>5.3$ relative to $z<4.5$ supports the former.  }
\label{fig:omega}
\end{figure*}

Error bars are calculated using the \citet{sto96} method (their
Equation 4), and do not include systematics such as cosmic variance.
We also apply this error estimation method to the BSRS dataset, which
results in much larger error bars than they published as a result of
them including only the uncertainty in their summed column densities.

\subsubsection{Low Ionization Species}

The resulting $\Omega$'s for the five species and different fields are
plotted in Figure~\ref{fig:omega}.  We also plot the total metal
density of the corresponding element, summed by counting the gas phase
mass of that species in the entire simulation box at the given
redshift ($\Omega({\rm species})$, black dash-dot line).

With no ionization correction in the {\it No Field} case, the ions
$\CII$, $\OI$, and $\SiII$ directly trace the increase seen in metal
density of $\sim \times 8$ from $z=8\rightarrow 5$.  Note, however,
that the ion density should be identical to the metal density if
absorption lines of all column density are counted, yet the ion
density is $\times 4-6$ less than the metal density.  Where are the
unaccounted metals?  The answer is that since the CDDs have slope
$\alpha>-2$, they are in high-column systems that are extraordinarily
rare.  Hence using a volume-weighted method as quasar LOSs to obtain a
handle on a mass-weighted quantity such as ion density is challenging,
more so for low ionization species, which preferentially trace high
overdensities.

The $\Omega(\OI)$ determined by BSRS favors the {\it No Field} case
over either ionization field, considering their sample from $z>6.2$
quasars.  This suggests that at least some of the $z\sim 6$
metal-enriched IGM is neutral.  That BSRS found four $\OI$ systems
(all with $N(\OI)>10^{13.7} \cms$) toward SDSS J1148+5251 and not more
than one in any of the other 8 high-$z$ LOSs; this is unlikely to be a
statistical fluctuation (0.18\% chance according to BSRS).  The {\it
  No Field} case produces a frequency of 9.2 $\Delta z^{-1}$ of such
$\OI$ lines, which is in agreement with the J1148 LOS and at least
$\times 20$ higher than either of our simulated ionization fields.
Hence our simulations agree with BSRS's interpretation that the J1148
LOS possibly probes a long stretch of IGM with incomplete
reionization, while the other LOSs of BSRS are consistent with being
completely reionized.  How one obtains a relatively neutral LOS along
the 200 comoving Mpc spanned by the J1148 LOS while most LOSs are
reionized is not easy to understand, but we cannot address this
directly with our current simulations owing to lack of volume.  If
reionization proceeds across such metal-enriched regions extending
below $z=6$, a dramatic decline in $\Omega(\OI)$ is expected by $z=5$.

The {\it No Field} case is barely favored by the $\Omega(\SiII)$ data
of BSRS, but $\Omega(\CII)$ appears closer to predictions of an
ionized IGM.  The fact that observed $\CII$, $\SiII$, and $\OI$ arise
together with similar profile shapes suggests that each of these ionic
species may be dominant, consistent with the {\it No Field}
assumption.  If so, the lack of consistency among the three low ions
may be a result of metal yields differing those of \citet{chi04} we
employ, which is one reason we provide alternative yields in Table
\ref{table:yields}.  $\Omega(\CII)$ and $\Omega(\SiII)$ show less
difference between the {\it No Field} and two ionizing cases than
$\Omega(\OI)$ in Figure \ref{fig:omega}, because the first two species
have higher ionization potentials that are not as easily ionized by
the addition of a field.

Overall, the low ionization species paint a scenario in which there
are still significant patches of neutral IGM gas at $z\sim 6$ along at
least some LOSs.  Hence in principle one could use such ions to trace
the patchiness of reionization.  We plan to investigate this
possibility using simulations that more accurately track the
ionization field using radiative transfer~\citep{fin09}.  

\subsubsection{High Ionization Species} \label{sec:omega_hiion}

$\Omega(\CIV)$ has been measured in several studies at $z>4.5$, with
differing results.  \citet{son01} finds $\Omega(\CIV)$ is consistent
with remaining constant from $z=2\rightarrow 5.5$, which was extended
by the measurements of \citet{sim06b} out to $z=5.4-6.2$.  Conversely,
the null detections by \citet{bec09} of $\CIV$ in the highest $S/N$
data at $z=5.3-6.0$ supports a significantly declining $\Omega(\CIV)$
at $z>4.5$.  The measurements of \citet{rya09} over the largest
pathlength ($\Delta X=25.1$) centered at $\langle z \rangle=5.76$ also
finds a similar $\Omega(\CIV)$ value.

Exploring a range of wind models, OD06 explained the near constancy in
$\Omega(\CIV)$ at $z=2-4.5$ by a decreasing global ionization fraction
of $\CIV$ counterbalancing the increasing enrichment of the IGM.
Therefore, the stability of $\Omega(\CIV)$ does not indicate that the
IGM must be enriched at $z>6$.  Instead, the ionization fraction
decreases with time for several reasons: (i) Decreasing physical
densities due to Hubble expansion, and (ii) the increasing {\it
  HM2001} ionization field strength from $z=5\rightarrow 2$ result in
$\CIV$ tracing higher overdensities and less of the carbon in the
diffuse IGM; while (iii) increasing energy input into the IGM
primarily from galactic outflows at later times push metals to hotter
temperatures where $\CIV$ does a poor job of tracing carbon (see
\citet{opp07} for a simple breakdown of this evolution).

OD06 predicted $\Omega(\CIV)$ rises by a factor of two from
$z=6\rightarrow 4.5$ using the {\it HM2001} field, although the new
cosmology we now use with a reduced $\sigma_8$ significantly curtails
high-$z$ SF resulting in a greater rise that agrees better with
\citet{bec09} and \citet{rya09}.  Figure \ref{fig:omega} shows that
$\Omega(\CIV)$ continues this growth trend between $z=8\rightarrow 5$,
for both ionization fields we consider.  For the {\it HM2001} field,
the $\CIV$ ionization correction falls by a factor of 10 (from 220 to
21) between $z=8\rightarrow 5$.  This is a stunning drop and occurs
because of the ionization behavior explored in \S\ref{sec:ionbehave}:
the uniform background grows stronger while physical densities decline
resulting in a greater ionization parameter allowing metals to be
ionized to $\CIV$ by $z=5$.  In essence, some of the same trends that
cause the $\CIV$ ionization correction to increase from
$z=5\rightarrow 2$ (OD06), cause it to decrease from $z=8\rightarrow
5$.

The {\it Bubble} model predicts a slower evolution in $\Omega(\CIV)$
at $z>4.5$, because this field does not evolve as much as the {\it
  HM2001} uniform field.  Metals around galaxies are ionized by a
nearly unevolving field between $z=8\rightarrow 4$, so an increase of
$\Omega(\CIV)$ is more likely to reflect an actual increase in IGM
metallicity near galaxies.  From $z=8\rightarrow 5$, $\Omega(\CIV)$
increases by $\times 14$ while $\Omega({\rm C})$ increases by $\times
9$.  As future $\Omega(\CIV)$ observations achieve higher redshift
above 6, we predict to see a decline corresponding to the evolution in
the global metal density if the {\it Bubble} field is
correct. Meanwhile, the physical and ionization conditions make $z\sim
4-6$ the ideal time for $\CIV$ to trace metals in the IGM for either
field, as the ionization correction reaches a minimum at these epochs.

$\Omega(\SiIV)$ mirrors the behavior for $\Omega(\CIV)$ for the two
fields we consider, with smaller differences.  The lower ionization
potential of $\SiIII$ causes $\SiIV$ to trace higher overdensities,
which can track some of the metals already injected into the IGM
at $z=8$ with the {\it HM2001} field.  The evolution of $\Omega(\SiIV)$
is nearly identical to that of $\Omega(\CIV)$ for the {\it Bubble}
field for the reasons mentioned above.

The trends in our simulated high-ionization $\Omega$'s appear
consistent with those that are observed.  Our predicted $\Omega(\CIV)$
values with either ionization field over-estimates the most recent
observations at $z>5.3$; however we argue that not all metals are
ionized by such a field, and that the observed value falling between
the extremes of the {\it Bubble} field and the {\it No Field} case
(i.e. no $\Omega(\CIV)$) is encouraging.  

\subsubsection{Systematic Uncertainties in $\Omega(\CIV)$}\label{sec:omegac4}


While the above agreement with $\Omega(\CIV)$ data for our ionizing
field models is nice, it is worth pointing out that there
are many possible systematics in the modeling that can impact
predictions.  Here we point out some of the ways our predictions
of $\Omega(\CIV)$ may be uncertain.

{\bf Ionization background:} Besides the possibility that not all
metals are ionized, there is the uncertainty about the nature of
the ionization field where it exists.  The intensity of the ionization
background at the $\CIII$ ionization potential is unobservable and
uncertain, yet is vital for the correct determination of the $\CIV$
ionization correction, as we stress in \S\ref{sec:ionbehave}.  The
carbon spectral ratio of the {\it Bubble} field makes $\Omega(\CIV)$
higher than the {\it HM2001} field; in fact, the {\it Bubble} field
produces a nearly optimally high value of $\Omega(\CIV)$ at $z=6$.
On the other hand, \citet{mad09} notes that reprocessing of radiation
between 3 and 4 Ryd by resonant line absorption of the $\HeII$ Lyman
series could reduce by $>$$\times 2$ the photons capable of ionizing
$\CIII$ to $\CIV$.  \citet{bec09} suggests this could be an explanation
for the strong evolution they find in $\Omega(\CIV)$ ($\times 4.4$
increase between $z=5.3-6.0$ and $z=2.0-4.5$) if $\HeII$ reionization
proceeds over this redshift range, increasing $\CIII$ ionizing
radiation, and resulting in a lower ionization correction for $\CIV$
at $z<4.5$.

{\bf Wind model:} While we have investigated a plausible wind model
for enriching the IGM, based on momentum-driven wind scalings, this
may not be the only possible outflow scenario that can reproduce
observations.  Varying wind speeds and mass loading factors in OD06
led us to some useful intuition when compared to $\CIV$ absorption
data: too high wind speeds overheat the IGM while too low wind
speeds do not sufficiently enrich filaments; also, too high mass
loading suppresses SF so much that too few metals are produced while
too low mass loading results in insufficient metals in the IGM.
But while constraints at $z\la 5$ are fairly tight, at higher $z$
the possibility of more exotic processes such as Pop III stars or
the lack of dust may impart fundamental changes to outflows.  Without
a more detailed physical understanding of the physics of outflow
driving, it is difficult to ascertain the systematic uncertainties
in our wind model.




{\bf Mini-halos and Population~III stars:} The very first stars and
galaxies are not resolved in our simulation, as they likely form
within mini-halos having virial temperatures below
$10^4$K~\citep[e.g.][]{abe02a}.  If such systems dominate the early
enrichment or ionization budget, our modeling will be incomplete.
However, \citet{wis08} suggests that they do not dominate the
ionization budget, and \citet{bro04} estimated that they increase the
global gas metallicity by only $\sim 10^{-4} \Zsolar$, which is
$\times 30$ less than the global gas metallicity at $z=6$ in our
simulations.  Hence unless they have much higher yields than expected,
it seems unlikely that metal production in mini-halos should be
important at $z=5-8$.

{\bf Unresolved galaxies:} For galaxies in halos with virial
temperatures above $10^4$K, our simulation nearly resolves the Jeans
mass at $z\la 10$.  Using the formula for Jeans length from \S6 of
OD06, we find a Jeans mass of $9\times 10^{7} \msolar$ at $z=10$
assuming $\delta=50$ and $T=10^4$K, which is nearly the dynamical mass
limit of a resolved galaxy in our simulation.  The flattening of the
stellar mass function at $M_*\la 10^7 \msolar$ reflects the effect of
the photoionization of regions containing forming galaxies (i.e. the
filtering mass).  For this reason, our resolution-converged fit to the
stellar mass function (Equation \ref{eqn:M*func}) suggests that we
are not missing a large population of lower mass galaxies in the
d16n512vzw box.


{\bf Yields:} Another uncertain factor that could alter $\Omega(\CIV)$
is the SNe yields.  We use the yields of \citet{chi04}, which do not
change much at low-$Z$ for carbon.  In our simulations the SFR
efficiency is so high in early galaxies that the average stellar
metallicity in star particles reaches $0.1 \Zsolar$ by $z=7$, which
minimizes the importance of low-$Z$ yields.  It has been suggested,
however, that patches of low-metallicity SF persist to late
epochs~\citep[e.g.][]{tor07, cen08}, which if true means our
simulations are over-mixing metals.  Integrating the carbon yields
over the various stellar IMFs listed in Table \ref{table:yields} does
indicate as much as a factor of two difference with the yields we use
at the higher end.  Note also that the yields calculated by
\citet{hir07} for rotating stars at solar metallicity can more than
double the total carbon yield versus non-rotating models.

{\bf Initial mass function:} Even if metal-free stars are unimportant,
it is possible that there is an overall shift towards a more top-heavy
IMF at high redshifts.  \citet{dav08a} finds the evolution in the
$M_*$-SFR calibration could be explained by an increasingly top-heavy
IMF from $z=0\rightarrow 2$, which suggests very few low-mass stars
if extrapolated to $z\sim 6$.  \citet{tum06} presents constraints
indicating a top-heavy IMF at early epochs that explain the frequency
of Galactic carbon-enhanced metal poor stars\footnote{This itself
is not a carbon yield constraint, and only an IMF constraint.}.
IMF variation concerns are mitigated in large part by noting that
our simulations match the observed UV LF, which indicate that they
are producing roughly the correct number of massive stars.  Hence
to first order the amount of carbon produced will be correct.  But
depending on the nature and form of an IMF variation, it remains
possible that the overall carbon yield per unit high-mass SF could
still vary somewhat.  As an example, the \citet{por98} yields using
a contribution only from 30-120 $\msolar$ stars produces half as
much carbon as the \citet{chi04} yields (Table \ref{table:yields}).

{\bf Cosmological parameters:} Given the uncertainties on cosmological
parameters in the 5-year WMAP results \citet{hin08}, the measurement
of $\sigma_8$ ($0.820\pm0.028$) imparts the greatest uncertainty in
regards to high-$z$ SF.  As mentioned in \S\ref{sec:galmass}, the
primary difference causing the factor of two increase in stellar mass
density at $z=6$ in the w-series used in DFO06 over the d-series
used here is $\sigma_8$ (0.90 compared to 0.83).  Similarly, we find
$\times 1.8$ fewer $z=6$ stars in a $16 \hmpc$ $2\times256^3$ test run
where we reduce $\sigma_8$ from 0.83 to 0.75 holding all other
parameters constant.  Given the error bars on $\sigma_8$, a factor of
two difference in SF and chemical nucleosynthesis at $z\ge 6$ is
possible, which will lead to similar uncertainty in $\Omega(\CIV)$ if
the metals are injected into the IGM the same way.  This is the main
reason why the d-series results in a larger increase in
$\Omega(\CIV)$ from $z=6.0\rightarrow 4.5$ compared to the published
w-series predictions in OD06.

To summarize, a number of possibilities exist to alter $\Omega(\CIV)$
by a factor of a couple.  An important constraint is the number of
weaker lines ($N(\CIV)<10^{14} \cms$), which our models may be
over-predicting given the comparison to the largest sample from
\citet{bec09} that can detect such absorbers at $z>5.3$; this could
be a sign that superwinds have a smaller filling factor in the IGM
than our simulations predict, or that more diffuse metals are less
ionized (see \S\ref{sec:masslimit}).  Lower carbon yields, possibly
from a top-heavy IMF, could potentially lower $\Omega(\CIV)$, while
a different ionization background is more likely also to decrease
$\Omega(\CIV)$ compared to the {\it Bubble} field.  The determination
of $\sigma_8$ also can potentially alter $\Omega(\CIV)$ by a factor
of two.  The good agreement seen in $\Omega(\CIV)$ evolution should
be taken as an encouraging but not definitive demonstration of the
validity of our models.

\begin{figure*}
\includegraphics[scale=0.80]{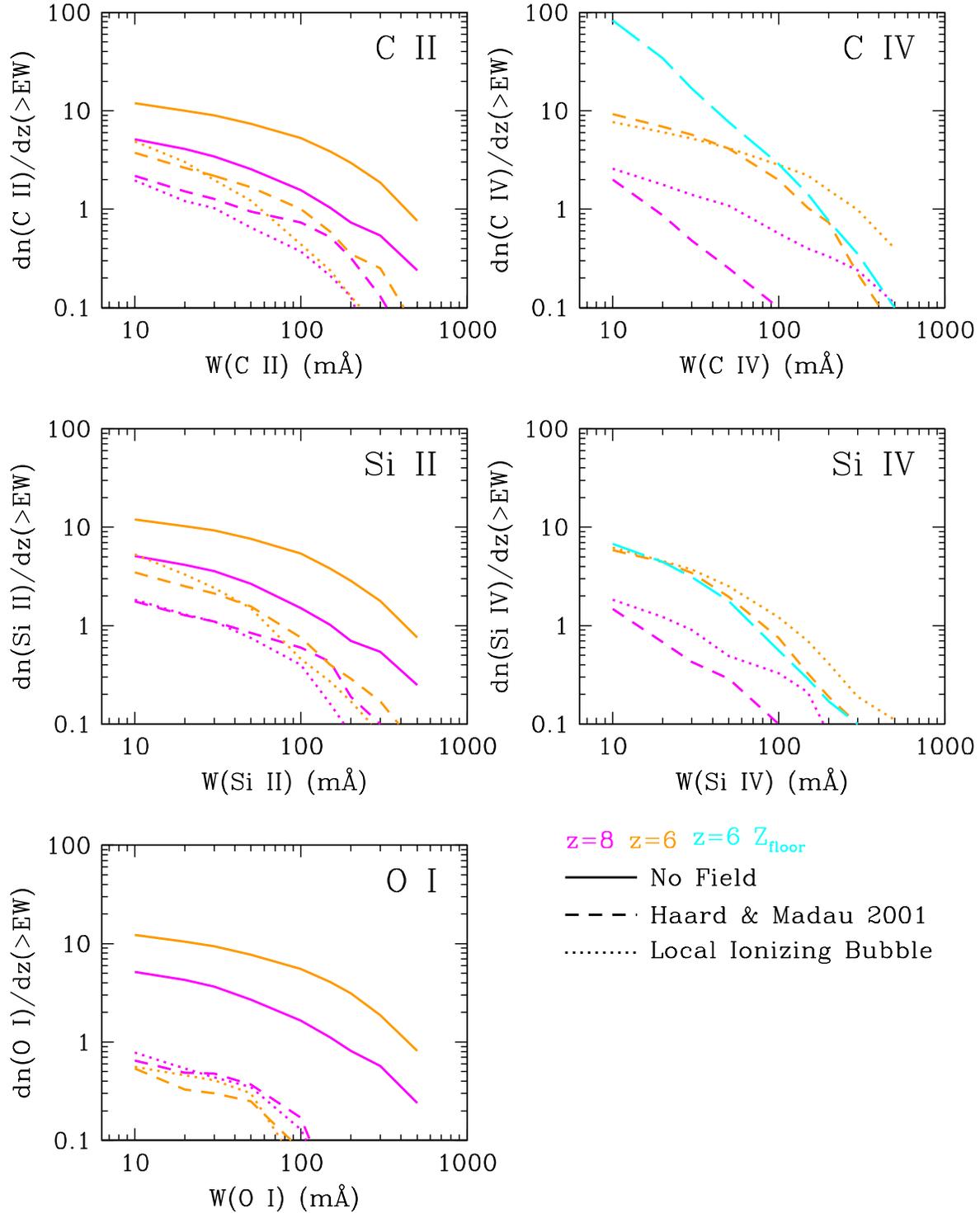}
\caption[]{Predicted cumulative equivalent width distributions for
  absorber systems of five species at $z=8$ (magenta) and $z=6$
  (orange).  The three ionization cases are shown, along with long
  dashed cyan lines for $\CIV$ and $\SiIV$ corresponding to the case
  where a uniform metallicity floor of $10^{-3} \Zsolar$ is applied to
  the entire volume at $z=6$ using the {\it HM2001} field.  The {\it
  HM2001} field makes more strong $\CII$ and more weak $\CIV$ systems
  compared to the {\it Bubble} field at $z=6$. $\OI$ is the best
  tracer of the {\it No Field} case.  $\CIV$ is ideal to trace a
  larger volume filling factor of metals, while this is not true for
  its sister ion $\SiIV$, which is mostly ionized to $\SiV$ in the
  same volume.  }
\label{fig:ew}
\end{figure*}

\subsection{Equivalent Width Distributions}\label{sec:ew}

We present cumulative rest equivalent width (EW) distributions at
$z=8$ and 6 for all five species in Figure
\ref{fig:ew}\footnote{Please contact the author for the tabular form
  of the predictions in these or any other figures.  We have EW and
  CDD distributions for any other desired redshift as well.}.  This is
intended to facilitate future observational comparisons, as well as
present a more realistic assessment of the discriminatory power of
high-$z$ metal absorber statistics from lower resolution near-IR
spectrographs.

Many of the trends already discussed earlier in this section are
reproduced here.  The average weaker ionizing flux of the {\it
HM2001} field at $z=6$ results in more strong $\CII$ absorbers and
fewer strong $\CIV$ absorbers compared to the {\it Bubble} field;
the opposite is true for weak absorbers.  The same trends are
apparent in $\SiII$ and $\SiIV$, and also at $z=8$ for $\CIV$,
although the {\it Bubble} field produces more high ionization
absorbers at all equivalent widths here.  $\OI$ best distinguishes
neutral and ionized gas.

We note that EWs, particularly at the high end, may be influenced by
the phenomenon of turbulent broadening.  In \citet{opp08b} we found
turbulent broadening necessary to reproduce the observed low-$z$
$\OVI$ $b$-parameter distribution.  The purported turbulence occurs
far below the mass scale resolved by our simulation, and could be
associated with energy injection by outflows~\citep{opp08b}.  Since
high-$z$ absorbers are universally young and associated with recent
outflows, turbulence may result in a higher EW (for saturated
lines), which would affect the high end EW distribution.
\citet{rau01} observed velocity differences in $\langle z\rangle\sim
2.7$ $\CIV$ profiles from lensed quasar LOS pairs at sub-kpc scales,
which indicate more turbulence than we expect in most low-$z$ $\OVI$
absorbers.  Fortunately, turbulence should not much affect the weak
end, since turbulent broadening will leave EWs mostly unaltered.
The other observables we have considered such as $\Omega(\CIV)$ all
deal with column densities, which should remain unaffected by
turbulent broadening as long as lines are not saturated.
$b$-parameters are the observable most likely affected by turbulent
broadening, and may explain some of the broad $\CIV$ profiles seen by
\citet{rya06} and \citet{sim06b}, although \citet{bec09} suggests that
these may be multi-component systems with smaller line widths if
observed at higher resolution.

We also consider the case of a metallicity floor of $10^{-3} \Zsolar$
added everywhere in the simulation box with the uniform {\it HM2001}
ionization field at $z=6$ for $\CIV$ (long dashed cyan line in the
upper panel of Figure \ref{fig:ew}).  Note that for a Chabrier IMF and
\citet{chi04} yields, obtaining this metal floor would involve turning
0.05\% of baryons into stars, equivalent to all the stars formed by
$z=7.5$, and distributing this uniformly among all baryons, which
seems implausible.  Hence this scenario requires a more exotic origin,
possibly resulting from very early fast winds powered by high-yield
VMSs.  Such a scenario is similar to that assumed in semi-analytical
enrichment models of \citet{mad01} \&~\citet{sca02}.  \citet{mad01}
calculated a complete filling factor of metals from pre-galactic
outflows in their most extreme cases, while \citet{sca02} found as
much as 40\% of the $z=6$ volume is enriched to $\ge 10^{-3} \Zsolar$
primarily by SN-driven outflows from early galaxies.

The observational signature of such a metal floor is dramatic at the
weak end of the $\CIV$ equivalent width distribution.  Today's
instruments should be able to integrate down to EW=30 m\AA~to
statistically distinguish between our standard simulation and a metal
floor: 16.9 $\Delta z^{-1}$ versus 5.7 $\Delta z^{-1}$.  We also show
this scenario in the CDD plot in the bottom left of Figure
\ref{fig:cdd} (long dashed cyan line).  The highest $S/N$ observations
yet at $z>5.5$ \citep{bec09} find no weak components, indicating that
a significant fraction of the IGM is not enriched far above $10^{-3}
\Zsolar$.  Data is currently inconclusive as to the extent of volume
enrichment at these low levels, but sitting on the brightest quasar at
$z>6$ for several nights on an 8-10 meter telescope should produce the
necessary $S/N$ of $\sim 40-50$ to adequately test this scenario.
$\CIV$ is by far the best species for testing such a scenario, because
it is the most effective tracer of metals around mean overdensity at
$z=6$; $\SiIV$ is ionized to $\SiV$ at the same overdensities and the
metallicity floor does affect the weak absorber frequency.

Finally, we note a contrast in the $\CIV$ distributions at high-$z$
to the $z=0-0.5$ EW distribution for $\OVI$.  The faint end
turnover of the latter shows very little sensitivity to the volume
filling factor of metals \citep{opp08b}.  Part of the difference
is that $\OVI$ is not tracing the lowest overdensities where metals
reside at low-$z$; the $\CIV$ contours overlap the metals at the
lowest overdensities as shown in Figure \ref{fig:ioncomp}.  This
is precisely the reason why $\CIV$ holds so much potential for
mapping the true extent of high-$z$ metals.

\subsection{Aligned Absorber Ratios}

\begin{figure*}
\includegraphics[scale=0.80]{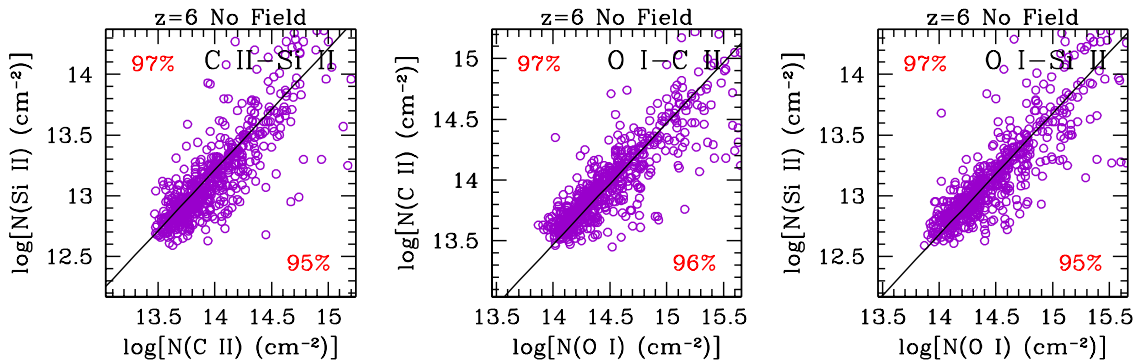}
\caption[]{{\it Aligned absorbers} of different ions are defined as
  having both components above 50 m\AA~and a velocity difference less
  than 15 $\kms$.  Here we present the {\it No Field} case, which at
  $z=6$ shows absorbers fall along a linear relation with some
  scatter; the lines show the expected relation if Type II SNe yields
  \citep[][$Z=10^{-3}$]{chi04} with no ionization corrections assumed.
  Alignment fractions are listed in the corners, where the percentage
  corresponds to the likelihood of finding an absorber of the species
  on the opposite axis aligned with the absorber species of the near
  axis.  The percentages exceed 95\% in every case, and remain
  similarly high when the velocity tolerance is reduced to 8 $\kms$,
  because they are tracing metals in the same neutral gas.  }
\label{fig:align_neutral}
\end{figure*}

Another observationally-accessible statistic is aligned absorber
fractions.  We consider absorbers of different ions to be aligned
if both {\it components} are stronger than 50 m\AA~and are
within 15 $\kms$ of each other ($\delta v \leq 15 \kms$).  We examine
the {\it No Field} first in Figure~\ref{fig:align_neutral}.  The
alignment fractions are listed in the corners, where the percentage
corresponds to the likelihood of finding an absorber of the species
on the opposite axis aligned with the absorber species of the near
axis.  For the neutral case, the alignment percentages are above
95\% in every case.  This lends support to the idea that the low
ionization species observed by BSRS, which appear aligned in every
case for $\OI$ observed above $z=5.8$, are tracing metals where
there is no ionizing flux above the Lyman limit.  These systems
occur far too frequently to be arising within the ISM of intervening
galaxies according to BSRS.  This could be evidence that patchy
reionization is still occurring at these redshifts, or alternatively
that dense cloudlets entrained in winds are self-shielded from the
ionizing background.  The latter scenario would predict that there be high
ionization species in the same systems, possibly offset,
indicating multi-phase outflows as observed
locally~\citep[e.g.][]{str02,mar05b} and seen in galactic-scale
simulations \citep[e.g.][]{fuj08}.  Therefore $\CIV$ and $\SiIV$
could be very useful additional constraints.

The high alignment fractions are helped by the fact that these three
species have similar EWs for typical Type II SNe yields.  Aligned
absorbers fall along the linear relations in each subplot of Figure
\ref{fig:align_neutral}, where the ratios in column densities equal
the ratios in Type II SNe \citet{chi04} yields.  Although we use a
component separation criterion of $\delta v \leq 15 \kms$, the
alignment statistics are nearly identical when $\delta v = 8 \kms$
and $\delta v = 32 \kms$.  This is generally the case in all alignment
statistics considered here, which indicates that the species are
tracing the same underlying gas.  This contrasts with, for instance,
low-$z$ $\OVI-\HI$ alignment fractions which are markedly larger
at higher $\delta v$ \citep{tri08, tho08a}, because $\OVI$ traces
physically distinct gas from the $\lya$-forest \citep{opp08b}.

\begin{figure*}
\includegraphics[scale=0.80]{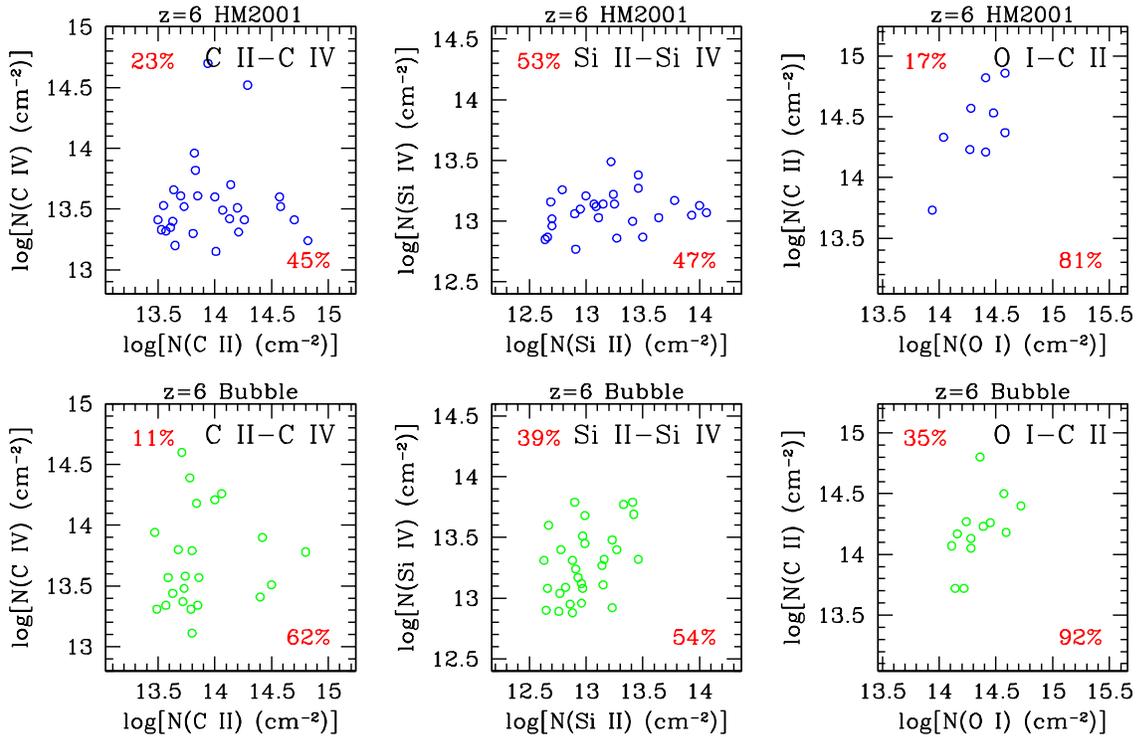}
\caption[]{Aligned absorber ratios and percentages displayed
  analogously to Figure \ref{fig:align_neutral} for the {\it HM2001}
  and {\it Bubble} ionization fields.  Symbols indicate components
  where both absorber species are $>50$ m\AA~and $\delta v<15 \kms$,
  while percentages correspond to the likelihood of finding an
  absorber over 50 m\AA~of the species on the opposite axis aligned
  with the $>50 $ m\AA!absorber species of the near axis.  Aligned
  ratios can help distinguish the shape and magnitude of the
  ionization field, especially in the case of $\CII-\CIV$.  The
  alignment fraction of $\CII$ with $\OI$ is one of the best ways to
  distinguish the three ionization cases, although these are rare if
  the IGM is ionized.}
\label{fig:align_ionize}
\end{figure*}

The alignment between low and high ionization species in Figure
\ref{fig:align_ionize} indicate many fewer aligned absorbers.  $\CII$
is aligned with $\CIV$ 45-62\% of the time, while $\CIV$ is found
aligned 11-23\% of the time with $\CII$.  Finding $\CII$ where $\CIV$
exists should occur much less frequently than the other way around
according to the visualizations in Figures \ref{fig:snaps60} and
\ref{fig:snapsevol}; the small filling factor of $\CII$ is more often
coincident with $\CIV$, while $\CIV$ can also trace the more extended
metals.  This is an ideal ratio to measure the shape of the ionization
background, since abundance ratios are irrelevant and the carbon
spectral ratio that is measured covers a large ionization range from
11.3 to 47.9 eV.  There exist stronger $\CIV$ components aligned with
$\CII$ for the {\it Bubble} field relative to the {\it HM2001}, due to
the greater hardness of the former field.  In the one component
\citet{rya06} could measure both at $z=5.7239$ in SDSS J1030+0524,
they found no aligned $\CII$.  A sample size of at least a dozen
strong $\CIV$ components exploring aligned $\CII$ should begin to
build enough statistics to constrain the shape and intensity of the
ionizing background.

$\SiII$ and $\SiIV$ alignment provides a similar opportunity as the
carbon species to measure the shape of the background, and the
alignment fraction is even higher than for carbon due to the lower
ionization potential of $\SiIII$ (33.5 eV) compared to $\CIII$.  BSRS
actually does find aligned $\SiIV$ in three of their components lying
close together in the $z=5.3364$ system of SDSS J0231-0728.  This
system indicates possible non-uniform ionization, because the very
aligned profiles of $\CII$, $\OI$, and $\SiII$ in components about 100
$\kms$ away from the aligned $\SiII-\SiIV$ is a strong indicator of
the {\it No Field} case, while $\SiIV$ requires an ionizing field.
BSRS mentions that aligned $\CIV$ is also present according to
\citet{bar03}, but the low resolution of that observation makes it
difficult to determine the component alignment; we expect it to be
aligned to the components exhibiting the $\SiIV$.  The presence of low
and high ions nearby suggests a galactic ISM as might be seen in a
damped Ly$\alpha$ system, as forwarded by BSRS.

$\CII$ is usually found with $\OI$, but the reverse is not true if
there is an ionization field above the Lyman limit.  The bluest
component of the $z=5.3364$ BSRS system mentioned above is consistent
with a non-detection of $\OI$ with the observed $\CII$ and $\SiII$,
supporting some sort of ionization field.  Finding the fraction of
$\CII$ absorbers with aligned $\OI$ is one of the most sensitive
indicators of the three different ionizing cases we explore.

Finally, we do not show the incidence of $\OI$ aligned with $\CIV$,
because this case rarely occurs under an ionizing field due to the
large difference in ionization potentials.  This is apparent in SDSS
J1148+5251 as \citet{bec09} finds no $\CIV$ despite the 4 $\OI$
absorbers (BSRS).

Overall, the alignment fractions offer another way to constrain the
nature of the ionization field.  Current data are in line with
expectations from our models, though this is not highly constraining.
Larger samples will elucidate the nature of individual absorbers,
and possibly allow one to measure the carbon spectral ratio which
could give clues as to the nature of ionizing sources.

\subsection{System Profiles}

\begin{figure} \includegraphics[scale=1.00]{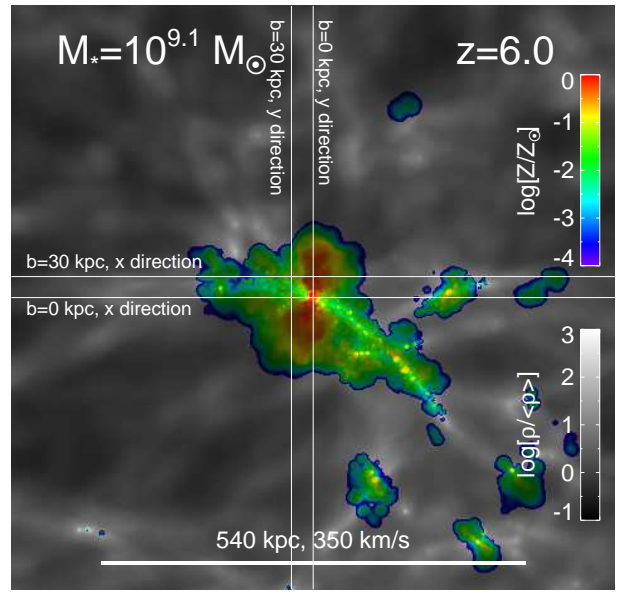}
  \caption[]{The $4\times 4 \hmpc \times 25 \kms$ region shown in
    Figure \ref{fig:snapsevol}~at $z=6$ with intersecting LOSs, of
    which we show the absorption line profiles in Figures
    \ref{fig:sysspec_b0} and \ref{fig:sysspec_b30}.  This snapshot
    spans 950 physical kpc, and the Hubble parameter is 651
    $\kms$ Mpc$^{-1}$.  }
\label{fig:snaps_z60_gal2}
\end{figure}

\begin{figure*}
\includegraphics[scale=0.7,angle=-90]{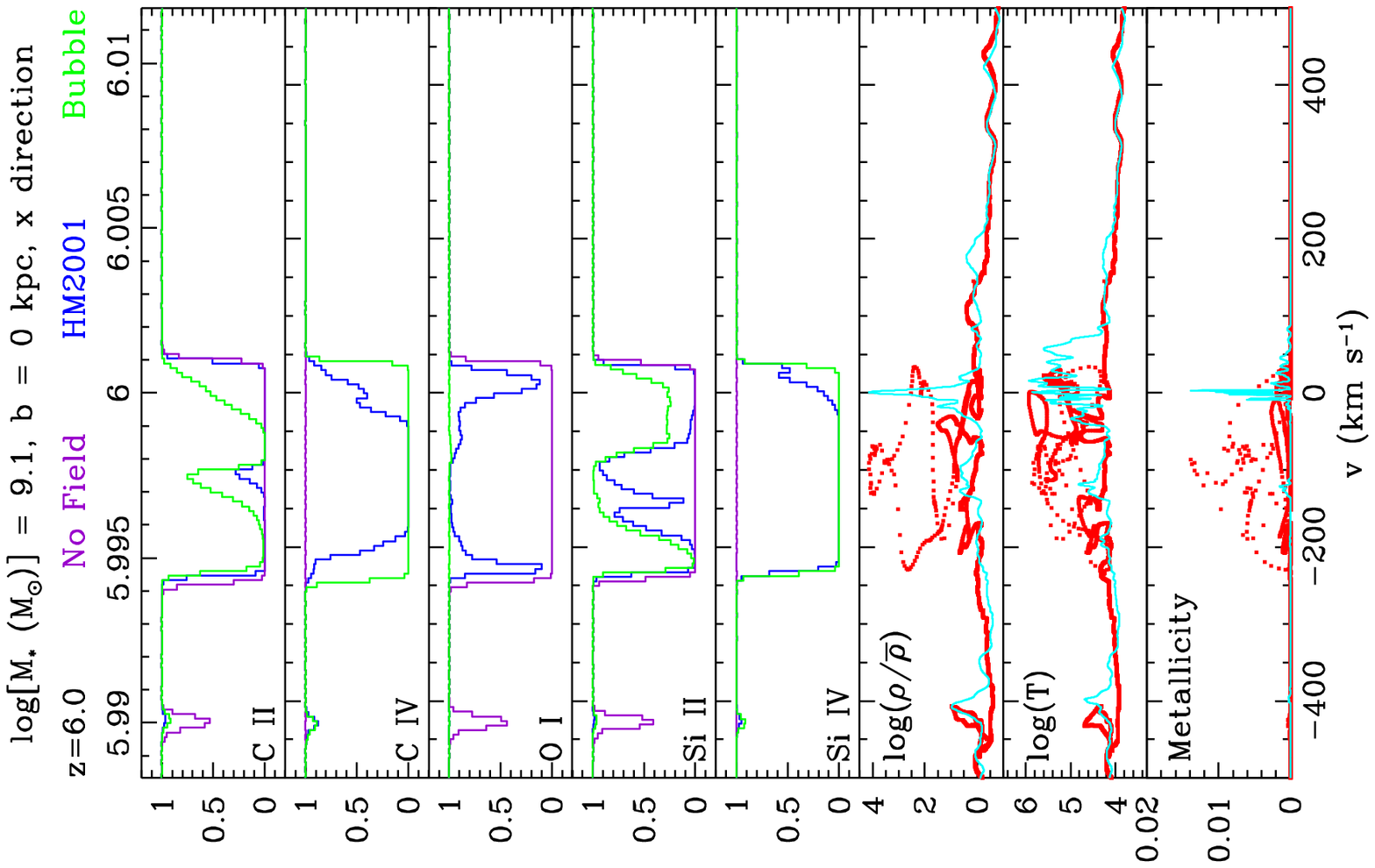}
\includegraphics[scale=0.7,angle=-90]{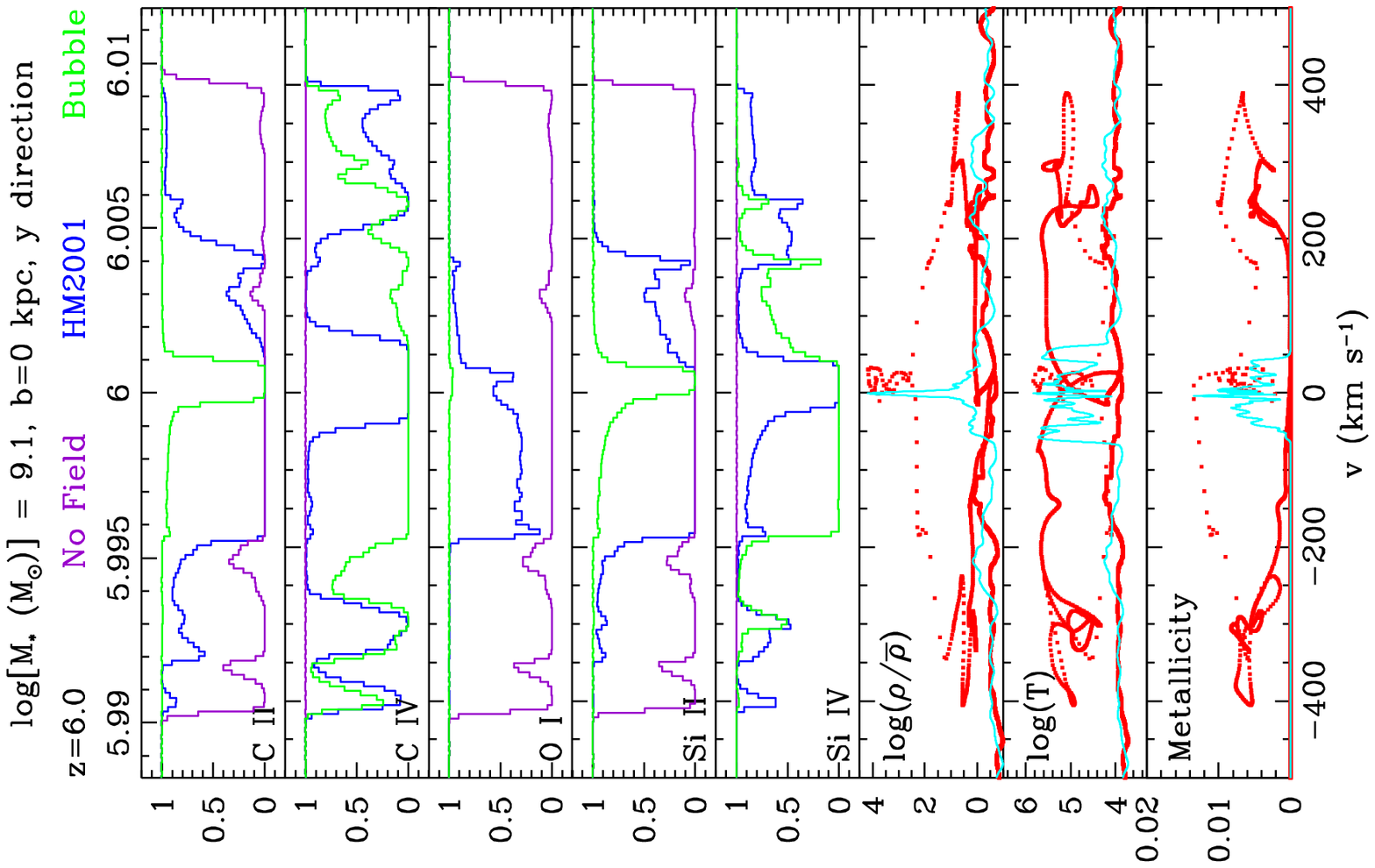}
\caption[]{The LOSs in Figure \ref{fig:snaps_z60_gal2} intersecting
  the center of the $M_*=10^{9.1} \msolar$ galaxy.  The $y$ LOS
  (right) shows more extended absorption line profiles ($\delta v\sim
  800 \kms$), because it is along the bipolar outflow axis, while the
  $x$ LOS (left) follows more closely the local filament.  No systems
  in the data or in our 70 simulated LOSs match such profiles, because
  intersecting such regions is extremely rare when considering quasar
  LOSs, which probe a volume-weighted sample.  The blue-shifted side
  of these profiles may be more appropriate when considering the
  absorption of the outflows and galactic ISM in the spectra of LBGs
  and high-$z$ galaxies.  Lower ionization species show broader
  profiles in the {\it HM2001} case versus the {\it Bubble} case.}
\label{fig:sysspec_b0}
\end{figure*}

Finally in this section, we consider the appearance of systems in
absorption line spectra.  As an illustration, we choose a series of
LOSs through the region indicated in Figure \ref{fig:snaps_z60_gal2},
which is the same region shown evolving in Figure \ref{fig:snapsevol}
centered on the second most massive galaxy in the simulation box.  We
choose two LOSs that go exactly through the center of this
$M_{*}=10^{9.1} \msolar$ galaxy and show the system profiles for our
five ions and our three fields in Figure \ref{fig:sysspec_b0}.  The
$x$-direction spectrum is more along the direction of the local
filament, while the $y$-direction is nearly along the axis of the
bipolar outflow.  Both show extremely strong systems, but the $y$
spectrum is remarkably wide (800 $\kms$) owing to peculiar
velocities imprinted by the enriching outflows.
Comparing the physical parameters of the gas in the bottom three
panels without peculiar velocities (cyan) and with peculiar velocities
(red) shows that the velocity structure dominates the system profiles
of both LOSs.  This galaxy has a maximum $\vw$ just over $500 \kms$,
therefore $\delta v\sim 400 \kms$ on each side of the $y$ spectrum is
sensible as winds are launched with an intentional spread of
velocities, plus the LOS is slightly offset from the outflowing axis.

None of the observed systems in the data (or in our 70 LOSs) have
profiles as strong and as wide as the systems in Figure
\ref{fig:sysspec_b0}, because not surprisingly lines of sight
directly though the centers of galaxies are rare.  If we assume
similar profiles arise within 3 kpc of the 10 galaxies in our
simulation box over $M_*=5\times10^{8} \msolar$, then the frequency
of intersecting such a system occurs once in $\Delta z = 2000$!  A
more relevant consideration for the spectra in Figure \ref{fig:sysspec_b0}
is to take the blue-shifted half and model the outflows in the
spectra of LBGs at lower redshift ($z\sim2-3$), where such galactic
superwinds have been observed to be common and observations are
more obtainable \citep{pet01, sha03}; this is beyond the scope of
this paper.  We note that the faster moving materials absorb more
in the lower ionization species with the weaker {\it HM2001}
background.

\begin{figure*}
\includegraphics[scale=0.7,angle=-90]{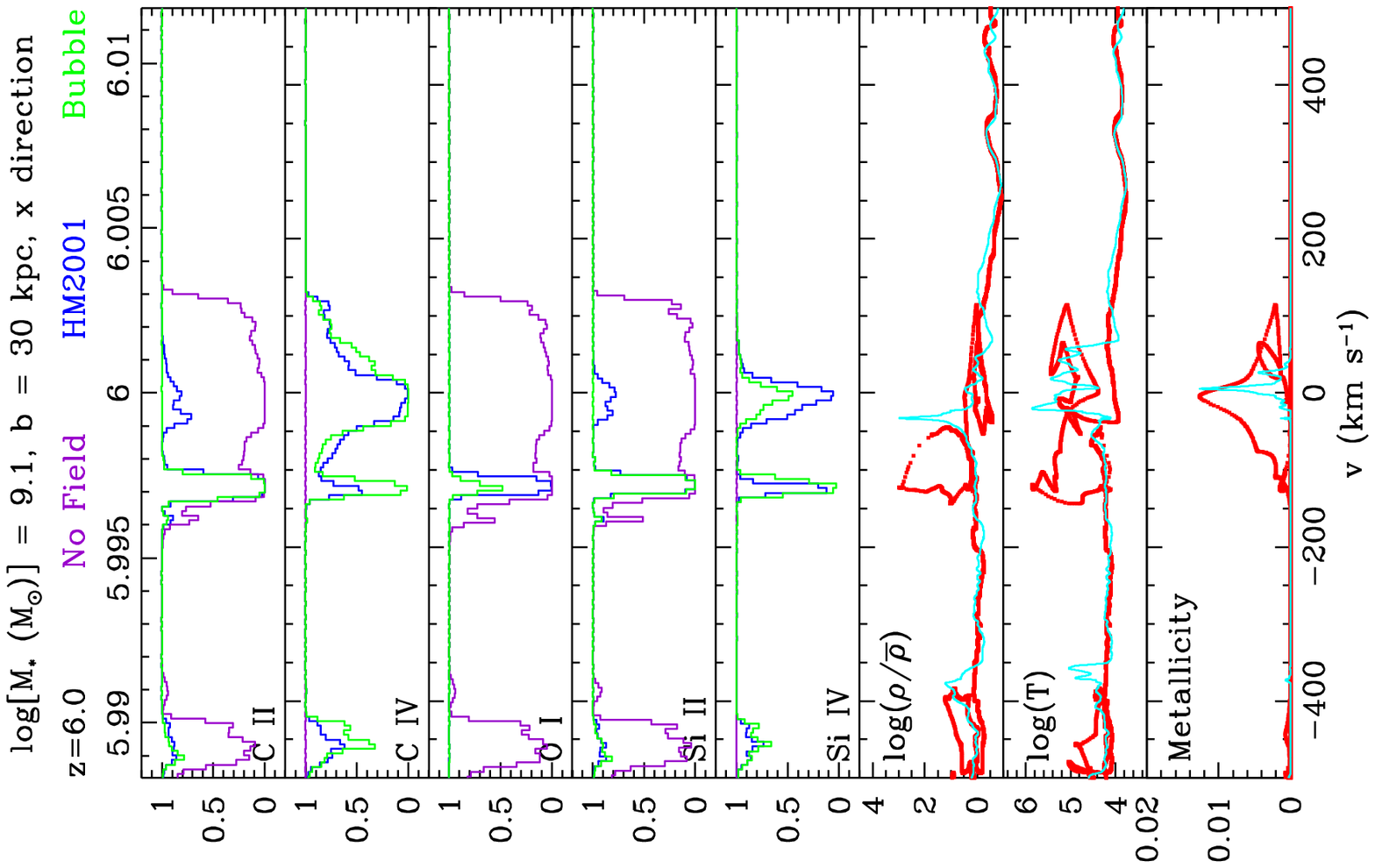}
\includegraphics[scale=0.7,angle=-90]{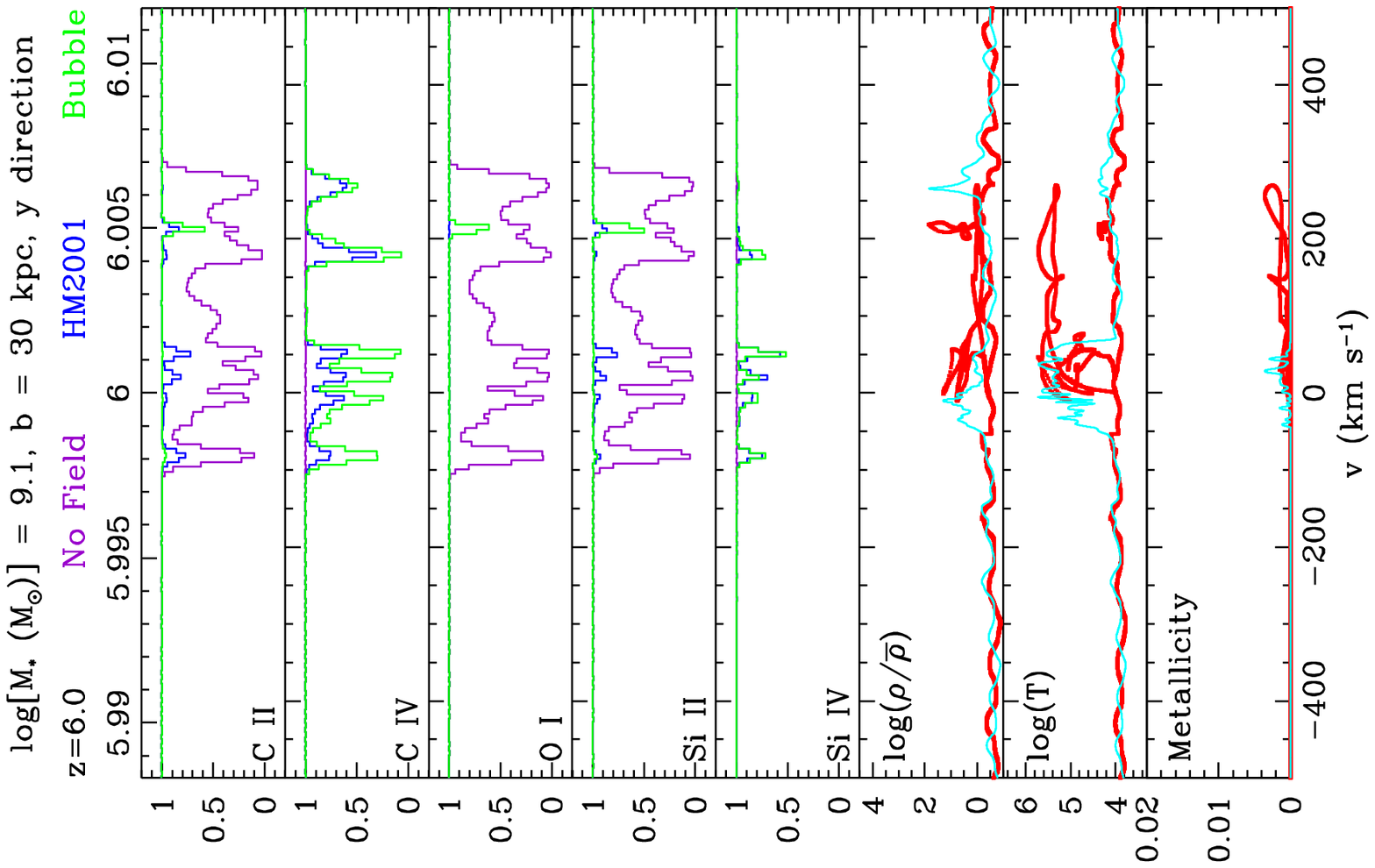}
\caption[]{The two LOSs illustrated in Figure
  \ref{fig:snaps_z60_gal2}~with impact parameters of 30 kpc.  As in
  Figure \ref{fig:sysspec_b0}~peculiar velocities dominate the system
  width, and the $y$ LOS (right) exhibits a more extended
  profile as this is close to the axis of the bipolar outflow.  These
  systems are more similar to those observed in quasar LOSs,
  and could be analogous to some of the broader systems observed by
  BSRS and \citet{sim06b}.  }
\label{fig:sysspec_b30}
\end{figure*}

More typical examples of quasar absorption line systems are shown
in Figure \ref{fig:sysspec_b30} where two LOSs have impact parameters,
$b$, of 30 kpc to the same galaxy.  Velocity profiles of absorbing
ions are still dominated by peculiar velocities, and the LOS closest
to the outflow axis ($y$ direction) produces a wider system.  This
system is perhaps most qualitatively similar to the two wide systems
observed in low ionization species by BSRS where $\CII$, $\SiII$,
and in one case $\OI$ show a total velocity spread of $\delta
v\sim250 \kms$.  This supports the idea of neutral/low-ionization
gas entrained in outflows, self-shielded from the ionization
background; this probably makes more sense for the $z=5.3364$ system
than patchy reionization extending to this redshift.  A similar
velocity spread is observed by \citet{sim06b} in his $\CIV$ $z=5.829$
system.

Although an alignment within 30 kpcs occurs once in $\Delta z=20$
(if we count the ten most massive galaxies in the d16n256vzw box),
we will argue in \S\ref{sec:galabs} that such systems are qualitatively
similar to systems observed around lower mass galaxies at lower
impact parameters.  The frequency of intersecting such qualitatively
similar systems as those shown in Figure \ref{fig:sysspec_b30} is
$\Delta z\sim 2.8$ if we consider all galaxies $M_*>10^{7} \msolar$,
and $d_{gal}<30 \sqrt\frac{M_{*}}{10^{9}}$.  Two such systems should
have been intersected by now given the combined pathlength of
\citet{bec09} \& \citet{rya09}, which compares well to three strong
$\CIV$ systems being observed.  Three differences when scaling to
lower stellar mass are (i) $\delta v$ of systems are smaller as the
outflow velocities are smaller, (ii) metallicities will be lower
due to the mass-metallicity relation, and (iii) the mass loading
factors should be higher at lower mass, which counterbalances (ii).

\begin{figure}
\includegraphics[scale=0.9]{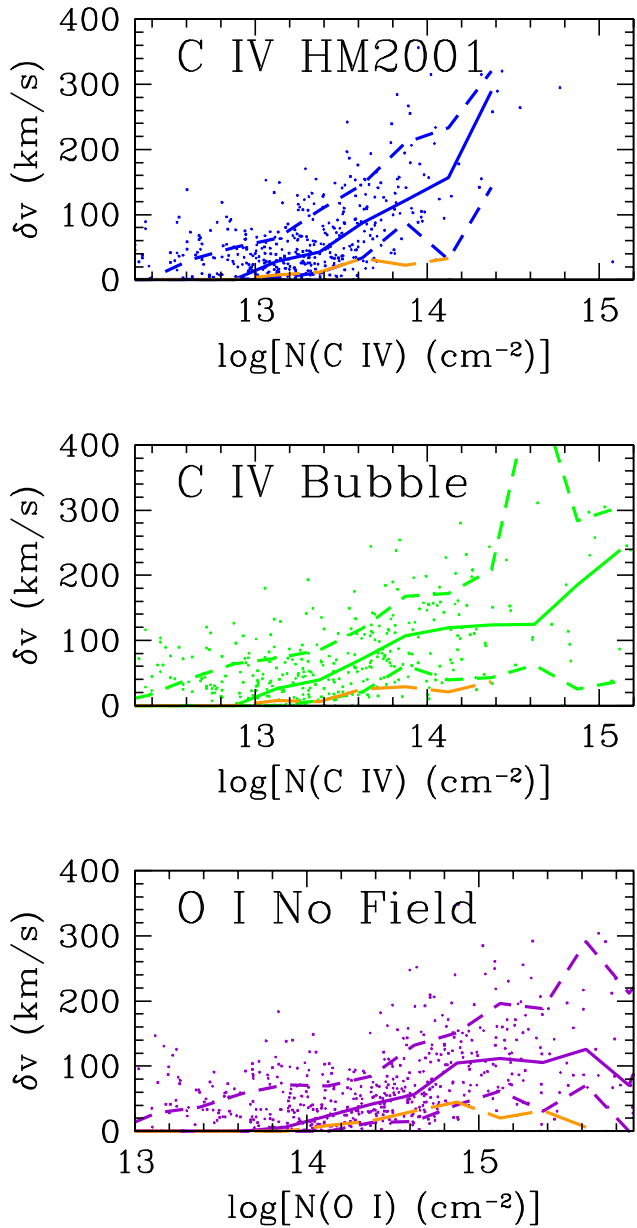}
\caption[]{System width ($\delta v$) versus column density for $\CIV$
{\it HM2001} and {\it Bubble} fields (top and middle), and the $\OI$
{\it No Field} case (bottom).  Solid lines indicate the median and
dashed lines show 1 $\sigma$ dispersions.  $\delta v$ is defined as
the velocity difference between the central velocities of the two
furthest components in a system.  The orange long-dashed lines show
$\delta v$ if we make our LOSs without peculiar velocities
(i.e. the spatial distribution contributes to the system width).  The
majority of the system width is due to peculiar velocities, where
outflowing metals broaden the profile.}
\label{fig:syswidth}
\end{figure}

In Figure \ref{fig:syswidth} we show how median $\delta v$ varies
as a function of column density in our 70 LOS $z=6$ sample.  Here
$\delta v$ is defined as the difference between the central velocities
of the two furthest components in a system ($\delta v=0$ when there
is a one component system ).  One $\sigma$ boundaries (dashed lines)
show the range of velocity spreads in each case.  This can be
compared to a sample of the same spectra without peculiar velocities,
shown as long dashed orange lines.  While this is not strictly a
fair comparison as systems without peculiar velocities will not
necessarily fall into the same column density bin, this nevertheless
clearly illustrates that the stronger observed systems are dominated
by peculiar velocities.  At high column densities, greater than
50\% of a system's width owes to the peculiar velocities from
outflows.  With future samples, such a plot can be used to infer
conclusions about the imprints of outflows on high-$z$ metal-line
profiles.  We do not make a direct comparison now, because of the
paucity of observed systems and differences in both $S/N$ and
resolution, which may affect such a comparison.

The main points here are that (i) we produce systems
qualitatively similar to those observed in the data at a frequency
that seems reasonable; and (ii) the velocity widths of the strongest
observed systems are likely to be dominated by peculiar velocities
associated with outflowing gas.  The latter suggests that plots
of $\delta v$ vs. column density may provide constraints on outflow
kinematics from high-$z$ galaxies.

\section{The Physical Environment of Absorbers}\label{sec:physenv}

As our simulations form galaxies and enrich the IGM together, we
are able explore the relationship between galaxies and metal
absorbers, along with their physical and environmental properties.
{\tt specexbin} tracks for each absorber the physical properties
of density, temperature, and metallicity; the environmental properties
of the distance and stellar mass of the nearest galaxy; plus the
last time an SPH particle was launched in a wind (i.e. its ``age").
We also track an SPH particle's SFR if it is in a galaxy, or $\vw$
at which it was launched if in an outflow.  Unless otherwise
mentioned, we use the {\it Bubble} model, because (i) the ionizing
flux from the local galaxy should dominate at the distance of the
metals given the reasonable assumptions for the transmission above
the Lyman limit, and (ii) this field produces a shallower column
density distribution, which seems to be supported by the data.
Note, however, that the {\it No Field} approximation may best explain
some absorbers, most notably those in the J1148+5251 LOS.  This may
help remedy the overestimated {\it Bubble} $\Omega(\CIV)$ if more
metals are at lower ionization states as we suggest in
\S\ref{sec:masslimit}.  We also address the physical environment
of the {\it HM2001} field, which may be most appropriate at $z<6$,
in \S\ref{sec:physenvuniform}

We concentrate on $\CIV$ absorbers at $z=6$, examining first their
physical state and then their evolutionary status, and bring them
together by exploring the high-$z$ galaxy-absorber connection.  We
consider an absorber's connection to the stellar mass of its
originating galaxy, which in our models is a good proxy for the
more observationally-accessible SFR.  This leads us to a suggestion
of possibilities of why we may produce too many weak absorbers at
high-$z$.  We then switch to lower ionization species represented
by $\CII$, and lastly extend our analysis back to $z=8$.

\subsection{Physical Conditions}\label{sec:physical}

\begin{figure*}
\includegraphics[scale=0.80]{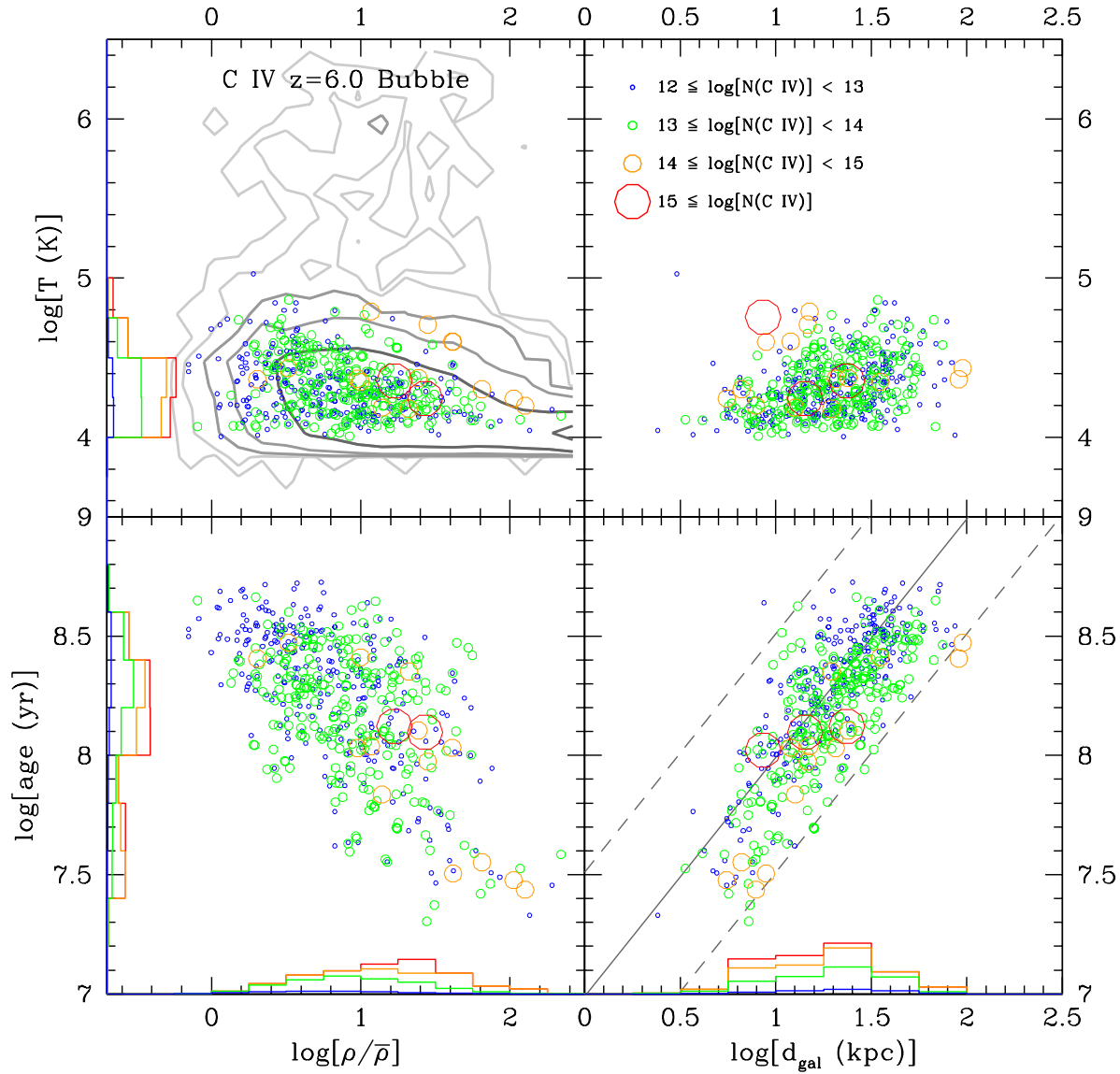}
\caption[]{{\it Physical conditions of $\CIV$:} $\CIV$ absorbers at
  $z=6$ with the variable {\it Bubble} field are plotted in planes of
  density, temperature, absorber age, and distance to the nearest
  galaxy.  Logarithmic contours at 0.5 dex steps in the $\rho-T$ phase
  space correspond to the metallicity-weighted density (darker
  contours are greater densities).  In the $d_{gal}$-age plane, an
  effective outflow velocity can be calculated by dividing the two
  quantities; we show a solid gray line corresponding to $v_{\rm
  eff}=100 \kms$, with the 30 and 300 $\kms$ indicated by the dashed
  gray lines to the left and right.  Histograms along the side show
  the summed $\Sigma N(\CIV)$, with each color corresponding to the
  column density range indicated by the key in the upper right panel;
  all absorbers where $N(\CIV)>10^{15} \cms$ are counted as
  contributing only $10^{15} \cms$ to this sum due to these lines
  being saturated.  Histograms show stats in 70 simulated LOSs, but
  points are plotted for absorbers in 30 LOSs.  $z=6$ $\CIV$ is
  predominantly photo-ionized, tracing a range between $\delta=0-100$.
  Stronger absorbers arise from higher overdensities and are generally
  younger.  Most absorbers are $30-300$~Myr old, lying $5-50$ kpc from
  their parent galaxy.  The average $v_{\rm eff}\sim 100\kms$ far from
  galaxies, which is half as much as the typical ejection speed $\vw$.
  }
\label{fig:c4_z6_4panel}
\end{figure*}

Figure \ref{fig:c4_z6_4panel} shows four absorber properties plotted
against each other for $z=6$ $\CIV$ absorbers in the {\it Bubble}
field.  In the upper left panel are the same type of logarithmic
metallicity contours considered in Figure \ref{fig:ioncomp}.  $\CIV$
has the smallest global ionization correction at $z\sim 5-6$, because
the ionization fractions of this species are ideally aligned with the
range of densities at which metals reside at this redshift
($\delta\sim 0-100$).  Histograms in the bottom left panel show that
stronger absorbers trace higher overdensities.

Temperatures indicate predominantly photo-ionized $\CIV$.  Outflows
typically heat gas to $10^{5-6}$~K around galaxies, but the gas
quickly cools to warm IGM temperatures due to efficient cooling in
this range.  This appears to contradict our previous work in OD06,
where we claimed a significant fraction of $z\sim 6$ $\CIV$ (43\%) is
collisionally ionized, but the difference arises because in OD06 we
quoted a mass-weighted fraction (from particles), while here we are
computing a volume-weighted quantity (from spectra).  For
$\delta<200$, our present models agree well with OD06, comparing
Figure 9 in OD06 with the upper left of Figure \ref{fig:c4_z6_4panel}.
Regions with $\delta>200$ are mostly forming stars at $z=6$, and the
frequency of metal systems where $N(\CIV)\ge 10^{13} \cms$ includes
SPH particles above the SF threshold is $0.4 \Delta z^{-1}$.  Of the
54 components with $N(\CIV)>10^{14} \cms$, only 12 have any traces of
SF.  Ironically, only one of these 12 lines has its line center
density at $\delta>200$; even when LOSs probe galactic environments,
the $\CIV$ absorbing gas is primarily intergalactic\footnote{The one
  absorber tracing primarily galactic gas has $\delta\sim 1000$ and
  $N(\CIV)=10^{17.5} \cms$!  This may be an unphysical amount of
  $\CIV$, but this one absorber comprises 88\% of the total
  $\Omega(\CIV)$ along all LOSs.  This example illustrates the reason
  for the discrepancy between the volume-derived $\Omega(\CIV)$ of
  $N(\CIV)=10^{13.0-15.0} \cms$ absorbers versus the mass-derived
  $\Omega(\CIV)$, as discussed in \S\ref{sec:omega}.}.  Hence our
models clearly predict that the vast majority of typical $\CIV$
absorbers are primarily intergalactic at $z=6$, and the rest mostly
trace halo outskirts.

The average carbon metallicities at $N(\CIV)=10^{13.0}$ and $10^{14.0}
\cms$ are $0.05$ and $0.10$ $\Zsolar$ respectively.  This is about
$\sim \times 15-20$ the IGM metallicities at the corresponding
overdensities, using the simulation-averaged metallicity-density
relation in Figure~\ref{fig:rho_Z}.  Hence $\CIV$ absorbers are
tracing a clumpy, inhomogeneous distribution of metals with a very
low volume filling factor, $\la 0.1\%$.  This is generic in our
models, and is even seen at the present day for $\OVI$ absorbers
~\citep{opp08b}.  The metallicity inhomogeneity is even more extreme
at $z=8$: $0.06 \Zsolar$ for $N(\CIV)=10^{13.5-14.0} \cms$, while
the simulation-averaged metallicity is $9\times10^{-4} \Zsolar$ (a
factor of $\sim 70$).  This shows that metal-line absorbers trace
an increasingly clumpy distribution of metals at early times going
back from $z=6\rightarrow 8$, a trend that extends to low-$z$
IGM metals as traced by $\OVI$ \citep{opp08b}.

In summary, $\CIV$ absorbers seen at $z\sim 6$ are mostly expected
to be true intergalactic absorbers, not gas within galaxies or
galactic halos.  Galaxy outflows reach a larger comoving volume at
early times (OD08), but in actuality they are enriching a small
fraction of the volume in a highly patchy fashion.

\subsection{Origin of IGM Metals}\label{sec:origin}

We now consider how an absorber's age relates to its environment,
as a way of characterizing in which regions IGM metals originate.
An age-density anti-correlation exists in $z=6$ $\CIV$, with a
significant spread, as shown in the bottom left panel of Figure
\ref{fig:c4_z6_4panel}.  This is the same trend we see in low-$z$
$\OVI$ \citep{opp08b} where stronger absorbers trace higher densities
that have been enriched more recently.  In the case of $\CIV$,
absorbers over $10^{14} \cms$ are more likely to trace metals ejected
on average 100 Myr ago, while weaker absorbers trace metals 100-300
Myr ago from earlier galaxies when they were at lower mass and
metallicity.

The age-density relation can be understood by considering the
age-galaxy distance ($d_{\rm gal}$, shown in physical kpc) relation
in the bottom right panel of Figure \ref{fig:c4_z6_4panel}.  Older
absorbers have had time to travel greater distances reaching lower
overdensities.  Taking $d_{\rm gal}$ divided by age, we can plot
the average velocity these winds need to travel to get as far as
they do, which we call the effective wind speed, $v_{\rm eff}$.
The gray solid line shows $v_{\rm eff}=100 \kms$ and the dashed
lines on the left and right sides are 30 and 300 $\kms$ respectively.
Comparing this to the average wind speed at launch $\vw =200-250
\kms$~(cf. Figure~\ref{fig:sfr_wind}), we see that outflowing gas
typically slows $20-50$\% by the time it reaches the IGM.  As argued
in OD08, this slowing is primarily hydrodynamical as outflows run
into a dense IGM, though gravity plays a non-trivial role.

The good age-$d_{\rm gal}$ correlation indicates that these absorbers
trace metals leaving galaxies for the first time.  This also is
expected from the evolutionary picture of feedback from $z=6\rightarrow
0$ presented in OD08, where we found that metals more often than
not recycle (i.e. return to a galaxy at later time and are shot out
in another wind/form into stars).  Metals recycle on a timescale
of 1-3 Gyr reaching a median farthest distance of $\sim 60-100$
proper kpc from their parent galaxies.  $\CIV$ is tracing metals
on their first journey into the IGM, because the ages and distances
in Figure \ref{fig:c4_z6_4panel} are nearly all less than the
recycling times and turn-around distances.  These absorbers primarily
reside in the IGM and not in halos, because the halo radius
\citep[from][]{nav97} is $\la 5$~physical kpc for all $z=6$ galaxies
in our simulation; nearly all absorbers are at distances above this
value in Figure \ref{fig:c4_z6_4panel}.  Hence $\CIV$ absorbers are
tracing IGM metals on their first journey outward from galaxies at
$z=6$.

\begin{figure*}
\includegraphics[scale=0.80]{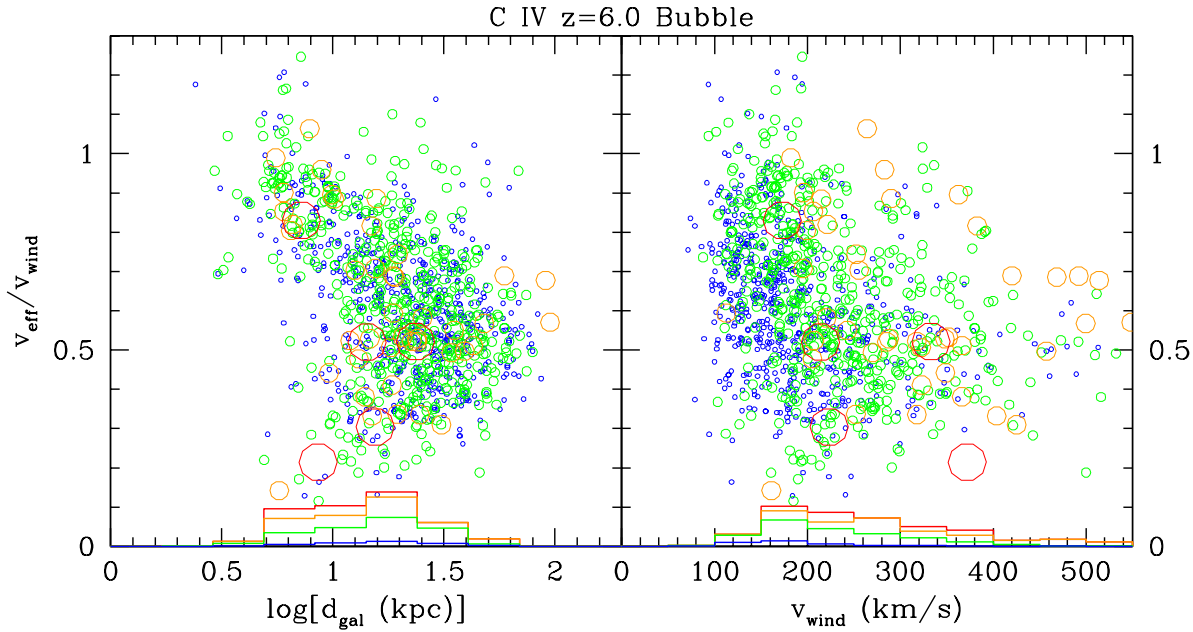}
\caption[]{The ratio of a wind particle's effective velocity
  ($d_{gal}$ divided by age) and launch velocity $\vw$ is plotted
  against $d_{gal}$ (left) and $\vw$ (right).  This ratio generally
  declines the farther it travels from a galaxy as hydrodynamic and
  gravitational forces slow outflowing metals.  Stronger absorbers
  arise from faster winds launched from more massive galaxies, but
  these winds are more likely to slow down faster as indicated by a
  declining $v_{\rm eff}/\vw$ at higher $\vw$.  The color key for
  absorber strength in Figure \ref{fig:c4_z6_4panel} applies here.  }
\label{fig:c4_z6_vw_vratio}
\end{figure*}

As another way of considering absorber origin, we plot the ratio
of the effective velocity to the launch velocity ($v_{\rm eff}/\vw$)
versus the distance from a galaxy (left panel) and the launch
velocity (right panel) in Figure \ref{fig:c4_z6_vw_vratio}.  This
ratio should not exceed one, given that wind particles are slowed
as they travel outward.  A few absorbers at greater than one are
mostly due to the rare instances when the neighboring galaxy is not
the originating galaxy for a wind.  The clearest trend is that
$v_{\rm eff/}\vw$ shows a general decline with $d_{\rm gal}$.  For
absorbers with $N(\CIV)=10^{13.5-14.5} \cms$, the strength of most
components observed by \citet{rya06} and \citet{sim06b}, the median
ratio is nearly unity at 3-5 kpc, 0.8 at 10 kpc, and 0.5 by 50 kpc.
Note that our outflow prescription~\citep[following][]{spr03a} turns
off hydrodynamic forces for a time that does not exceed that required
to cover 4~physical kpc (at $z=6$) at $\vw$.  This shows that, as
soon as hydro forces are turned on, outflows begin slowing down as
they travel further from their parent galaxy.

Several other physically interesting trends are also evident in the
right panel of Figure \ref{fig:c4_z6_vw_vratio}.  First, stronger
absorbers arise from more massive galaxies with greater $\vw$; more
massive galaxies enrich the IGM with higher metallicity outflows
due to the mass-metallicity relationship at high redshift ($Z\propto
M_{gal}^{0.3}$; DFO06).  Second, winds from massive galaxies slow
down quicker as indicated by lower $v_{\rm eff}/\vw$ at higher
$\vw$; this arises because massive galaxies tend to be surrounded
by denser gas (OD08).

\subsection{The Galaxy-Absorber Connection}\label{sec:galabs}

\begin{figure}
\includegraphics[scale=0.80]{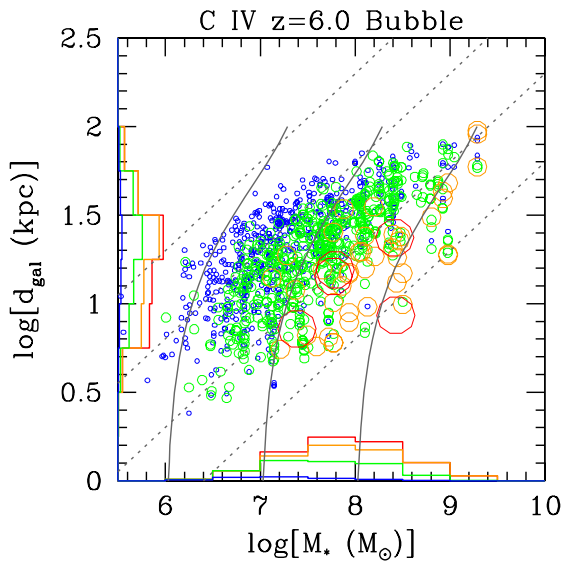}
\caption[]{{\it Galaxy-absorber connection plot:} $d_{gal}$
vs. $M_{*}$ elucidates the fundamental and close connection between
galaxies and absorbers at high-$z$.  It reveals a strong correlation
between $N(\CIV)$ and $d_{gal}$ for a given mass galaxy.  The ionizing
conditions are identical in the {\it Bubble} field along the gray
dotted lines, explaining much of this correlation.  Evolutionary
tracks (solid gray lines for winds emanating from a $M_* = 10^6$,
$10^7$, \&~$10^8 \msolar$ galaxy) tracing an absorber back to its
parent galaxy assuming a constant $v_{\rm eff}=100 \kms$ and secular
galaxy growth.  Currently observable $z=6$ $\CIV$ absorbers are
launched from galaxies between $M_*=10^{6.5-8.0} \msolar$, but are
observed around galaxies with $M_*=10^{7.0-8.5} \msolar$.  The color
key for absorber strength in Figure~\ref{fig:c4_z6_4panel}~applies
here.  }
\label{fig:c4_z6_mgal_dgal}
\end{figure}

While we have explored numerous relations between absorber physical
and environmental parameters at $z=6$, there is as of yet no
over-riding correlation to explain $\CIV$ absorber strengths, as each
parameter exhibits considerable scatter versus $N(\CIV)$.  Here we
argue that the {\it fundamental relation} governing metal absorbers is
the stellar mass, $M_*$, of the neighboring galaxy versus the distance
to that galaxy, $d_{\rm gal}$, and most other correlations are merely
a consequence.  The neighboring galaxy is defined as the galaxy with
the greatest impinging flux on an absorber (i.e. the nearest galaxy
weighted by the inverse square law).  Figure \ref{fig:c4_z6_mgal_dgal}
shows the $M_*-d_{\rm gal}$ relation.  As is evident, the relation
between $d_{\rm gal}$ and $N(\CIV)$ at a given $M_*$ shows a strong
correlation.

We plot gray dotted lines corresponding to the relation $d_{gal}
\propto \sqrt{M_{*}}$ representing the constant flux levels resulting
from the {\it Bubble} field due to the inverse square law.  If all
physical conditions are equal, absorbers along these lines should be
identical.  Indeed, this seems to hold reasonably well.  Hence the key
deterministic relation for absorber strength is distance from galaxy
of a given stellar mass, more so than any of the physical parameters
explored in the previous sections.  The paucity of absorbers below the
bottom-most gray dotted line is only a selection effect poorly
sampling these regions in a volume-averaged measurement; very strong
absorbers exist here if LOSs are fortunate enough to sample them
(e.g. the extreme systems displayed in Figure \ref{fig:sysspec_b0}).

Absorbers are closely related to galaxies at high-$z$ moreso than any
other time because (i) metals are on their first journey into the IGM,
(ii) they are still in close proximity to their parent galaxies, and
(iii) galaxy evolution is relatively simple to understand at high-$z$
as galaxies are mostly evolving in isolation.  We use these
assumptions to construct the solid gray ``evolutionary tracks'' that
metals take into the IGM, assuming constant $v_{\rm eff}= 100 \kms$
and an exponentially increasing $M_*$ with time, which approximately
falls out of the integration of Equation \ref{eqn:SFR_M*}.  Although
both of these assumptions are not exactly true, this treatment reveals
some important behaviors.

First, ``isochrones'' travel parallel to the x-axis of Figure
\ref{fig:c4_z6_mgal_dgal} by construction if $v_{\rm eff}$ is assumed
to be independent of $M_{*}$, meaning the absorbers of the same age
are found at a similar distance.  This of course an
oversimplification, because $v_{\rm eff}$ slows down as time passes
and winds with higher $\vw$ slows down faster.  However, this explains
why there are very few weak absorbers ($N(\CIV)<10^{13} \cms$) around
$M_*>10^{8} \msolar$ galaxies; metals simply have not had time to
travel far enough to the low-density regions where weak $\CIV$
absorbers would arise.

Second, tracing the evolutionary tracks back to the bottom reveals the
mass of the launch galaxy for each metal absorber.  DFO06 plotted the
mass-metallicity relationship of $z=6$ and $z=8$ galaxies, finding
that gas metallicity increases by $\sim 0.3$ dex per decade of $M_*$.
Although the over-riding trend for absorber strength depends on the
ionizing flux in the {\it Bubble} field, a more subtle trend (hardly
visible) is that absorbers grow stronger rightward along the same
dotted gray line primarily as a result of the mass-metallicity
relationship.  However, this effect is about half as strong as
expected ($\sim 0.15$ dex), and should be counter-balanced by
increased mass loading factors of smaller galaxies.  Also, consider
that $v_{\rm eff}$ is not really constant, which limits us from
over-interpreting this plot.

The only candidate observed high-$z$ galaxy-$\CIV$ absorber connection
is along the J1030+0524 LOS, where \citet{bec09} notes the observed
systems at $z=5.7- 5.9$ may be related to an excess of $i$-band
dropouts observed by \citet{kim08} along this quasar LOS.  In our
models, most absorbers between $N(\CIV)=10^{13.0-15.0} \cms$ are
associated with galaxies where $M_*=10^{7.0-8.5} \msolar$, which
have corresponding 1350 \AA~AB magnitudes ranging between $-15.4$
and $-18.9$ (see the luminosity range bounded by green dotted lines
in Figure \ref{fig:lumfunc}), and SFRs from $0.05-1 \msolar {\rm
yr}^{-1}$.  The closest of the J1030 absorbers lies about 50
arcseconds away from a galaxy, i.e. an impact parameter of 300 kpc,
which is too far for our winds to reach (e.g. $v_{\rm eff}\ga 1000
\kms$ for 300 Myr); plus the magnitudes of these galaxies appear
to be brighter than any of the galaxies in the d16n256vzw simulation
box.  This may be an extended overdense region of space where
smaller, currently unobservable galaxies are primarily responsible
for the enrichment.


\subsubsection{HM2001 Background Case} \label{sec:physenvuniform}

\begin{figure}
\includegraphics[scale=0.80]{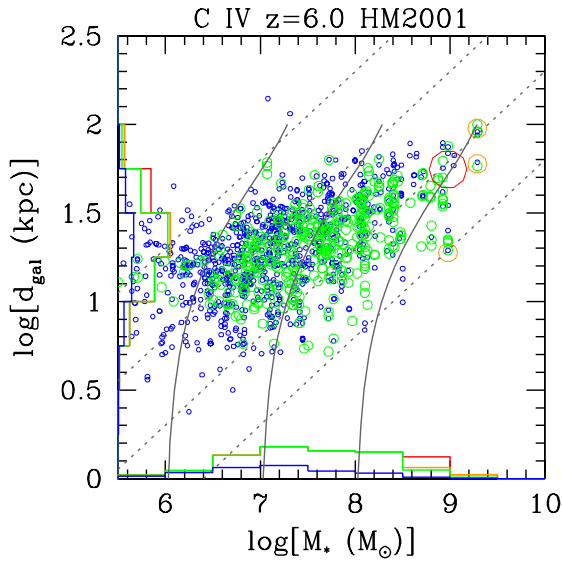}
\caption[]{The galaxy-absorber plot for the uniform {\it HM2001}
  model, which is analogous to Figure \ref{fig:c4_z6_mgal_dgal} for te
  {\it Bubble} model.  Here the gray dotted lines do not have any
  significance, but are shown for reference.  Fewer strong absorbers
  exist near galaxies and more weak absorbers exist far from galaxies
  due to the weaker ionizing field in such environments.  One
  important difference with the {\it Bubble} field is that strong
  absorbers are only associated with $M_{*}\sim 10^{9} \msolar$
  galaxies, and these absorbers trace enriched low-density regions
  further from galaxies.  The color key for absorber strength in
  Figure~\ref{fig:c4_z6_4panel}~applies here.}
\label{fig:c4_z6_mgal_dgal_uniform}
\end{figure}

The uniform {\it HM2001} background case may be more appropriate below
$z=6$, and actually better explains the observed frequency of strong
$\CIV$ absorbers at $\langle z \rangle=5.75$, although it is more
discrepant in over-estimating weaker absorbers.  We show the
galaxy-absorber connection plot in Figure
\ref{fig:c4_z6_mgal_dgal_uniform} for this field and note the
following differences.  First, the distance and mass of the
neighboring galaxy matters less for the characteristics of these
absorbers; the dotted lines do not have any meaning, but are shown for
relative contrasts with Figure \ref{fig:c4_z6_mgal_dgal}.  As this
field is weaker near galaxies, there are fewer strong absorbers
because carbon is in lower ionization states.  For the same reason,
there are more weak $\CIV$ absorbers at greater distances from
galaxies, because these metals are otherwise ionized to $\CV$ by the
{\it Bubble} field.

A potentially accessible distinguishing characteristic of the two
fields involves components with $N(\CIV)> 10^{14.0} \cms$, which are
only found around $M_{*}\sim 10^{9} \msolar$ galaxies in the uniform
case.  These absorbers arise in the situations like those illustrated
in Figure \ref{fig:snaps_z60_gal2}, where a bipolar outflow copiously
enriches the low density IGM outside of filaments where the {\it
HM2001} $\CIV$ ionization correction is lowest.  The strongest
absorbers actually arise from overdensities less than 10, which is the
opposite of the behavior in the {\it Bubble} model.

\subsubsection{Where Are The Weak $z=6$ $\CIV$ Absorbers?} \label{sec:masslimit}

Both our backgrounds produce a significant amount of weak absorbers
($N(\CIV)=10^{13.0-14.0} \cms$), contributing non-trivially to
$\Omega(\CIV)$ (30\% in the {\it Bubble} case and 70\% in the {\it
HM2001} case).  The two observational $z=6$ $\CIV$ CDD constraints
shown in Figure \ref{fig:cdd} (bottom right) suggest a possibly
shallower distribution than either of our models produce, with the
$\CIV$ mass density coming almost exclusively from strong systems.
The \citet{bec09} data point is a conservative upper limit, and already
provides a powerful constraint that we interpret as indicating a
significant volume of the IGM is not enriched far above $Z=10^{-3}
\Zsolar$.  However, if we consider that they observe no $\CIV$
absorbers between $N(\CIV)=10^{13.5-14.0} \cms$ and assume their
completeness is 75\% over this column density range (this seems
reasonably conservative using their Figure 3), the difference with our
models is even more stark.  We predict 3.3 and 3.9 such absorbers
should be detected at $z=5.65$ for the {\it Bubble} and {\it HM2001}
fields respectively over the \citet{bec09} adjusted pathlength of
$\Delta z = 1.65$.

A possible explanation is that the small galaxies in our simulation do
not produce $\CIV$ absorbers for some reason.  The evolutionary tracks
in Figure \ref{fig:c4_z6_mgal_dgal} can be used as a stellar mass cut
for which galaxies produce $\CIV$ systems.  For example, if the
originating $M_*$ must be $\ge 10^{7} \msolar$, then only absorbers
right of the middle evolutionary track contribute.  Systems below
$N(\CIV)=10^{14} \cms$ are more than halved, while leaving the
stronger end unaffected if applying this cut.  The power law slope now
approaches $\alpha=-1$ (i.e. equal numbers of absorbers in low and
high density CDD bins).

Why would low-mass haloes not produce $\CIV$ absorbers?  A possible
explanation is that the highly mass-loaded outflows from small
galaxies self-shield metals leading to them being observed in lower
ionization states.  Thus the over-excess of $\CIV$ absorbers and
$\Omega(\CIV)$ in either ionization background may be counter-balanced
by the BSRS low-ionization systems, which are underestimated when
applying these backgrounds to all metals.    

A second possibility for us overestimating weak $\CIV$ absorbers is
that our outflows are traveling too far into the IGM.  For example,
there exist few very weak, $N(\CIV)< 10^{13} \cms$, absorbers around
galaxies $M_*>10^8 \msolar$ in Figure \ref{fig:c4_z6_mgal_dgal}
because winds have not had time to enrich regions at $d_{gal}\ga 100$
kpc.  If wind speeds are even lower, metals will populate less of the
$M_*-d_{gal}$ phase space at large $d_{gal}$, where they otherwise
would produce weaker $\CIV$ absorbers.

Our discussion in this section on the galaxy-absorber connection
may be currently inaccessible to observations and somewhat speculative,
as only a handful of absorbers at $z\sim 6$ are known thus far and
identifying the nearest galaxy may have to await a next-generation
facility.  Our purpose is to emphasize the $M_*-d_{\rm gal}$ relation
as the key one for understanding the origin and subsequent evolution
of high-$z$ metal-line absorbers.  As these early galaxies push a
majority of their metals into the IGM, the metals are still close
enough that the high-$z$ galaxy-absorber connection tightly establishes
the absorber properties.  This is a fundamental prediction of our
model, and if proved correct, offers a novel way to probe feedback
processes in the reionization epoch.



\subsection{Low Ionization Species}

\begin{figure*}
\includegraphics[scale=0.80]{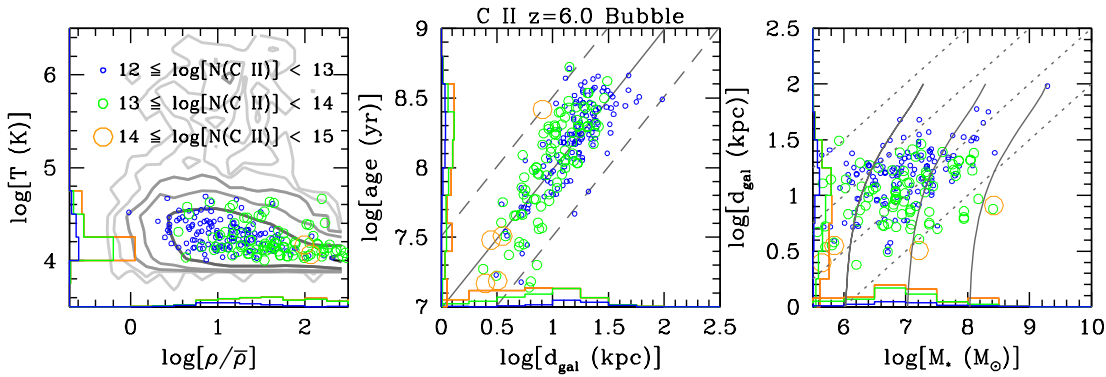}
\caption[]{Physical conditions of $\CII$ absorbers at $z=6$; compare
  to corresponding panels for $\CIV$ in Figure
  \ref{fig:c4_z6_4panel}~and Figure \ref{fig:c4_z6_mgal_dgal}.  $\CII$
  is more dependent on density than $\CIV$ due to rapidly rising
  ionization fractions at higher densities.  $\CII$ absorbers are more
  apt to trace metals immediately outflowing from galaxies, as well as
  the outflows that slow and stall closer to galaxies.  Stronger
  $\CII$ absorbers are found closer to galaxies, and originate from
  galaxies having on average $\times 10$ less stellar mass than that
  for $\CIV$ absorbers.  }
\label{fig:c2_z6}
\end{figure*}

We now examine the physical conditions of $\CII$ absorbers, as a
representative of lower ionization species.  In \S\ref{sec:ionbehave}~we
showed $\CII$ traces higher overdensities closer to galaxies;
therefore we might expect this species to trace metal-enriched
outflows at an earlier stage.  This turns out to be only partially
true.  In Figure \ref{fig:c2_z6}~we plot $\CII$ absorbers in $\rho-T$
phase space demonstrating that $\CII$ absorbers do not trace the
same extent of this phase space as $\CIV$.  Our visualization in
Figure \ref{fig:snaps60} shows $\CII$ tracing less volume than
$\CIV$.  Another difference is there is a stronger correlation with
density for $\CII$, much more so than for $\CIV$.  The reason for
this is that the $\CII$ ionization fraction falls off rapidly toward
lower overdensities as shown in Figure \ref{fig:ioncomp}~while the
$\CIV$ ionization fraction peaks at overdensities corresponding to
intergalactic metals.  $\CII$ absorption strength depends much more
on the physical state (density) than $\CIV$, which depends more on
its environmental state (proximity to and mass \& metallicity of
its parent galaxy).  The ionization field has little effect here,
because density is the primary determinant for $\CII$.

$\CII$ tracing higher overdensities translates into this species
tracing regions closer to the launch galaxy, as the range in $d_{gal}$
in Figure~\ref{fig:c2_z6}~is less than that for $\CIV$.  $\Omega(\CII)$
arises from absorbers at the same ages traced by the $\CIV$ absorbers
(compare y-axis histograms in the center panel of Figure
\ref{fig:c2_z6}~and bottom left panel of \ref{fig:c4_z6_4panel}).
$\CII$ is tracing older metals that remain closer to their parent
galaxies by virtue of either lower $\vw$ or more hydrodynamic
slowing.  The primary reason is the lower $\vw$: the average $\vw$
of a $\CII$ absorber is 150 $\kms$ versus 250 $\kms$, while the
$v_{\rm eff}/\vw$ ratio is not different than that for $\CIV$; Hence
$\CII$ is arising from slower winds around less massive galaxies.
Despite $\CII$ tracing a different set of winds, 62\% of $\CII$
absorbers over 50 m\AA~are aligned with $\CIV$ in the {\it Bubble}
field.  Rare strong $\CII$ absorbers trace some of the youngest
outflows ($10-30$ Myrs), which are sometimes still within the galactic
halos of $z=6$ galaxies.



In the right panel of Figure \ref{fig:c2_z6}, $\CII$ strength is
determined more plainly by $d_{gal}$ independent of $M_*$, which
is a result of this ion's greater dependence on density compared
to $\CIV$.  The galaxy-absorber connection is {\it not} as important
for $\CII$ absorber strength.  If $N(\CII)$ is mostly independent
of $M_*$, the selection effects of LOSs probing a random volume
finds the average $\CII$ absorber around a smaller galaxy.  The
median $M_{*}$ of the $\Sigma N(\CII)$ histogram in the bottom of
this panel is $10^{6.9} \msolar$, corresponding to a SFR of 0.03
$\msolar {\rm yr}^{-1}$.  $\CII$ is probing a different set of
metals than $\CIV$, which itself is tracing faster winds arising
from more massive galaxies ($10^{7.9} \msolar$) enriching a larger
volume.


\subsection{Evolution to $z=8$}

\begin{figure*}
\includegraphics[scale=0.80]{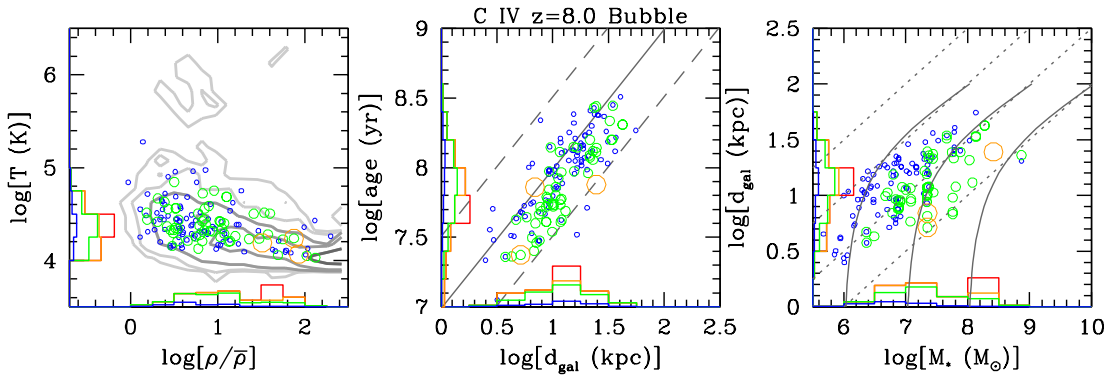}
\caption[]{Similar to Figure \ref{fig:c2_z6}, but for $\CIV$ at $z=8$.
The color key for absorber strength from
Figure~\ref{fig:c4_z6_4panel}~applies here.  See previous figure
captions for an explanation of all contours and lines.  We modify the
evolutionary tracks in the right panel to reflect $v_{\rm eff}=200
\kms$ and double the SFR from Equation \ref{eqn:SFR_M*}~reflecting
galaxy evolution as found in DFO06.  Absorbers still trace the
location of metals well in this {\it Bubble} model, as shown by the
overlap with metal contours in the $\rho-T$ phase space of the left
panel.  The calculated $v_{\rm eff}$ is higher on average in the
central panel than at $z=6$.  The evolutionary tracks in the right
panel indicate that stronger absorbers arise from more massive
galaxies, as at $z=6$.}
\label{fig:c4_z8}
\end{figure*}

We are already delving into the hypothetical here by considering
the physical and environmental parameters of absorbers and galaxies
not even discovered yet, so we take it one step further by considering
what $\CIV$ absorbers trace at $z=8$.  If our suggestion that $\CIV$
absorbers are tracing metals on their first journey into the IGM,
metals should be observed even closer to galaxies at higher redshift.
This indeed is the case in Figure~\ref{fig:c4_z8} for $z=8$ absorbers;
the absorbers occupy slightly higher overdensity (left panel) than
at $z=6$ due to winds not reaching mean overdensities yet, while
absorbers are closer to their parent galaxies as a consequence of
having less time to travel (middle panel).  These absorbers have
$v_{\rm eff}$ more similar to $\vw$ signifying these outflows are
just leaving their galaxy.  In fact, the ratio $v_{\rm eff}/\vw$
for the absorbers between with $N(\CIV)=10^{13.5-14.5} \cms$ for
$d_{gal}=10-20$ kpc is 0.73 at $z=8$ versus 0.62 at $z=6$.  Such
absorbers at $z=8$ trace outflowing gas more directly, whereas at
$z=6$ some absorbers at this distance trace outflows that have
already stalled or are returning to galaxies.

The right panel displays our galaxy-absorber connection plot with
dotted lines corresponding to constant {\it Bubble} ionization flux,
while gray lines are evolutionary tracks assuming $v_{\rm eff}=200
\kms$ and the SFR is doubled from Equation~\ref{eqn:SFR_M*}~as in DFO6 at
$z=8$ versus $z=6$.  The more rapid assembly of galaxies at this epoch
is noticeable by greater curvature in these evolutionary tracks; more
galactic evolution likely has occurred in an absorber's associated
galaxy despite the younger age of the Universe.  The mass-metallicity
holds at this redshift (DFO06) and appears to play a role in the
increasing $N(\CIV)$ for more massive galaxies.


\section{Summary}

We explore high redshift metal-line absorbers using a state-of-the-art,
$2\times512^3$ particle \gad~cosmological simulation enriching the
IGM with star formation-driven galactic outflows.  Our galactic
outflows follow the predicted relations of momentum-driven winds,
where outflow velocity scales with galaxy velocity dispersion and
the mass loading factor scales inversely, leading to higher mass
loading from small galaxies.  Our purpose is to reproduce IGM
metal-line observations at $z\sim 5-6$ while providing predictions
for future observations out to $z=8$, as well as to investigate the
physical nature of high-$z$ metal absorbers and its connection to
galaxies.  We explore five species ($\CII$, $\CIV$, $\OI$, $\SiII$,
\&~$\SiIV$) using three different ionizing backgrounds: fully
neutral, fully reionized, and a patchy bubble model based on the
flux of the nearest galaxy.  Our key results and predictions are
as follows: \\

(i) Metal lines already discovered at $z>5$ are primarily tracing
the diffuse IGM, not gas in galaxies.  Some $\CIV$ lines already
discovered could be tracing metals at overdensities not much higher
than the mean density of the Universe.

(ii) We can reproduce the observed trends of $\Omega(\CIV)$ at
$z\sim 6\rightarrow 4.5$ with steady enrichment of the IGM filling
only 1\% of the IGM with metals by $z=5$.  The recent samples of
\citet{bec09} and \citet{rya09} that find lower $\Omega(\CIV)$s
suggests that the filling factor could be even less.  We also propose
that patchy ionization conditions where some metals are ionized by
their local galaxy and others are not ionized may best explain the
observed trends in both $\CIV$ and lower ionization absorbers.  The
determination of $\Omega(\CIV)$ is also highly dependent on the
shape of the ionizing field (especially above 47.9 eV), SNe yields,
the $\sigma_8$ parameter, and the number of $N(\CIV)<10^{14} \cms$
absorbers.

(iii) The global gas-phase metallicity increases by a factor of $9$
from $z=8\rightarrow 5$.  The evolution of $\Omega(\CIV)$ generally
reflects this change, but the exact amount of evolution depends on
the form of the ionization background. The uniform \citet{haa01}
evolving background results in an increase in $\Omega(\CIV)$ of
$\times 91$, while the {\it Bubble} field shows only a $\times 14$
increase reflecting the actual increase in gas-phase metals.  The
greater ionization intensity around 47.9 eV in the latter background,
which is a result of the dust attenuation law we apply along with
the proximity of metals to galaxies, results in $\times 8$ greater
$\Omega(\CIV)$ at $z=8$.

(iv) $\CIV$ is the ideal tracer for metals at low overdensities at
$z=6$, and its detection holds the best option at high-$z$ to
determine the volume filling factor of metals.  $\CIV$ has the
lowest global ionization correction at $z\sim 5-6$ (i.e.
$\Omega(\CIV)/\Omega($C$)$ is at its maximum).  Adding a metallicity
floor of $10^{-3} \Zsolar$ will have dramatic effects on the numbers
of weak $\CIV$ absorbers, just below thresholds of currently published
results.  Integrating for several nights on a single bright $z>6$
quasar could determine if a significant portion of the IGM volume
is enriched to this level.  The results of \citet{bec09} already
indicate that a significant volume of the IGM is unlikely to be
enriched much above $10^{-3} \Zsolar$, and possibly there exists
too many weak $\CIV$ absorbers in our model without this floor.

(v) The velocity profiles of systems are dominated by the peculiar
velocities of outflowing gas.  We reproduce the large $\delta v$'s
($>200 \kms$) observed in strong $\OI$ and $\CIV$ systems.  The Hubble
expansion of spatially distributed metals alone is far too small to
create these observed system profiles.

(vi) The galaxies responsible for the enrichment of the IGM have
stellar masses in the range of $10^{7.0-8.5} \msolar$ and
corresponding rest-frame UV magnitudes of $-15.4$ and $-18.9$ (1350
\AA~AB) when considering $\CIV$, and a factor of ten less when
considering $\CII$.  Most absorbers should still be near their
originating galaxies (5-50 physical kpc for $z=6$ $\CIV$ and 2-20 kpc
for $\CII$ absorbers).  We stress the need to identify such galaxies,
which will require the likes of JWST and 30-meter telescope
facilities, in order to study the high-$z$ galaxy-absorber connection.
Identifying the environments of strong $\CIV$ components ($\ge 10^{14}
\cms$) can constrain the ionizing source: within $30$ kpc of
$M_*=10^{7.0-9.0} \msolar$ galaxies for the {\it Bubble} field, and at
$\ge 30$ kpc around $M_*\sim 10^{9} \msolar$ galaxies for the {\it
HM2001} field.

(vii) The abundance of $\OI$ absorbers observed by \citet{bec06} in
the SDSS J1148+5251 sight line is consistent with metals residing in a
neutral IGM (as also argued by those authors).  We cannot produce such
a frequency of $\OI$ within our models with ionization extending above
the Lyman limit.  The excellent alignment with $\CII$ and $\SiII$ in
this dataset also strongly supports this.  The fact that most sight
lines show much less $\OI$ suggests the possibility that reionization
is still inhomogeneous over large scales at $z\ga 6$.  It is unlikely
to be an effect of inhomogeneous enrichment, since smaller galaxies
with relatively low clustering lengths are primarily responsible for
enriching the IGM at these epochs.  Another possibility for broad
$\OI$ systems observed down to $z=5.3$ are metals entrained in
outflows, which remain shielded from the ionization background as they
travel into the IGM.

(viii) Aligned absorber ratios are a powerful tool to determine the
physical parameters of the metal-enriched IGM and the nature of the
ionization field.  Aligned absorbers with both components above 50
m\AA~are rare at $z=6$, unless the IGM is neutral.  This makes the
\citet{bec06} finding of frequent low-ionization aligned absorbers all
the more supportive of either a neutral (probably recombined) portion
of the IGM, or metals entrained in self-shielded outflows as the
travel into the IGM.\\

Our analysis in OD08 indicated that most metals injected into the IGM
recycle back into galaxies on a typical timescale of $1-2$ Gyrs.
Hence at $z>5$, metals should primarily be on the outward leg of their
first journey into the IGM.  This indeed turns out to be the case when
we look in detail at the evolutionary state of absorbers.  This has
consequences for the galaxy-absorber connection, in that this
relationship is closest at high-$z$ because metals are still in
relatively close proximity of their parent galaxies.  Absorbers trace
gas on an outward trajectory still holding signatures of their
galactic outflow origin, while possibly being primarily irradiated by
their parent galaxy's radiation field.  The properties of $\CIV$
absorbers depend most on their evolutionary relation to their parent
galaxy (proximity, mass, \& metallicity), while lower ionization
species depend more on the physical state of the gas (density,
primarily).  This is a result of the different $\rho-T$ phase space
traced by these ions.

Another tenet of OD08 is that metals travel a relatively constant
physical distance ($\sim 100$ kpc) from their parent galaxies,
meaning early outflows have the greatest chance of enriching a large
volume of the Universe.  However this journey takes significant
time as wind outflows travel at an average of $\sim 100 \kms$ taking
$\sim 1$ Gyr to get to this distance, or more than the age of any
stars at $z=6$.  Therefore, the volume of the Universe enriched to
significant levels (i.e. 0.01-0.1 $\Zsolar$) grows by an order of
magnitude over the relatively brief time (575 Myr) covered between
$z=8\rightarrow 5$.  This era can be called the ``Age of Volume
Enrichment.''  However metals do not exhibit the expansionist zeal
of the Mongol Empire when it comes to enriching the IGM; the vast
majority of the IGM (as much as 90\%) still remains unenriched in
our simulations run to $z=0$ \citep{opp08b}.

The outlook for the study of very high-$z$ metal lines in the IGM
looks promising as long as sources (quasars and GRBs) are discovered
to provide light bulbs for the near-IR spectrometers that will be
on-line in the near future.  Relatively low resolution and $S/N$
observations can provide key constraints as to the filling factors
of metals, the velocities and mass loading factors of the winds
that put the metals there, and of course the ionization state of
the IGM.  In contrast to metal lines between $z=5\rightarrow 2$,
we expect significant evolution in metal-line properties between
$z=8\rightarrow 5$, a prediction that may have already been observed
by \citet{bec09} and \citet{rya09}.  Further down the line is the
study of the intriguing high-$z$ galaxy-absorber connection, as
this relation is closer than at lower redshift.  The JWST and 20-30
meter class telescopes are required to study this realm as galaxies
at $M_*<10^8 \msolar$ with SFRs below 0.5 $\msolar {\rm yr}^{-1}$,
which are primarily responsible for most IGM metal lines at $z\ge
6$.  To sum it all up, absorption line spectroscopy applied to the
early Universe should yield significant insights into early galaxy
formation, and whether such galaxies enrich the IGM in a manner
similar to lower redshift systems, or in some exotic fashions not
considered here.  We suspect the former.

\section*{Acknowledgments}  \label{sec: ack}

Very useful and encouraging conversations with George Becker, Max
Pettini, Emma Ryan-Weber, and Rob Simcoe were fundamental to this
paper.  We also thank them for reading over this paper and providing
important improvements.  We much appreciate the referee's comments,
which added significantly to the paper.  We additionally thank Karl
Gordon, David Fanning, and Berkeley Zych.  The simulations were run on
the Intel 64 Linux Cluster Abe Supercluster at the National Center for
Supercomputing Applications.  The authors thank R. Kennicutt for
gracious hospitality at the Institute of Astronomy in Cambridge,
England where much of the work on this paper was done.  Support for
this work was provided by NASA through grant number HST-AR-10946 from
the SPACE TELESCOPE SCIENCE INSTITUTE, which is operated by AURA,
Inc. under NASA contract NAS5-26555.  Support for this work, part of
the Spitzer Space Telescope Theoretical Research Program, was also
provided by NASA through a contract issued by the Jet Propulsion
Laboratory, California Institute of Technology under a contract with
NASA. Support was also provided by the National Science Foundation
through grant number DMS-0619881.

\label{lastpage}

\end{document}